\documentclass[useAMS,usenatbib]{mn2e}

\usepackage{amssymb, amsmath}
\usepackage{graphicx}
\usepackage{longtable}
\newcommand{\apj}{ApJ}
\newcommand{\apjs}{ApJS}
\newcommand{\apjl}{ApJL}
\newcommand{\aap}{A{\&}A}

\newcommand{\mnras}{MNRAS}
\newcommand{\aj}{AJ}
\newcommand{\araa}{ARAA}
\newcommand{\pasp}{PASP}
\newcommand{\nat}{Nature}
\newcommand{\apss}{Ap\&SS}

\title[Nuclear Stellar Cusps in LIRGs]{The Build-Up of Nuclear Stellar Cusps in Extreme Starburst Galaxies and Major Mergers}

\author[Haan et al.]{S. Haan$^{1,2}$\thanks{Email: Sebastian.Haan@csiro.au}, L. Armus$^{2}$, J.A. Surace$^{2}$, V. Charmandaris$^{3,4}$, A.S. Evans$^{5,6}$, T. Diaz-Santos$^{2}$,\newauthor J.L. Melbourne$^{7}$, J.M. Mazzarella$^{8}$, J.H. Howell$^{2}$,  S. Stierwalt$^{5}$,  D.C. Kim$^{6}$, T. Vavilkin$^{9}$,\newauthor D.B. Sanders$^{10}$, A. Petric$^{7}$, E.J. Murphy$^{11}$,  R. Braun$^{1}$, C.R. Bridge$^{7}$, H. Inami$^{12}$\\
$^{1}$CSIRO Astronomy and Space Science, ATNF, PO Box 76, Epping 1710, Australia\\
$^{2}$Spitzer Science Center, California Institute of Technology, Pasadena, CA 91125, USA\\
$^{3}$IESL/Foundation for Research and Technology - Hellas, GR-71110, Heraklion, Greece\\ and Chercheur Associ\'{e}, Observatoire de Paris, F-75014, Paris, France\\
$^{4}$Department of Physics and Institute of Theoretical and Computational Physics, University of Crete, GR-71003, Heraklion\\
$^{5}$Department of Astronomy, University of Virginia, P.O. Box 400325, Charlottesville, VA 22904, USA\\
$^{6}$National Radio Astronomy Observatory, 520 Edgemont Road, Charlottesville, VA 22903, USA\\
$^{7}$Department of Astronomy, California Institute of Technology, Pasadena, CA 91125, USA\\
$^{8}$Infrared Processing \& Analysis Center, MS 100-22, California Institute of Technology, Pasadena, CA 91125, USA\\
$^{9}$Department of Physics and Astronomy, SUNY Stony Brook, Stony Brook, NY 11794, USA\\
$^{10}$Institute for Astronomy, University of Hawaii, Honolulu, HI 96822, USA\\
$^{11}$The Observatories, Carnegie Institute of Washington, 813 Santa Barbara Street, Pasadena, CA 91101, USA\\
$^{12}$National Optical Astronomy Observatory, 950 N. Cherry Ave, Tucson, AZ 85719, USA
}
\begin{document}


\pagerange{\pageref{firstpage}--\pageref{lastpage}} \pubyear{2013}

\maketitle

\label{firstpage}

\begin{abstract}

Nuclear stellar cusps are defined as central excess light component in the stellar light profiles of galaxies and are suggested to be stellar relics of intense compact starbursts in the central $\sim$100--500~pc region of gas-rich major mergers. Here we probe the build-up of nuclear cusps during the actual starburst phase for a complete sample of Luminous Infrared Galaxy systems (85 LIRGs, with 11.4$<$log[L$_{IR}/L_{\odot}$]$<$12.5) in the GOALS sample. Cusp properties are derived via 2-dimensional fitting of the nuclear stellar light imaged in the near-infrared by the Hubble Space Telescope and have been combined with mid-IR diagnostics for AGN/starburst characterization. We find that nuclear stellar cusps are resolved in 76\% of LIRGs (merger and non-interacting galaxies). The cusp strength and luminosity increases with far-IR luminosity (excluding AGN) and merger stage, confirming theoretical models that starburst activity is associated with the build-up of nuclear stellar cusps. Evidence for ultra compact nuclear starbursts is found in $\sim$13\% of LIRGs, which have a strong unresolved central NIR light component but no significant contribution of an AGN. The nuclear near-IR surface density (measured within 1~kpc radius) increases by a factor of $\sim$5 towards late merger stages. A careful comparison to local early-type galaxies with comparable masses reveals (a) that local (U)LIRGs have a significantly larger cusp fraction and (b) that the majority of the cusp LIRGs have host galaxy luminosities (H-band) similar to core ellipticals which is roughly one order in magnitude larger than for cusp ellipticals. 
\bigskip
\end{abstract}


\section{Introduction}
\label{intro}

While extreme starburst galaxies and major mergers represent only a small fraction of galaxies observed today, observational and theoretical arguments suggest that most massive galaxies must have evolved through at least one active starburst and merger phase, transforming spirals into massive ellipticals and triggering the most rapid and intense star-formation known in the universe. 
The last decades have brought significant progress in understanding the formation of elliptical galaxies with the following picture emerging: During a merger process, tidal torques drive rapid inflows of gas into the centers of galaxies while violent relaxation acts on stars present in gas-rich progenitor disks producing the remnant spheroid \citep[e.g.,][]{Bar92}. However, the highest phase-space density material can be found in the galactic centers, which comes from gas dissipation in the final, merger-induced central starburst imprinting a central "extra light" component or stellar "cusp" into the surface brightness profiles of merger remnants \citep[e.g.,][]{Kor92, Mih94, Hop08}.\par 

In particular the advent of the Hubble Space Telescope (HST) has brought sufficient angular resolution to reveal the nuclear light profiles in large samples of early-type galaxies and merger remnants. Evidence for a nuclear cusp build-up has been found in a significant fraction of these early-type galaxies and merger remnants \citep{Lau92, Cra93, Jaf94, Fer94, Lau95, Fab97, Sei02, Gra03, Rot04, Cot07, Hop08a, Kor09}. Moreover, these observations suggest that ellipticals can be separated into two different types of nuclear brightness profiles, namely cusp (characterised by a steep inner slope) and core ellipticals (flat profile towards the center).  Follow-up studies have shown a strong dependence on the optical luminosity of the host galaxy, leading to a dichotomy between core and cusp galaxies, with core galaxies dominating at the highest luminosities and cusp galaxies at the lowest \citep{Fab97, Qui00, Rav01, Gra03, Rot04, Lau07}. The finding that giant elliptical galaxies are slow rotating and low-luminosity elliptical galaxies rapidly rotating \citep[see e.g. ][]{Dav83, Ben88, Ems07, Capp07} suggests that the formation of cores and cusps is linked to the dynamics of the merger process or progenitors \citep[see for recent review][]{Kor09}.

The most popular explanation is that cusp galaxies are produced in dissipative, gas-rich (‘wet’) mergers, while core galaxies are produced in dissipationless, gas-poor (‘dry’) mergers \citep{Fab97}.  Several studies suggest that merger progenitors of core ellipticals may have been even different from all galaxies seen today \citep[e.g.,][]{Bui08, Wel08, Naa09}, and are very unlikely cusp ellipticals \citep{Kor09}. Recent observations also indicate that the central stellar light deficit $L_{def}$ in core ellipticals is correlated with the mass of the central stellar BH \citep{Kor09b} and that core and cusp galaxies might follow a different M$_{BH}$-M$_{spheroid}$ relation \citep{Gra12}. This would suggest a slightly different mechanism linking the formation of the central massive black hole with its host galaxy evolution.\par 
However, until recently there has been little observational evidence on the build-up of these cusps during the actual merger-induced starburst stage. While previous ground-based observations \citep[e.g.][]{Wri90, Jam99, Rot04} have revealed that a significant fraction of merger remnants seems to exhibit radial profiles similar to elliptical galaxies based on their large-scale appearance (at radii of several kpc), we are now able to link for the first time the stellar light distribution and merger-induced starburst in the central kpc for a large sample of nearby active starburst mergers.\par

In particular one galaxy population, the Luminous Infrared Galaxies (LIRGs: i.e., $L_{IR} \geq 10^{11} L_{\odot}$\footnote{Infrared Luminosity L$_{IR}$=L(8--1000$\mu$m) based on IRAS 12$\mu$m, 25$\mu$m, 60$\mu$m and 100$\mu$m flux densities as defined in \cite{San96}}), provides a perfect laboratory to test current theories about the formation of nuclear cusps and elliptical galaxies. Although very rare at low-redshift, LIRGs are $\sim$1000 times more abundant and dominate the total IR energy density emitted at red-shifts of z$\sim$1--2 when star formation in the universe was at its peak and most of the galaxies stellar mass is building up \citep{Elb02, LeF05, Cap07, Bri07, Mag09}. Multi-wavelength imaging surveys of local LIRGs and Ultra LIRGs (ULIRGs, $L_{IR} \geq 10^{12} L_{\odot}$) have shown that with increasing IR luminosity, the star formation rate (SFR), the fraction of sources hosting an Active Galactic Nuclei (AGN), and the fraction of sources with interaction features significantly increases \citep{Arm89, Vei95, Vei97, Kim98a, Kim98b, Mur99, Mur01}.  Moreover, the projected nuclear separation in sources with double nuclei is significantly smaller for ULIRGs than for LIRGs \citep{Haa11}, confirming the picture that most ULIRGs evolve at a later merger stage than LIRGs.\par 

However, due to high column densities of gas and dust, it is almost impossible to reveal the circum-nuclear regions and to identify the nuclei in LIRGs at optical wavelengths. High resolution rest-frame near-infrared (NIR) imaging of their nuclear region with HST has revealed that the majority of (U)LIRGs with $L_{IR} \geq 10^{11} L_{\odot}$ have double nuclei, nearly two times more than those identified in the rest-frame optical light \citep{Haa11}. This, without even considering issues such as sensitivity and spatial resolution, implies strong limitations on the ability to detect the true nuclear structure and merger stage of LIRGs at higher redshifts (z $\gtrsim$ 1.5), even if observed in the NIR. Although (U)LIRGs at high redshift likely differ in some of their properties from local (U)LIRGs (e.g. in  merger fraction, size of starburst, SFR efficiencies etc.) due to higher gas-fractions and smaller galaxy separations at earlier epochs, high-resolution observations of local ULIRGs still provide the best laboratory to study the nature of these galaxies.


Despite recent progress in simulating the formation of nuclear cusps via merging of gas-rich progenitor galaxies \citep[e.g.][]{Hop09a, Hop09b}, several important questions are still open: 1)  How much stellar mass is typically built up in nuclear cusps and what are the critical time-scales for cusp formation? 2) Is there a link between the strength of cusps in LIRGs and the current rate of star formation in these galaxies? 3) Is the fraction of cusps in LIRGs the same as in elliptical galaxies? 
Do LIRGs display a similar bimodality between the presence of cusps and the stellar mass of the host galaxy and what are the implications for the formation and evolution of elliptical galaxies and spheroids? 4) Are nuclear cusps present only in merger-induced starburst galaxies or can cusps also form in isolated galaxies? If so, what is the mechanism that triggers the gas-dissipation and nuclear starburst in non-interacting galaxies with no apparent interaction features in the NIR and optical light \citep{Haa11, Kim12}?\par 

The Great Observatories All-sky LIRG Survey \citep[GOALS; see][]{Arm09} provides an excellent sample to address these questions, combining high-resolution multi-wavelength HST observations (NICMOS/WFC3 H-band, ACS I- and B-band) with mid and far-IR diagnostics (Spitzer IRS, IRAC, MIPS, HERSCHEL PACS, SPIRE) for a complete sample of (U)LIRGs in the local universe. Moreover, a wealth of additional multi-wavelength data is available, ranging from X-ray (Chandra), UV (GALEX) to radio (JVLA, ATCA) and sub-mm observations (CARMA, ALMA).\par
Our previous GOALS HST NICMOS study \citep{Haa11} has shown a tremendous increase of the bulge compactness as a function of merger stage, suggesting that cusp formation might play a significant role in these galaxies. Here we combine our previous HST NICMOS H-band sample with new WFC3 H-band observation for the remaining galaxies of the high-luminosity LIRG sample  $L_{IR} \geq 10^{11.4} L_{\odot}$, and resolve their nuclear cusps using 2-dimensional fitting of the ``Nuker'' profile \citep{Lau95}. This allows us to study the cusp properties in detail over a large range of merger stages and IR luminosities.\par

The outline of this paper is as follows: In \S~\ref{sec:obs} we describe the sample, the HST observations and the data reduction. The NIR light fitting procedure and the parametrization of cusps are specified in \S~\ref{sec:nuker}. The derived results are presented in \S~\ref{sec:res}, including the distribution of cusps, their possible correlation with IR luminosity and AGN activity, the build-up of cusps as a function of the merger stage sequence, as well as possible systematic errors and dependences. In \S~\ref{sec:dis} we derive estimates of the stellar mass build-up in these cusps  and constrain the time-scales required to form them based on their current SFR and evolution as a function of merger stage. Moreover, we carefully compare the (U)LIRG cusp properties to those of local spheroidal galaxies and discuss the implications for formation scenarios of elliptical galaxies as well as possible mechanisms for cusp formation in non-interacting galaxies. The last point of our discussion is on the ultra-compact nuclear starburst and estimates of their SFR densities in the central 100~pc of (U)LIRGs. A synopsis of the main findings and conclusions is provided in \S~\ref{sec:sum}.

\section{Sample and Observations}
\label{sec:obs}

The GOALS NIR imaging program targets the nuclear regions of all systems in the IRAS Revised Bright Galaxy Sample \citep[RBGS;][]{San03} with log[$L_{IR}/L_{\odot}]>11.4$. The RBGS is a complete sample of all extragalactic objects with $f_\nu(60\mu m) \geq 5.24$
Jy, covering the entire sky surveyed by IRAS at Galactic latitudes $\vert b\vert > 5^\circ$. The RBGS contains 201 LIRGs, of which 88 have luminosities of log[$L_{IR}/L_{\odot}]>11.4$, the luminosity at which the local space density of LIRGs exceeds that of optically-selected galaxies.
These galaxies are the most luminous members of the GOALS sample and they are predominantly mergers and strongly interacting galaxies \citep{Haa11}. Most of the single spiral galaxies or widely separated, weakly/non-interacting pairs in GOALS are filtered out by this luminosity threshold. The redshift range of our sample is $0.01 < z < 0.05$ and hence the galaxies are bright and well-resolved due to their large angular size. The final range in far-IR luminosity is 11.4$ < $log[$L_{IR}/L_{\odot}] < 12.5$. An overview of our our sample is given in Table~\ref{tab:obs}.\par 

Our previous study \citep{Haa11} included only the HST NICMOS/NIC2 images since NICMOS failed during the execution of this program, and not all targets in the complete sample of LIRGs were observed. Here we present new results based on the analysis of the complete HST NIR sample, including new HST WFC3 images for the rest of our sample. The HST WFC3 data have been obtained using the F160W filter for 13 LIRGs within a total time of 13 orbits (program 11235, Surace P.I.) and are combined with the 88 pointings of 73 LIRGs of \cite{Haa11} that already have high-quality archival NICMOS data (see Tab.~\ref{tab:obs} for an overview of our sample). Note that the LIRG system Arp240 was only partially observed with NICMOS (NGC~5257) and has been completed with WFC3 which observed the other merger component (NGC~5258). The WFC3 data were collected with a field of view of $136\arcsec \times 123\arcsec$ and a pixelsize of $0.13\arcsec$.  The data reduction and calibration have been done using the standard HST pipeline. Additional corrections were made to the individual frames to adjust for bias offsets and we recombined the images using the STIS software.\par 

A World Coordinate System (WCS) correction has been performed to align with the HST ACS images, which were registered using star position references from 2MASS. We carefully applied a WCS transformation to our NICMOS and WFC3 images given the corrected HST ACS images using corresponding reference points in both images, such as stars not catalogued, bright star cluster knots, or similar unresolved features.  

\section{Deriving the nuclear properties using the NUKER profile with GALFIT}
\label{sec:nuker}
\subsection{The NUKER profile}
\label{subsec:nuker}

Fitting the light profile of galaxies is far from trivial and the choice of the appropriate profile fit function depends on the structural components, physical scales, and the galaxy types of interest.  For this study we have chosen the Nuker profile which was introduced by \cite{Lau95} to fit the nuclear profile of nearby galaxies observed with HST and is a double power law with the parametric form
\begin{equation}
I(r)=2^{(\beta-\gamma)/\alpha} I_b(\frac{r_b}{r})^{-\gamma} [1+(\frac{r}{r_b})^\alpha]^{(\gamma-\beta)/\alpha}
\label{eq:nuker}
\end{equation}
where $\beta$ is the outer power law slope, $\gamma$ the inner slope, $r_b$ is the ``break radius''  between the outer and inner profile, and $\alpha$ moderates the sharpness of the transition (more rapid transition with larger $\alpha$).\par 

Although the Nuker profile is very well suited to describe the central region of galaxies, it does not  link the outer profile with a physical representation of large-scale galaxy features (such as bars, bulge,disk). However, the typical radii of cusps are 100--1000~pc (on average 250~pc), which is well within the limits of the Nuker profile. While single and double S\'{e}rsic profiles have been applied to measure the cusp in ellipticals in some studies as well \citep[e.g.,][]{Gra03, Hop09a}, a combination of S\'{e}rsic profiles would lead to degeneracies in fitting more complicated structures such as a combination of disk, bulge, and cusp which is more typical in LIRGs. We have tested a combination of multiple S\'{e}rsic profiles, but found that fitting simultaneously all structural components of a galaxy would reveal reasonable results only in very few LIRGS without complicated structure. 
The argument against a full decomposition of the entire galaxy's structure is that every component fit is strongly dependent on the outcome of the other components, which would a) make the interpretation of the multiple S\'{e}rsic profiles difficult and b) would not allow a coherent measurement of the cusp over the large morphological variety of LIRGs. The main advantage of the Nuker profile is that it is not an  extrapolation of the outer galaxy profile and does not depend on the large-scale structure of a galaxy unlike a single or combination of multiple S\'{e}rsic profiles.  

%

\subsection{GALFIT Fitting and Cusp Parameter Estimates}
\label{subsec:galfit}

To derive detailed information about the structural properties of LIRGS we performed a 2-dimensional decomposition of all galaxies in our sample using GALFIT \citep{Pen10}. 
In this paper we focus on an accurate measurement of the central light component, in particular the nuclear slope as parametrized by the Nuker profile to obtain the cusp and core properties of LIRGs. This approach focuses on an accurate parametrization of the very center ($<1~kpc$), in particular the stellar cusp, which appears at much smaller scales than those of our previous HST NICMOS study \citep{Haa11} where we used multiple S\'{e}rsic component fits \citep{Ser68} to decompose the galaxy primarily into bulge and disk component (this is not possible with a Nuker profile). 
The surface brightness profile which is fitted in equation \ref{eq:nuker} is defined as 
\begin{equation}
\mu_b=-2.5\;\mathrm{log}_{10}(\frac{I_b}{t_{exp}\Delta x \Delta y}) + \mathrm{mag} zpt,
\end{equation}
where the image exposure time $t_{exp}$ is in seconds, the pixel scale $\Delta x$ and $\Delta y$ in arcsec and magzpt is the magnitude zeropoint. In all there are a total of 9 parameters to fit ($x_{center}, y_{center},\mu, r_b, \alpha, \beta, \gamma$, and the inclination and position angle of the galaxy).\par

Nearly 15\% of the galaxies in our sample are not suitable for fitting their nuclear profiles which is caused by several circumstances: highly disturbed center, dust obscuration in edge on galaxies, and dominant strong point sources. We performed a careful analysis of each galaxy and identified the cases where a nuclear fit would not work or might lead to large uncertainties. However, most of these galaxies are selected out due to their geometrical alignment (edge on) and hence don't introduce a physical selection bias. This results in a representative sample of 81 galaxies for which we have obtained nuclear NUKER profiles. \par

Running GALFIT on an image requires some initial preparation. The desired fitting region and sky background must be provided, and the HST instrumental Point Spread Function (PSF) image, bad pixel mask (if needed), and pixel noise map must be generated. The integrated galaxy counts per second are converted to an apparent Vega magnitude \citep[see more details in][]{Haa11}, the required HST PSF image is simulated with the TinyTim code \citep{Kri97} using a pixel scale identical to that of the observations, and the uncertainty images are generated by the HST reduction pipeline and used as weighting images required for GALFIT to perform the $\chi ^2$ minimization. The PSF image is utilized to fit the unresolved component in the centers of these galaxies, which can originate either from a central AGN or nuclear star formation. \par

After these preparation steps we carried out an iterative fitting process since GALFIT requires initial guesses for each component it fits and uses a Levenberg-Marquardt downhill-gradient algorithm to determine the minimum  $\chi ^2$ based on the input guesses. At first we fitted Nuker profiles that are centered on the galaxies' nuclei (typically 1-2 nuclei per image). For merger systems that have of two or more galaxies, we take advantage of GALFITs ability to fit simultaneously multiple components at different spatial regions. Initial parameter guesses for the center, position angle, inclination and sky-background have been estimated with S\'{e}rsic profiles \citep{Haa11}, but have been kept free for the new model fit. A PSF fit has been tested for all sources but only applied in the final fit for those where a GALFIT solution was found. The center coordinates were fixed when it was necessary (e.g. in cases where PSF and Nuker profile centers drift too far away due to starburst region or other components that might mimic a point source). In several cases we have tested a range of fitting regions to obtain the best fit of the transition between outer and inner profiles. The outer radius used for the fitting is typically 1--3~kpc, which is roughly 3--10 times the cusp radius, which is defined here as the break radius of the Nuker law. One important point to be cautioned about when using the Nuker profile is to ensure that the transition parameter $\alpha$ is large enough to separate inner and outer profile because, for example, a low $\alpha$ value ($\alpha<$2) can be reproduced by simultaneously having a high $\gamma$ and a low $\beta$, which is a serious possibility for degeneracy \citep[see][]{Pen10}. Therefore, in cases where $\alpha$ behaved unreasonably, either $\alpha<2$ or $\alpha>10$, we tested several values in the range of $2<\gamma<10$, and fixed $\alpha$ at the value for which the best solution was found. Fitted models were selected only as long as the model parameters were all well behaved, i.e. requiring that all parameters such as break radius, inner/outer slope values, and transition parameter converged within the range of reasonable values. The best fit was chosen based on the $\chi^2$ for each model and the quality of the residual image. Since this study focuses only on the nuclear parametrization we have not attempted to find a perfect solution for the entire galaxy, for which the Nuker profile fitting is likely not suited in any case. All in all we have tried to find the most robust fit for the nuclear slope that can be applied in a  homogeneous way among all sources in our sample using one main component (plus a PSF-component) rather than splitting the model, e.g., in three or more S\'{e}rsic profiles (cusp, bulge, disk, bar, etc.) which introduces many more free parameters and possible interpretations.

\subsection{Example of a strong cusp galaxy: The nuclear NIR properties of NGC~1614}
\begin{figure}
\begin{center}
\includegraphics[scale=0.6]{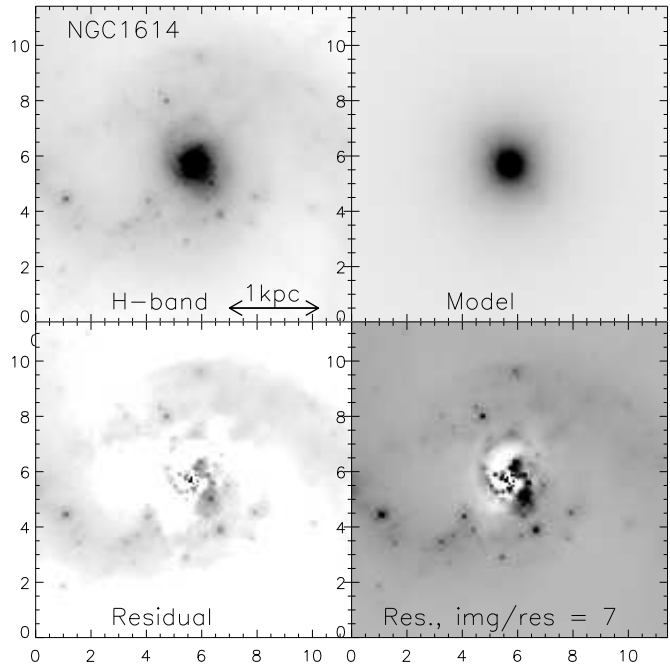}
\medskip
\caption{GALFIT output images for NGC~1614 (scaled with the square root of intensity). Top left: Region of NICMOS image that is fitted. Top right: Galfit model. Bottom left: Residual image (shows the difference between model and NICMOS image) with the same brightness level as the NICMOS image. Bottom right: Residual image scaled to its maximum brightness (the brightness ratio of NICMOS image to scaled residual map is shown as well at the bottom) to highlight the faint diffuse emission, spiral structure, and clusters remaining in the model-subtracted data. The axes are in units of arcsec and the figures are oriented in the observational frame.}
\label{fig:NGC1614_galfit}
\end{center}
\end{figure}

\begin{figure}
\begin{center}
\includegraphics[scale=0.51]{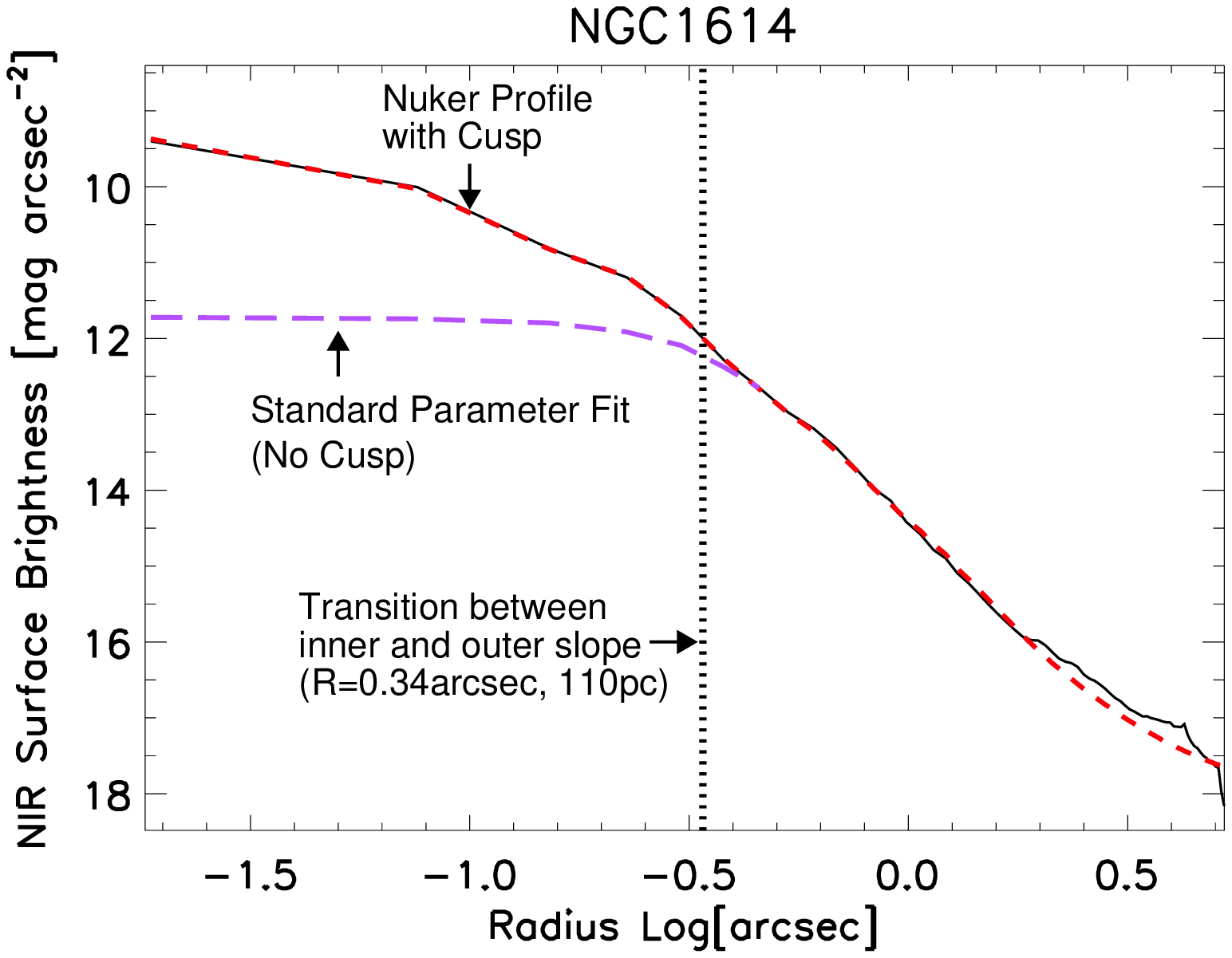}
\includegraphics[scale=0.51]{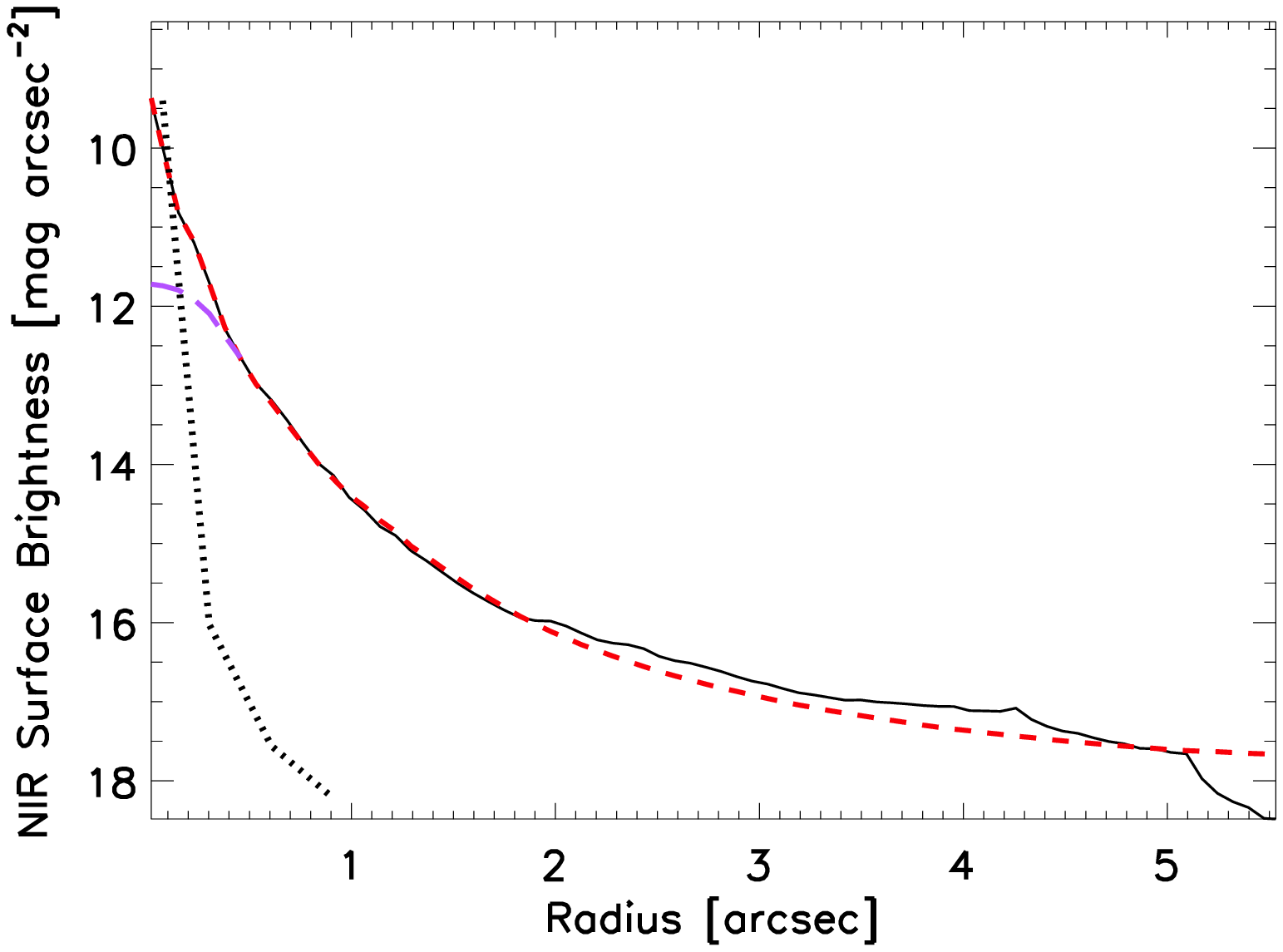}
\caption{The radial profile of NGC~1614 as an example for a strong cusp slope. Top (logarithmic radial scale): The solid black line is the actual data as observed with HST NICMOS at 1.6$\mu$m. The long dashed line shows a fit of the outer envelope with no cusp component, while the short dashed line (as well as the data) shows an additional inner power-law component as fitted with GALFIT in 2-D \citep{Pen10} with a slope  of $\gamma$=1.36. The transition between the outer disk and the additional cusp component occurs at a radius of $\sim 0.34$\arcsec (110~pc). Bottom (linear radial scale): The dotted line indicates the spatial resolution (NICMOS PSF) which is scaled to the maximum of the measured Nuker profile.}
\label{fig:NGC1614_profile}
\end{center}
\end{figure}

Cusps manifest themselves as strong central excess light on top of the normal light profile of a galaxy (i.e., stellar bulge, disk and often a bar). A typical example of a strong cusp galaxy is NGC~1614 whose HST and GALFIT residual image are presented in Fig.~\ref{fig:NGC1614_galfit} and the radial light profile in Fig.~\ref{fig:NGC1614_profile}.
This galaxy exhibits a cusp with a slope of $\gamma$=1.36 and a break radius of $\sim 0.34$\arcsec (110~pc) where the transition between the outer disk and the additional cusp component occurs. The size of the cusp is similar to the spatial size of Pa$\alpha$ emission (see \S~\ref{subsec:paschen}) which is a strong indicator of on-going star-formation on the cusp scale. This demonstrates that the stellar cusp is linked to the current star formation in this galaxy. The residual image reveals a nuclear spiral structure and a large number of circum-nuclear cluster, but typically at less than one magnitude lower surface brightness than the cusp. An image of the cusp component together with several other examples can be found in App.~\ref{app:b}.

\section{Survey Results}
\label{sec:res}

\subsection{Cusp-to-Core Ratio and Cusp Luminosities}
\label{subsec:dist}

The NIR luminosity\footnote{All luminosities derived from our HST NIR images  (i.e. L$_{cusp}$, L$_{PSF}$, L$_{nuc}$, and L$_{Nuker}$ refer to the H-band solar luminosity.} of the cusp L$_{cusp}$ is computed by subtracting the integrated magnitude  with $\gamma$ set to zero (equal to the extrapolated outer profile for the center) from the total integrated magnitude $M_{Nuker}$ within the cusp radius $r_b$,
\begin{equation}
M_{cusp}=\int_0^{r_b} M_{Nuker} - \int_0^{r_b} M_{Nuker,\gamma=0}.
\end{equation}
Except for $\gamma$, all parameter values are the same between both integrated magnitudes. Therefore, the cusp luminosity is defined as the residual light of the fitted Nuker profile after subtraction of the reference profile which is here the inward extrapolation of the outer power law fit of the Nuker profile. This calculation of the cusp luminosity is more robust and accurate than just fitting the light rising above a fitted outer envelope (e.g. using a S\'{e}rsic profile), which is complicated in our galaxies by the mix of multiple outer components (bulge, bar, disk) as discussed in \S~\ref{subsec:nuker}.
Note that the luminosity of the unresolved component $L_{PSF}$ is separated from this computation. Upper limits for core galaxies ($\gamma \sim 0$) have been derived based on the uncertainty in $\gamma$ (typically $d\gamma\sim\pm0.1$), but this does not take into account the spatial resolution limit and the possibility of an unresolved cusp (see \S.~\ref{subsec:systest}).\par 

\begin{figure}
\begin{center}
\includegraphics[scale=0.46]{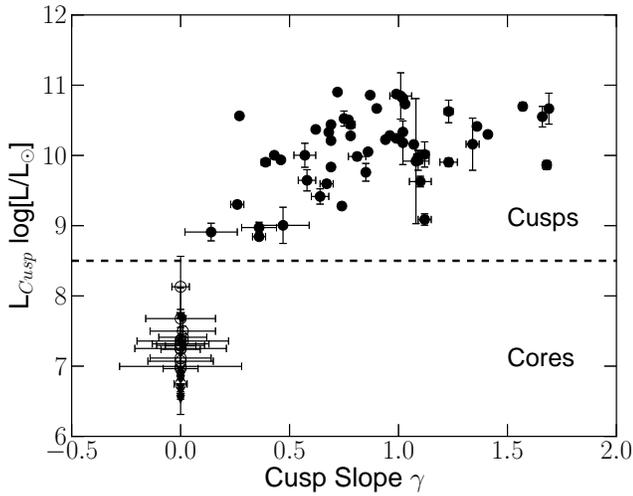}
\caption{The NIR  (1.6~$\mu$m) cusp luminosity L$_{cusp}$ as a function of the cusp slope $\gamma$.  Upper luminosity limits for sources with $\gamma \sim 0$ have been derived based on the uncertainty in $\gamma$ (typically $d\gamma\sim\pm0.1$).}
\label{fig:gamma_lcusp}
\end{center}
\end{figure}

\begin{figure}
\begin{center}
\includegraphics[scale=0.45]{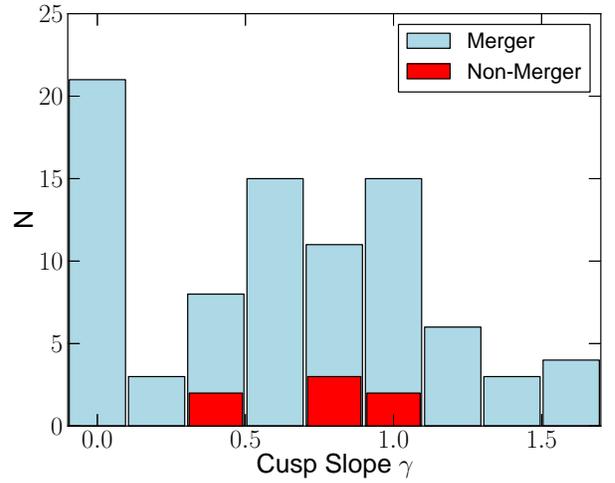}
\includegraphics[scale=0.45]{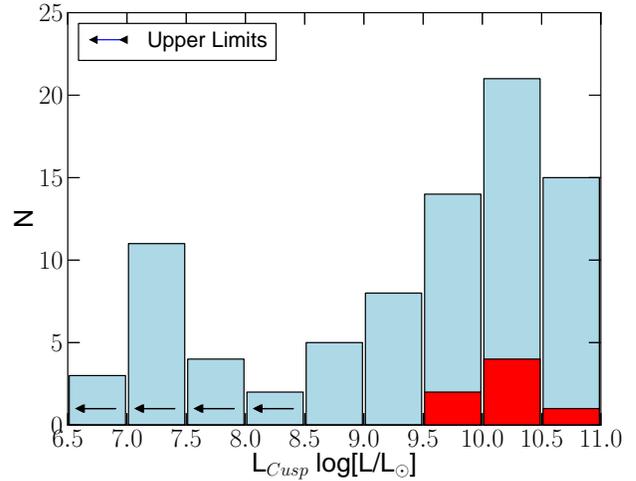}
\caption{Top: Histogram of the cusp slope ($\gamma$) of our sample of interacting (blue) and non-interacting LIRGs (red), indicating a dichotomy between core ($\gamma\sim0$) and cusp galaxies ($\gamma>0.3$). Bottom: Histogram of the cusp luminosity with upper limits for core galaxies (left arrows).}
\label{fig:hist_gamma}
\end{center}
\end{figure}

Fig.~\ref{fig:gamma_lcusp} shows the relation between cusp slope and cusp luminosity for 77 galaxies with good fit models. Only those models are accepted for which all parameters such as break radius, inner/outer slope values, and transition parameter converged within the range of reasonable values (see \S~\ref{subsec:galfit}). The cusp luminosity increases as a function of cusp slope in the interval of $0<\gamma<1$, and no significant increase is found for $\gamma>1$. Fig.~\ref{fig:hist_gamma} shows the distribution of the slope ($\gamma$) in our sample ranging from $\gamma=0$ to $\gamma=1.7$. The histogram reveals one peak at $\gamma\sim 0\pm0.1$ (core galaxies) and the other (broader) peak at $\gamma\sim0.8\pm0.4$. We use the distribution in $\gamma=0$ to define core ($\gamma<0.3$) and cusp galaxies ($0.3>\gamma$), as referred hereafter, with a a cusp/core ratio 3.2:1. Out of our sample of 77 galaxies with good profile fits, 59 (76\%) have nuclear cusps. The naming conventions vary in the literature, in general cusp galaxies are defined to have steep power-law surface-brightness profiles in the centre, while core galaxies have surface-brightness profiles with a flat central core. All derived cusp properties are listed in Table~\ref{tab:galfit}. We find no correlation between the total NIR luminosity of the host galaxy and the NIR cusp luminosity or cusp slope.\par
 
The cusp luminosity distribution follows the two peaks as seen in the $\gamma$ distribution. Based on the optical and NIR morphology (tidal tails, merger components) we have classified our LIRG sample into major merger and non-major merger (no major companion, tidal tails or strong disturbance, see \cite{Haa11}. Our results are shown in Fig.~\ref{fig:hist_gamma}. While there are only 7 apparent non-mergers in our HST F160W sample, they all have cusp luminosities above  log[$L_{cusp}/L_{\odot}]=9.5$, with an average luminosity of log[$L_{cusp}/L_{\odot}]=10.3\pm0.3$. The merging LIRGs on the other hand, have a much larger range in cusp luminosities, and 24\% have only upper limits with cusp luminosities below log[$L_{cusp}/L_{\odot}]=8.5$. We discuss the possible reasons for the difference in these distributions in \S~\ref{subsec:ellipticals}.


\subsection{Relationship between Current Star-formation and Cusp Strength}
\label{subsec:lir}
\begin{figure}
\begin{center}
\includegraphics[scale=0.46]{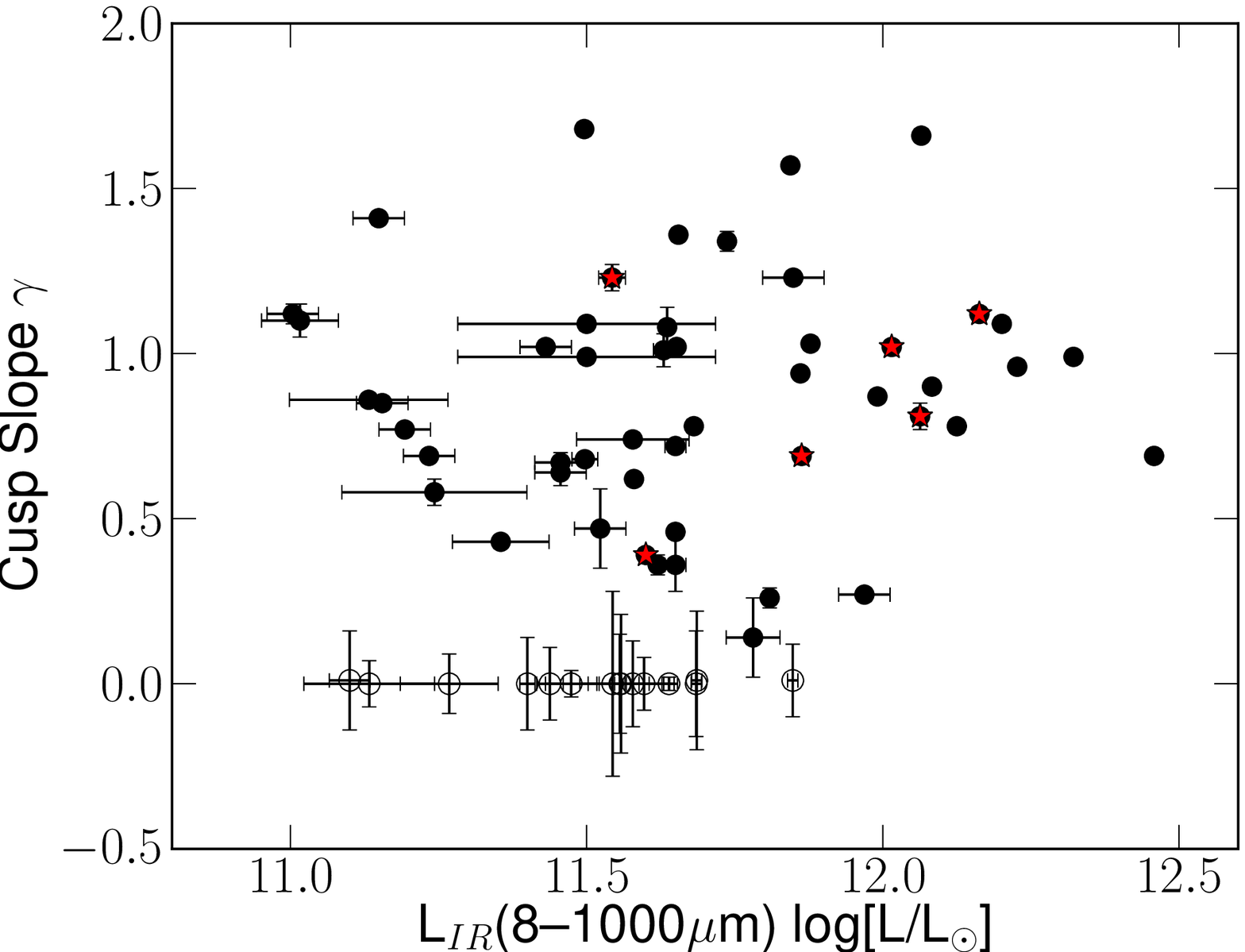}
\includegraphics[scale=0.46]{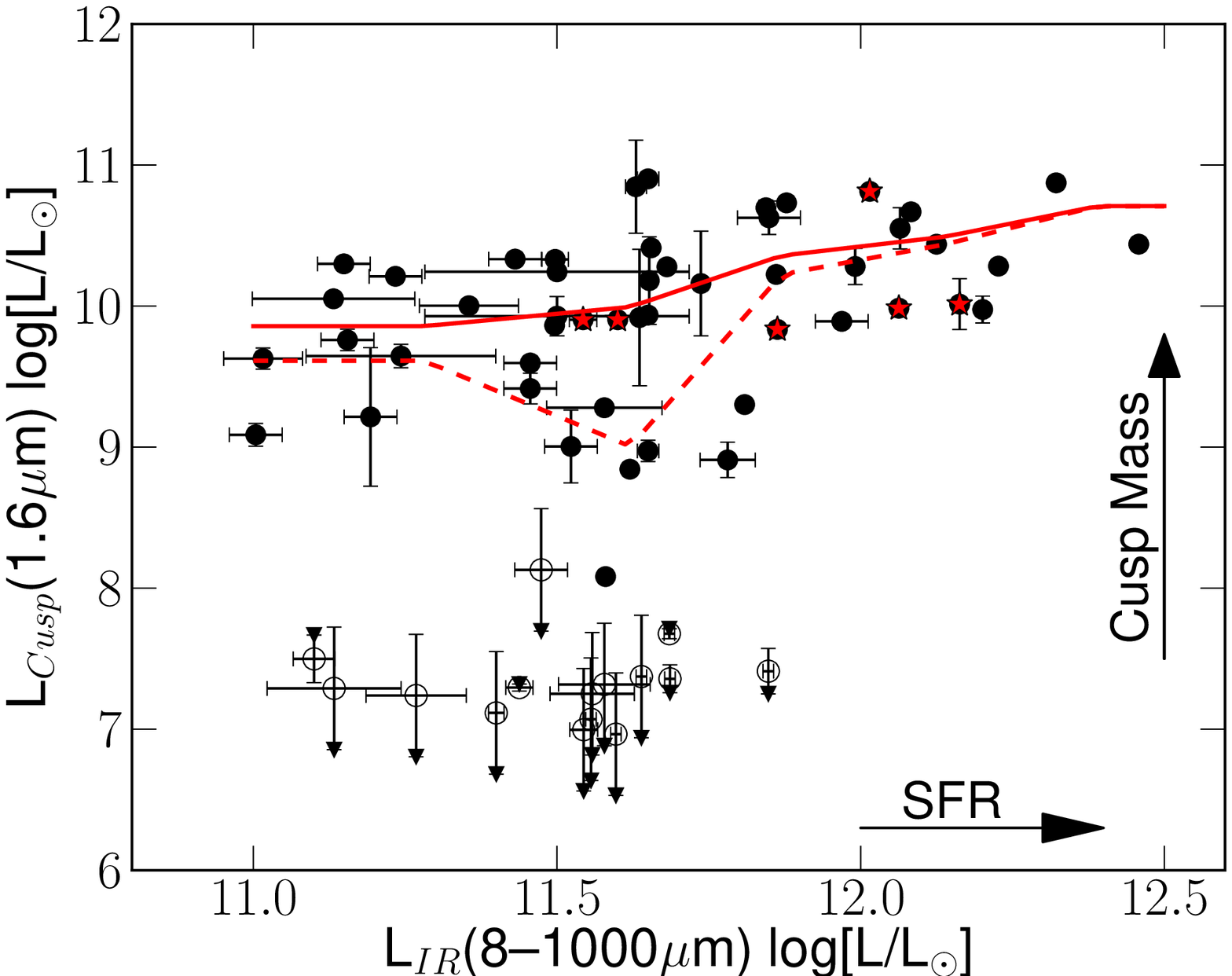}
\caption{The cusp slope ($\gamma$, top) and the cusp luminosity (bottom) as a function of IR luminosity. For core galaxies (arrow down marker) the upper luminosities limits are plotted (based on their uncertainty in $\gamma$ which is typically $\pm0.1$). The red solid (dashed) line shows the interpolated mean (median) cusp luminosity $L_{cusp}$ as a function of IR luminosity. Note that galaxies with dominant AGN contribution to the L$_{IR}$ (EQW PAH 6.2$\mu$m $<$0.27, diamond marker) have been excluded in the mean values.}
\label{fig:ir_gamma}
\end{center}
\end{figure}

To test if the central starburst activity in LIRGs is related to the build-up of cusps, we have plotted the measured cusp properties as a function of far-IR luminosity as shown in Fig.~\ref{fig:ir_gamma}. 
The total IR luminosity L$_{IR}$ is a very good tracer of the star-formation rate in LIRGs with star formation rates of several tens to hundreds of M$_\odot$~yr$^{-1}$. 
However, this is only applicable if most of the L$_{IR}$ originates from starbursts rather than AGN. Therefore we have selected for this comparison only galaxies that have no dominant AGN contribution to the L$_{IR}$ (PAH EQW $>$ 0.27, see \cite{Pet11}). A second criterion for this comparison is that the starburst must originate from the center of the galaxies as indicated by the peak of the Spitzer MIPS 24$\mu$m emission (which is the case for 90\% of galaxies in our sample). The exclusion of LIRGs with non-nuclear mid-IR emission is essential since the exact cause of the enhanced emission in these galaxies is not clear \citep[see e.g.][]{Mir98, Cha02, Cha04, Ina10, Haa11, LeF12} and is likely not directly associated with the cusp build-up.\par 

Since most of our LIRGs are merger systems we calculated the individual L$_{IR}$ luminosities for each merger component as described in the following \citep[see also][]{Dia13}: First we performed an aperture photometry on the MIPS 24$\mu$m images as a proxy of L$_{IR}$ based on the projected distances between the sources. We obtained the total flux of individual galaxies by using a large aperture for isolated sources, and for those in groups, an aperture radius of slightly less than half the distance between the closest sources. These derived fluxes agree within 5\% with values previously measured by \cite{How10}. For the overlapping galaxies we included in their luminosity uncertainties a factor that accounts for the small apertures used to obtain their photometry and the relative luminosity between the sources. In some cases the sum of the flux of all individual components does not add up to the total flux measured by \cite{How10} for the integrated systems, but the difference is typically not larger than 10\%. Hence we scaled up the individual luminosities proportionally to match up the integrated system luminosity. For merger systems with very small nuclear separation distances and both nuclei covered roughly equally within the 24$\mu$m emission, we have assigned each nuclei half of the L$_{IR}$ luminosity and updated the uncertainties accordingly. The individual L$_{IR}$ luminosities estimated this way and their uncertainties are listed in Tab.~\ref{tab:galfit}.\par

Fig.~\ref{fig:ir_gamma} shows the cusp NIR luminosity L$_{cusp}$ as a function of L$_{IR}$ for each galaxy. While there is a large scatter at low IR luminosities, the average L$_{cusp}$ is significantly larger at higher L$_{IR}$ luminosities. We note that a similar behaviour is also found if we use the L$_{IR}$ luminosities of the combined merger system. The increase in cusp luminosity can be attributed to the increase in the slope $\gamma$ and the bulge radius. However, independent of the cusp parameters, we also find that the central NIR surface luminosity within 1~kpc radius increases as a function of L$_{IR}$. Its average value in systems with log[L$_{IR}/L_{\odot}$]$>$11.9 is 2.5 times higher compared to the corresponding lower luminosity average. Note again that the unresolved component is not included here. 
The scatter at low IR luminosities is in agreement with the fact that these galaxies are typically a mix of mergers and non-mergers, which means they might have already gone through a phase of cusp formation or, alternatively, the cusp formation process is intrinsically different between merger-induced and non-merger induced starbursts. In contrast, high IR luminous LIRGs (log[L$_{IR}/L_{\odot}$]$>$11.9) are predominantly major mergers.  Note that galaxies with dominant AGN contribution to the L$_{IR}$ have been excluded in the statistical analysis but are marked in Fig.~\ref{fig:ir_gamma} for completeness.
Our results show that in particular above log[L$_{IR}/L_{\odot}$]=11.9, all sources have strong cusps ($L_{cusp}\gtrsim 10^{10}$ L$_\odot$). The average cusp luminosity increases about a factor of 5 from log[L$_{IR}/L_{\odot}$]=[11.0--11.5] (mean $L_{cusp}/L_{\odot}=[0.9\pm0.9]\times10^{10}$, median $L_{cusp}/L_{\odot}=0.52\times10^{10}$) to log[L$_{IR}/L_{\odot}$]=[12.0--12.5] (mean $L_{cusp}/L_{\odot}=[3.8\pm1.9]\times10^{10}$, median $L_{cusp}/L_{\odot}=3.3\times10^{10}$). This corresponds, on average, to an increase in the NIR cusp luminosity of $\sim 3\times10^{10}$~L$_\odot$. The statistical significance is tested via the Kolmogorov–Smirnov (KS) test which confirms that the distribution of the cusp luminosity between low and high IR luminous LIRGs is not the same (at a probability level of more than 99\%). In \S~\ref{subsec:cuspmass} we will estimate the associated build-up of stellar mass in these cusps and discuss the implications for star-formation rates and time-scales.

\subsection{Comparison between nuclear stellar structure and mid-IR spectral diagnostics}
\label{subsec:agn}
To study whether the central starburst and cusp growth in LIRGs is accompanied by an AGN activity, we compare our AGN characterization based on mid-IR diagnostics with the unresolved central component (PSF) in the HST NIR images.
The mid-IR AGN characterization is based on the Equivalent Width (EQW) of the 6.2~$\mu$m PAH feature, with lower values indicating an excess of hot dust emission which is usually associated with AGN activity \citep[see e.g.][]{Gen98, Stu00, Arm07, Des07}. \cite{Pet11} estimated that AGN are responsible for $\sim$12\% of the total bolometric luminosity of local LIRGs based on several mid-IR line diagnostics measured with the Infrared Spectrograph on Spitzer. \cite{Lee12} have found that half of the (U)LIRGs optically classified as non-Seyferts show AGN signatures in their NIR spectra and that the contribution of buried AGNs to the infrared luminosity is 5-10\%, which is based on AKARI NIR spectroscopy and mid-IR SED fitting for 36 (U)LIRGs. The fraction of AGN-dominated sources stays roughly constant as the merger progresses, while the fraction of composite sources (i.e. those with a weak AGN that does not dominate the emission in the MIR) increases \citep{Sti12}.\par
\begin{figure}
\begin{center}
\includegraphics[scale=0.46]{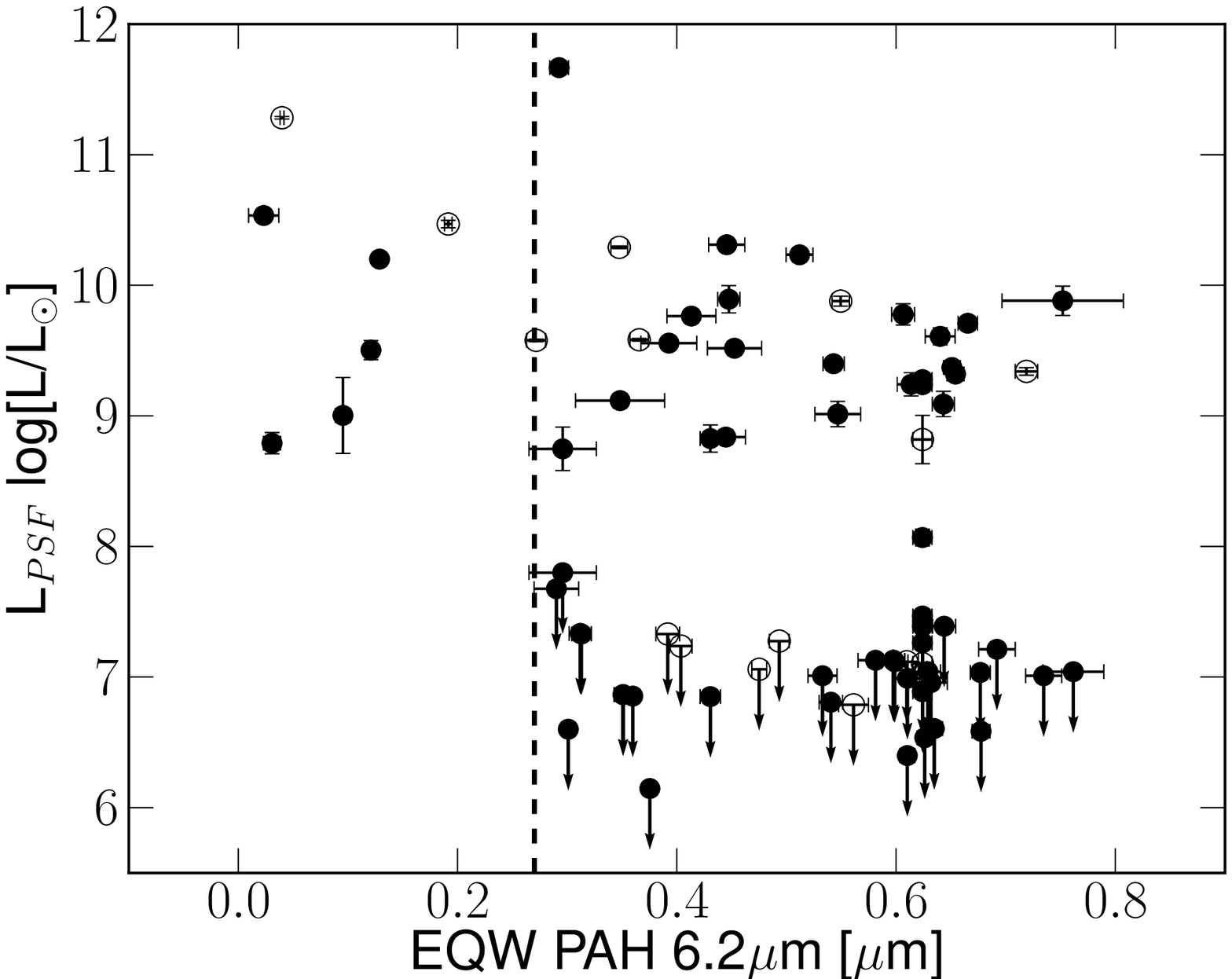}
\includegraphics[scale=0.46]{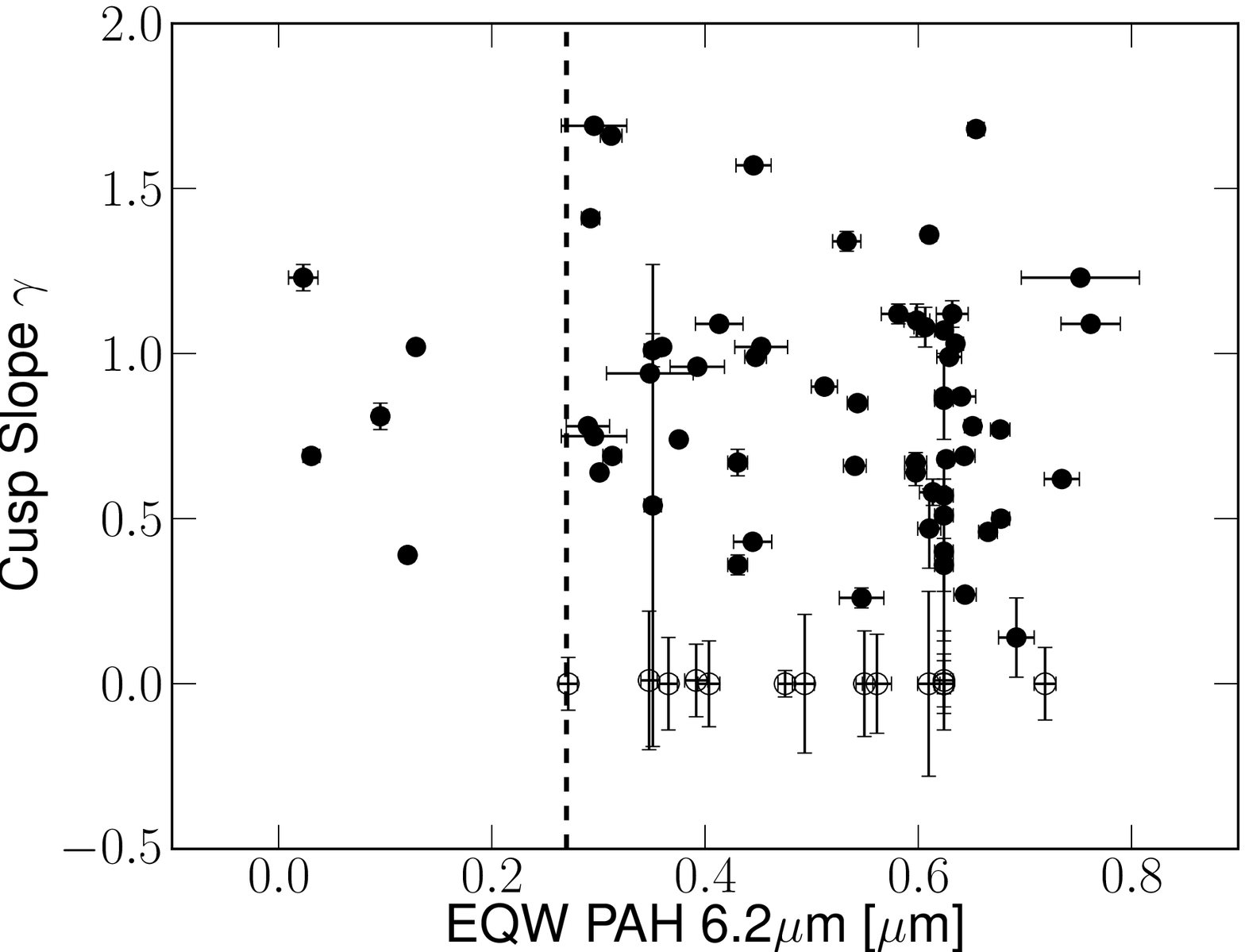}
\caption{Top: The fitted central unresolved component (HST 1.6$\mu$m PSF) as a function of the 6.2~$\mu$m PAH EQW. Small EQW indicate a dominant AGN contribution to the IR luminosity (dashed line, EQW $\lesssim$0.27~$\mu$m) while the center of galaxies with larger EQWs are a mix of starburst and AGN light (dashed line, 0.27~$\mu$m$<$EQW$<$0.54~$\mu$m) or dominated by starbursts (EQW$>$0.54~$\mu$m). The arrows at the bottom indicate that these are upper limits (uncertainty) of the PSF luminosity for galaxies with no significant PSF measured by GALFIT. Bottom: The cusp slope as a function of the 6.2~$\mu$m PAH EQW.}
\label{fig:eqw_psf}
\end{center}
\end{figure}

In principle, the presence of a strong unresolved component in the HST NIR images ($<$150pc, corresponding to the mean HST resolution for our sample) indicates a possible central AGN. However, such a PSF can potentially also originate from a very compact starburst \citep{Sco00}.
In Fig.~\ref{fig:eqw_psf} we investigate if the PAH EQW \citep[measured by][]{Sti12} is correlated with the PSF component of the HST NIR images. For composite and starburst dominated sources (PAH EQW $>$ 0.27$\mu$m) the fraction of PSF and non-PSF sources is roughly the same with no apparent increase of PSF luminosity as a function of decreasing  PAH EQW. Note that the characterization as AGN (EQW$<$0.27) , composite (0.27~$\mu$m$<$EQW$<$0.54~$\mu$m) and starburst (EQW$>$0.54~$\mu$m) IR dominated sources has been defined by \cite{Pet11}, but is not meant as a stringent classification. There is a small increase of sources from starburst dominated (EQW$>$0.54~$\mu$m) to composite sources (0.27$<$EQW$<$0.54~$\mu$m) with the fraction of PSF sources increasing from 40\% to 50\%, respectively. However, this change is small in comparison to the AGN-dominated IR sources (PAH EQW $<$ 0.27~$\mu$m), with all of them exhibiting a strong unresolved central component in the NIR (see Fig.~\ref{fig:eqw_psf}), which confirms the presence of a strong AGN in the center of these sources (ESO~203-IG001, AM~0702-601North, 2MASX~J08370182-4954302, UGC~05101, ESO~286-IG019, IC~5298, NGC~7674). However, the PSF luminosity is roughly the same for AGN dominated, composite and starburst sources, which implies that the presence of an AGN does not significantly affect the properties of the H-band PSF.  
\par

One interesting question is whether there is any relationship between the strength of the cusp and the presence of an AGN. The bottom panel of Fig.~\ref{fig:eqw_psf} shows the cusp slope $\gamma$ as a function of the EQW of the 6.2$\mu$ PAH feature emission. While we find no trend between EQW and cusp strength (in terms of $\gamma$ and L$_{cusp}$) for starburst and composite sources (EQW$>$0.27~$\mu$m), all of the sources with a dominant AGN in the mid-infrared  (EQW$<$0.27~$\mu$m and unresolved central NIR component) have cusps ($\gamma>0.3$ and  log[$L_{cusp}/L_{\odot}]>9.6$) rather than cores ($\gamma<0.3$). 

\subsection{The Nuclear Unresolved NIR Light Component: Evidence for Ultra Compact Nuclear Starbursts?}
\label{subsec:compsb}
A significant fraction of our LIRG sample ($\sim$13\%) has a substantial unresolved component (HST PSF measured with GALFIT) but no detected AGN contribution to the mid-IR (based on mid-IR diagnostics with PAH EQW $>$ 0.6, see \S~\ref{subsec:agn} and Fig.~\ref{fig:eqw_psf}). Interestingly all of them are LIRGs (log[11.4 $<L_{IR}/L_{\odot}]<12$) rather than ULIRGs ($L_{IR}/L_{\odot}]>12$). Although there are 7 ULIRGS with an unresolved component, all of them have PAH EQW $<$ 0.52 and hence we cannot rule out the possibility of an AGN contribution. The mean distance of LIRGs with unresolved component and PAH EQW $>$ 0.6  is 133~Mpc which is even slightly smaller than for the entire sample (mean distance: 153~Mpc), hence a possible distance-dependency can be likely ruled out.  All of these sources (ten) have an unresolved NIR component (log[L$_{PSF}/L_{\odot}$]$>$8.5) with a mean luminosity of log[L$_{PSF}/L_{\odot}$]=9.4.  This population of LIRGs spans a large range in far-IR luminosity (mean log[L$_{IR}/L_{\odot}$]=11.44), merger stages (from non-interacting to late stage merger), a large range in the nuclear slope $\gamma$ (including core and cusp galaxies), and a large range in distance. The most likely explanation is that the unresolved nuclear component originates from a very compact starburst that has either build-up a high stellar central density (assuming that most of the NIR light originates from old and intermediate aged stars) and/or is currently active (if most of the NIR light is due to radiation from young stars).  \par

To estimate the stellar density in these galaxies, we have derived their NIR luminosity surface densities $\Sigma_{NIR-center}$, which is given by (L$_{PSF} + $L$_{Nuker})/(\pi R^2_{PSF})$, where $R_{PSF}$ is the spatial limit of the PSF component of HST NICMOS and WFC3 ($\sim$0.15\arcsec) converted in kpc, L$_{PSF}$ is the NIR PSF luminosity and L$_{Nuker}$ the integrated light of the Nuker profile within the central aperture (0.15\arcsec).  The mean stellar surface density of our compact starburst sample is (2.4$\pm$1.2)$\times 10^{5}$ M$_\odot$~pc$^{-2}$ (within a mean nuclear radius of $\lesssim$100~pc) if we assume a standard mass-to-light ratio in the H-band $\Upsilon_{H}=0.3\pm0.15$. The applied $\Upsilon_{H}$ and its uncertainty is derived from STARBURST99 simulations \citep{Lei99} using Geneva High and Padova AGB tracks, Kroupa IMF, and instantaneous star formation with a mass of $10^6$~M$_\odot$ and solar metallicity. The lower limit of $\Upsilon_{H}=0.15$ is derived from the assumption that the NIR is dominated by young stars (4~Myrs) while the upper limit of $\Upsilon_{H}=0.45$ is a mix of young, intermediate (50~Myrs), and old stellar populations (few 100~Myrs) contributing equally to the NIR light. 
We note that a constant $\Upsilon_H$ for the inner 1~kpc does not take into account local variations due to not direct stellar photospheric emission (i.e. UV light from massive stars that has been extinguished by dust). The derived stellar surface density is similar to the maximum stellar surface density $\Sigma_{max}$ found in a large range of dense stellar systems, including  globular clusters, massive star clusters in nearby starbursts, nuclear star clusters in dwarf spheroidals and late-type disks, ultra-compact dwarfs, and galaxy spheroids spanning the range from low-mass bulges and ellipticals to massive ``core'' ellipticals \citep{Hop10}. NGC~3256 is the nearest LIRG (39~Mpc) of those 10 compact starburst sub-sample, and subsequently has the largest measured stellar surface density with $\sim$1$\times 10^6$ M$_\odot$~pc$^{-2}$ (within the nuclear radius of 28pc). If we assume an average SFR of $\sim$220~M$_\odot\;yr^{-1}$ for our galaxies, which is based on the SFR-L$_{IR}$ luminosity relation \citep{Ken98}, and that at least 10\% of the total star-formation originates from the compact starburst, we estimate a maximal time-scale of 100~Myrs to build up the stellar unresolved component in these compact starbursts.

\subsection{Nuclear Properties along the  Merger Stage Sequence}
\label{subsec:ms}

Most of the LIRGs in our sample are major mergers. However, seven LIRGs are isolated undisturbed galaxies, suggesting that mechanisms other than strong gravitational interactions may trigger the IR luminosity in those isolated LIRGs (e.g. minor mergers, high gas fraction and stellar mass, cold gas accretion, or other internal dynamical processes). Therefore, these two sub-classes of LIRGs, major mergers versus isolated galaxies, are treated separately in our studies. 
The non-merger/merger separation as well as the merger classification scheme is based on the optical and NIR light distribution (see Table~\ref{tab:galfit} for merger classification). While the HST NICMOS/WFC3 H-band images reveal the unobscured nuclei and stellar components, the HST ACS I- and B-band images provide a larger FOV for the detection of companion galaxies and are very sensitive for tracing tidal tails and other interaction features which are essential to identify the merger stage. This provides a distinct advantage over using a simple projected nuclear separation to infer the merger stage, which suffers from degeneracies when exploring a large range in stages. To establish a merger classification we carefully separated at first our sample into galaxies that exhibit interaction features (e.g. tidal tails, bridges) or have nearby companions, and those without any interaction features or nearby companions (undisturbed single galaxies), excluding the latter from our merger sequence analysis. The mergers are classified into six different stages \citep{Sur98a, Sur98b, Haa11, Kim12}: 1 - separate galaxies, but disks symmetric (intact) and no tidal tails, 2 - progenitor galaxies distinguishable with disks asymmetric or amorphous and/or tidal tails, 3 - two nuclei in common envelope, 4 - double nuclei plus tidal tails, 5 - single or obscured nucleus with long prominent tails, 6 - single or obscured nucleus with disturbed central morphology and short faint tails. In sum, these stages may be broadly characterized into pre-merger (1), ongoing merger (2,3,4), and post-merger LIRGs (5,6).  \par 

Table~\ref{tab:ms} represents the fraction of cusp galaxies ($\gamma>0.3$) as a function of merger stage. One puzzling result is the difference in the abundance of cusps between merger and non-merger LIRGs as already indicated in \S~\ref{subsec:dist}. In detail, all of the non-merger (MS 0) and 86\% of separated galaxy pairs that show no interaction signatures (MS 1) have a cusp, which is  larger than the average fraction of cusp galaxies in interacting galaxies $\sim$70\%). 
There might be several reasons for this difference and we will discuss several possibilities in more detail in \S~\ref{subsec:non-merger}, but note that this comparison may be statistically biased due to the small number of non-interacting LIRGs in our sample.\par

\begin{figure}
\begin{center}
\includegraphics[scale=0.2]{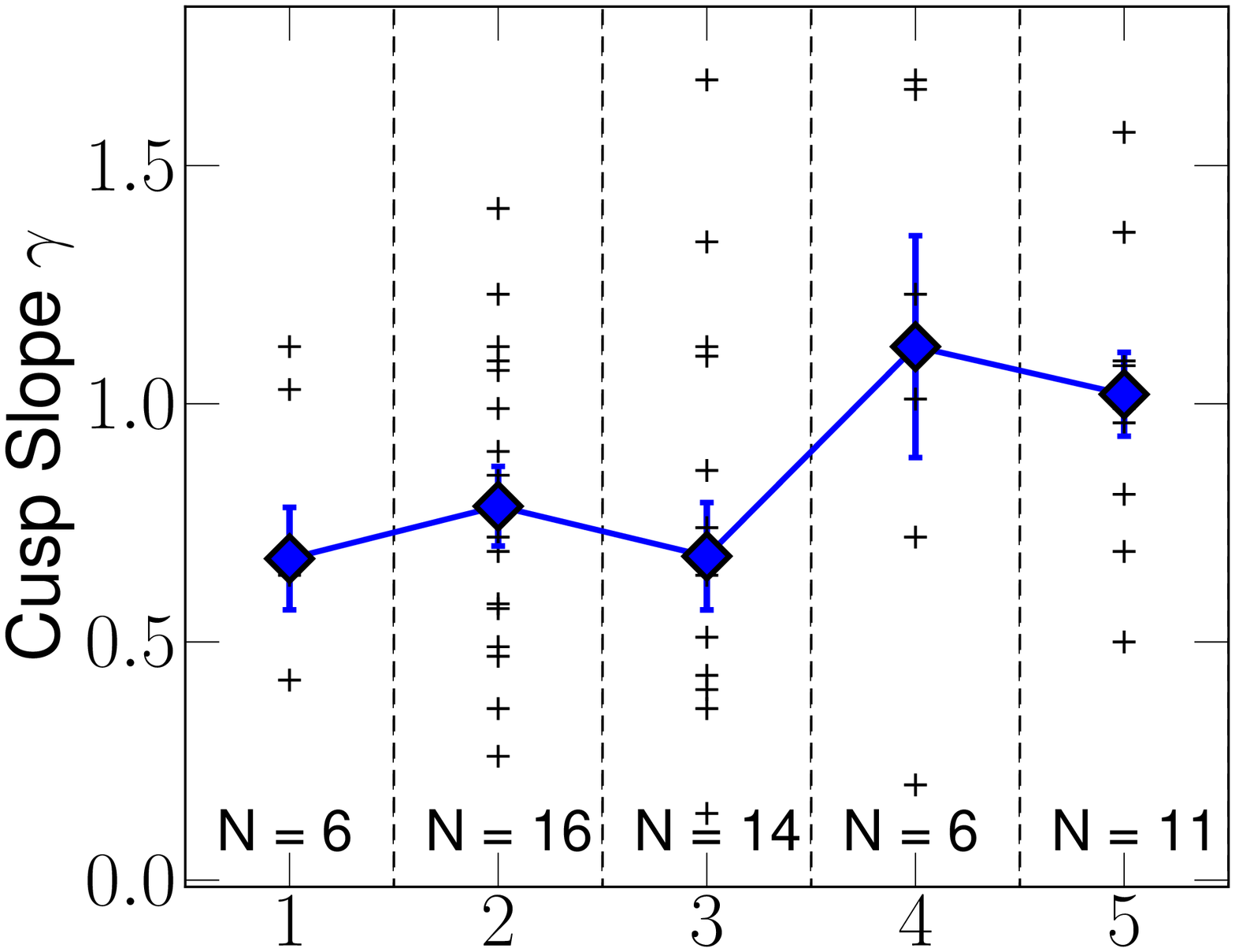}
\includegraphics[scale=0.2]{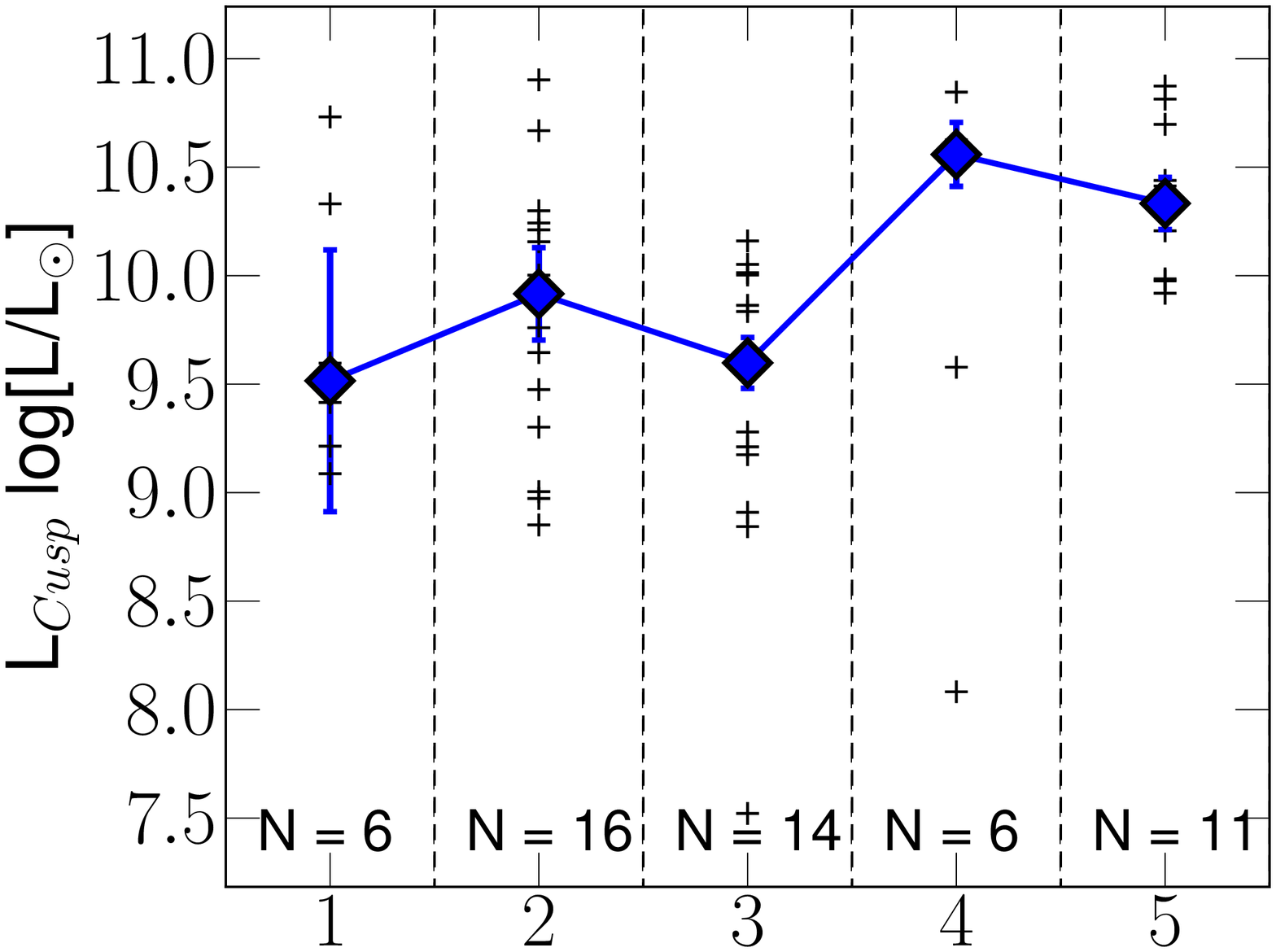}
\includegraphics[scale=0.2]{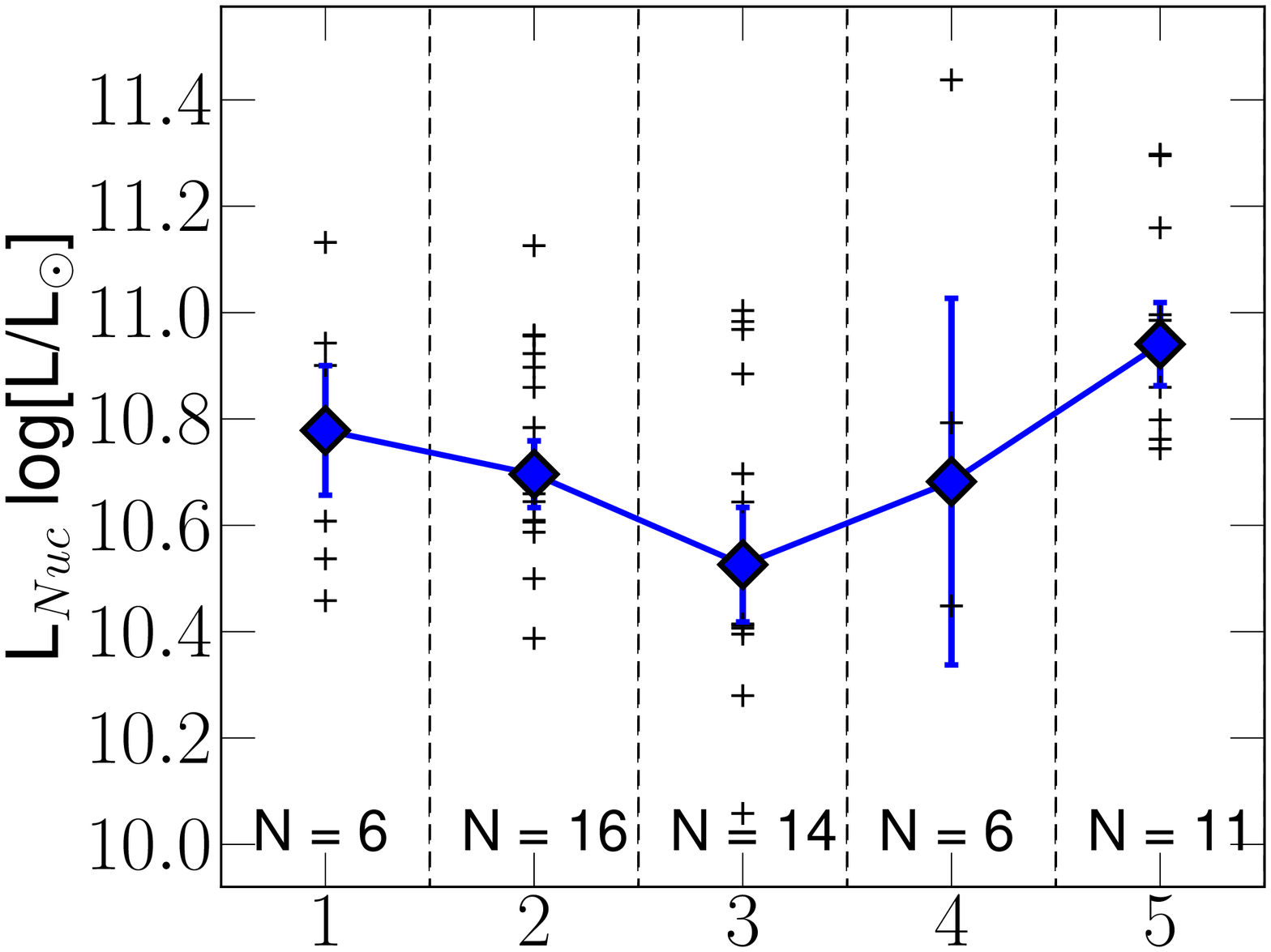}
\includegraphics[scale=0.2]{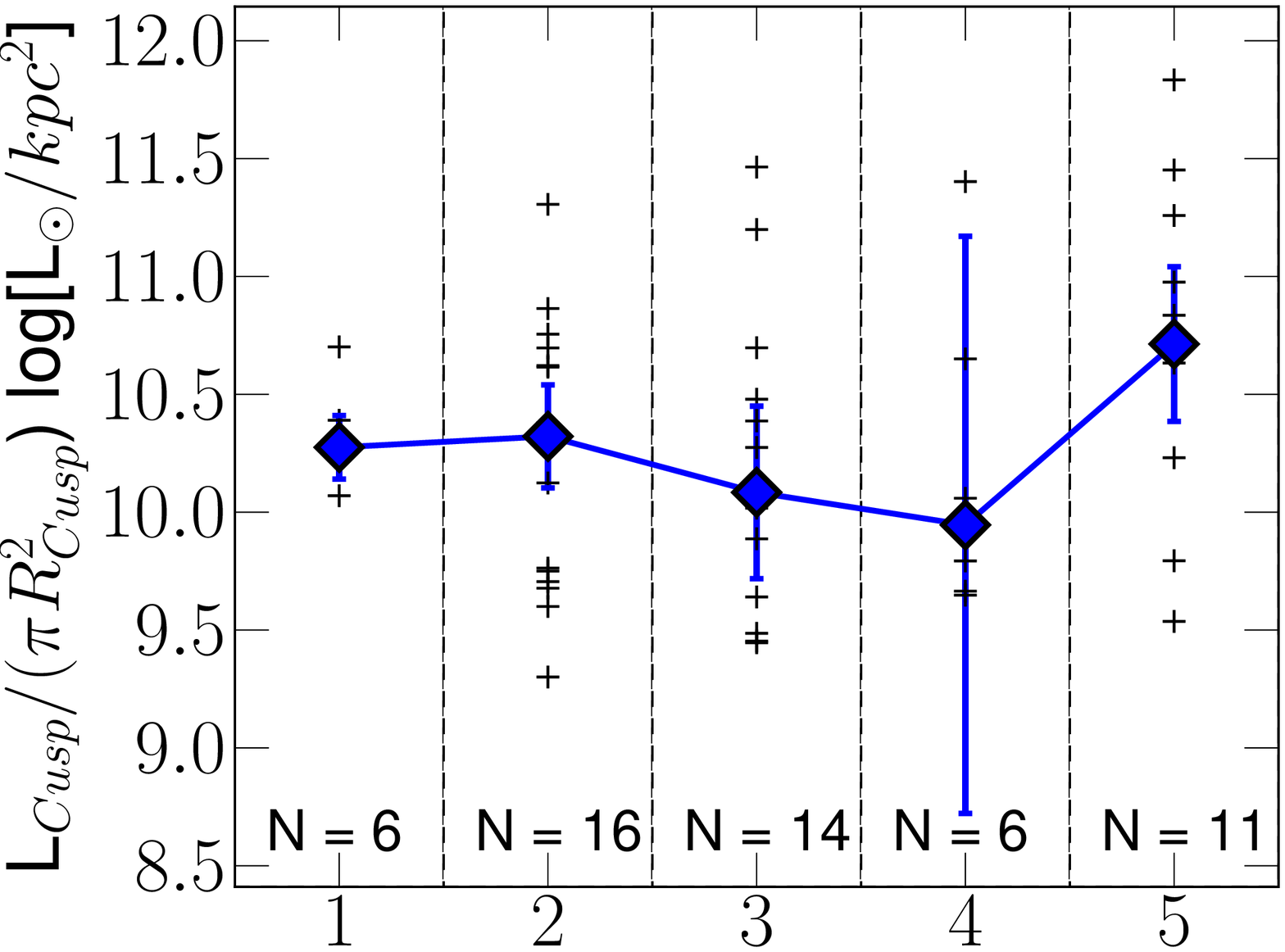}
\includegraphics[scale=0.2]{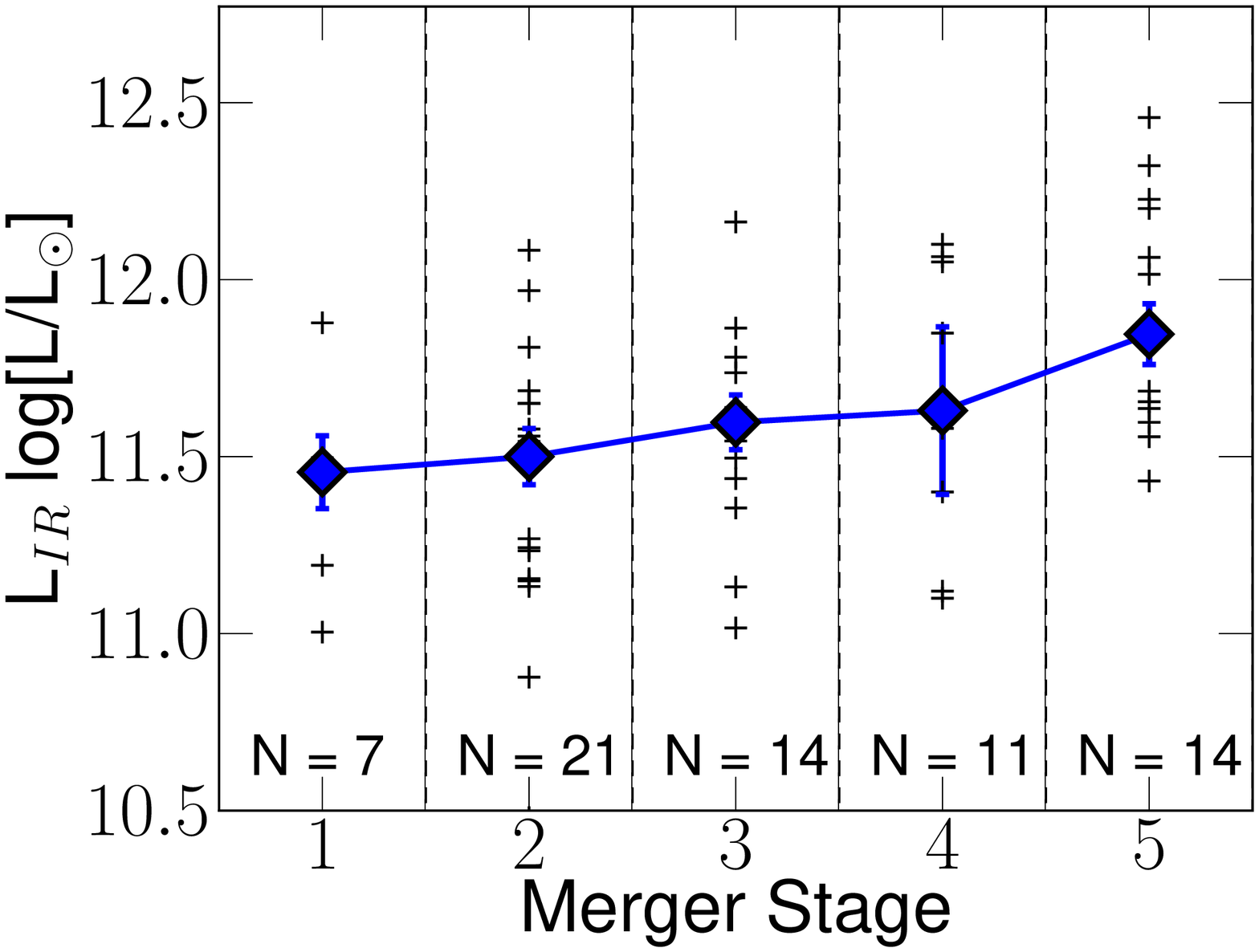}
\includegraphics[scale=0.2]{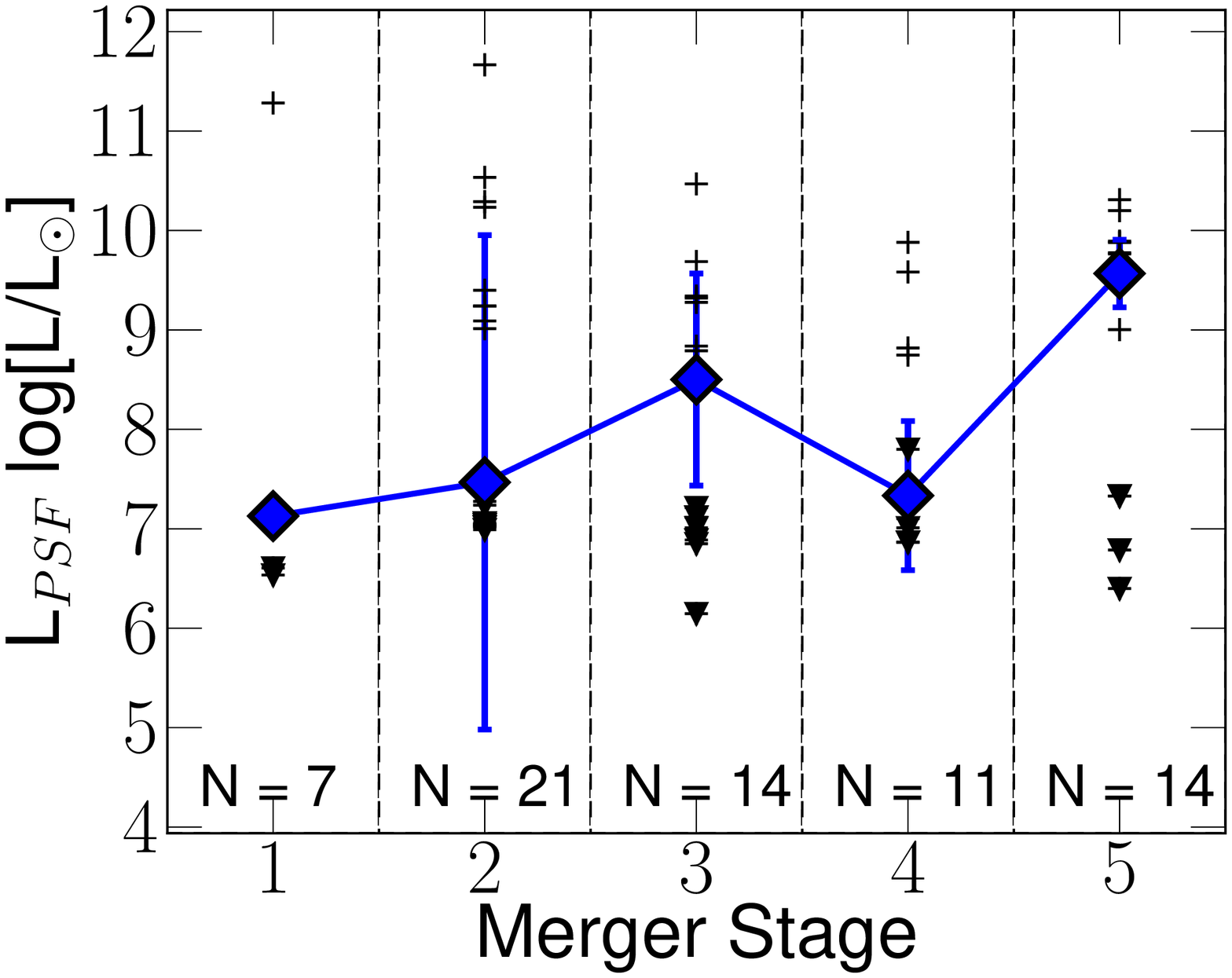}
\caption{The cusp properties as a function of merger stage for all sources with $\gamma>0.1$. Top left: the cusp slope ($\gamma$), top right: the cusp luminosity L$_{Cusp}$, mid left: the total NIR luminosity within 1~kpc radius (excluding the central unresolved PSF component), mid right: the cusp surface density L$_{cusp}/(\pi R_{cusp}^2)$, bottom left: the total IR (L$_{IR}$) luminosity of the galaxy, and bottom right: the luminosity of the central unresolved component (PSF). The number of sources in each merger stage bin is given at the bottom. The median values (with the standard error of the mean as blue errorbar) are indicated with blue diamond markers and connected with a solid line.}
\label{fig:ms_cusp}
\end{center}
\end{figure}

In Fig.~\ref{fig:ms_cusp} we show a number of nuclear properties as a function of merger stage (excluding MS 0): The cusp slope $\gamma$, the cusp luminosity L$_{cusp}$, the cusp surface density L$_{cusp}/(\pi R_{cusp}^2)$, and the total nuclear luminosity L$_{nuc}$ within 1~kpc radius. Most of these properties show evidence for an increase towards the final merger stage (MS~5). A small deviation is found for MS 1, which is probably due to the fact that this merger stage (galaxy pairs that show no interaction features) is likely a mix of merger and non-merger induced starbursts, as already discussed previously. The median value of the cusp slope $\gamma$ increases by a factor of 1.5 from MS~2 to MS~4/5, and almost a factor of two if core galaxies ($\gamma<0.3$) are taken into account. The statistical significance is tested via the Kolmogorov–Smirnov (KS) test which confirms that the distribution of $\gamma$ between pre/mid-merger and late-stage merger bin are not the same (at a probability level of 92\%). Also the cusp luminosity L$_{cusp}$ shows a peak at MS~4/5 with an increase of a factor of three (from log[$L_{cusp}/L_{\odot}$]=10.0 at MS~2 to log[$L_{cusp}/L_{\odot}$]$\sim$10.5 at MS~4/5) with a KS probability level of $>$99\%. The cusp surface density is significantly larger at the late merger stage (MS~5) of about a factor of 5--10 in comparison to earlier merger stages (MS 1--4) with a KS probability level of 99\%. 
This trend in cusp surface density is confirmed by the increase in the total nuclear luminosity measured within 1~kpc radius towards the late merger stage (increase factor of 4--5, KS probability level of 99\%). This measurement does not depend on the cusp radius which is important to test since the cusp radius has in general larger systematic fit uncertainties than the cusp slope as shown in \S~\ref{subsec:systest}. Since L$_{nuc}$ is defined within a 1~kpc radius, it is equivalent to the central surface brightness density within a unit circle area of 1~kpc radius. The  increase of total nuclear and cusp luminosity towards the late merger stage (MS5) is in agreement with the increase of the nuclear bulge density as shown in \cite{Haa11}. Note that for all of these measurements the central unresolved component (PSF) has been excluded.\par 

For completeness we present in the bottom panel of Fig.~\ref{fig:ms_cusp} the luminosity of the unresolved component L$_{PSF}$ and the luminosity L$_{IR}$ of individual merger components as a function of merger stage for all our sources, both cusp and core galaxies. The median value of the luminosity of the unresolved component L$_{PSF}$ (including also upper limits of non-PSF sources), which can be either an AGN or a very compact starburst, shows an increase of more than two orders of magnitude towards late merger stage, albeit with a large variation in each merger bin. The L$_{IR}$ luminosity of the merger components shows an increase along the merger sequence, in particular at the late merger stage. These two findings are in agreement with previous studies of LIRGs which have indicated an increase of AGN activity and IR luminosity of the combined merger system along the merger sequence \citep[e.g.,][]{Arm89, Vei95, Vei97, Kim98a, Kim98b, Mur99, Mur01}.

\subsection{Nuclear Morphology of Star-Formation: Comparison with HST Pa$\alpha$ emission line images}
\label{subsec:paschen}
To investigate whether cusps are associated with a distinct star-formation morphology, such as e.g. nuclear star-formation rings or spirals, we have searched the HST archives for existing NIR, narrow-band images (much preferred over optical data due to nuclear dust effects).  Four of our LIRGs have existing HST Pa$\alpha$ imaging data, previously discussed in \cite{Alo06}. All of them exhibit a cusp with a slope $\gamma>0.7$. \cite{Alo06} classified the nuclear morphological Pa$\alpha$ features as seen in their sample of 32 LIRGs into five categories: a)  compact ($<1$~kpc) nuclear Pa$\alpha$ emission, b) nuclear mini-spiral (inner 1--2 kpc), c) nuclear star-forming rings, d) large-scale (several kpc) Pa$\alpha$ emission with HII regions located in the spiral arms, and e) large-scale Pa$\alpha$ emission in highly inclined dusty systems. Since all overlapping galaxies have strong cusps, one might expect that all have the same morphological Pa$\alpha$ features; i.e. if there is a direct relationship between the strength of cusp and the current star-formation morphology.\par 
We find that all four galaxies have a strong Pa$\alpha$ emission component on roughly the same scale as the cusp radius. Two galaxies (NGC~3256 and NGC~3690) show an additional extended Pa$\alpha$ emission with HII regions located in the spiral arms. However, the morphology of the central Pa$\alpha$ component ranges from concentrated Pa$\alpha$ emission (NGC~3256 and NGC~3690) to nuclear star-forming ring (NGC~1614) and nuclear minispiral (MCG +12-02-001). 
This variety in morphology of the star-formation component (Pa$\alpha$ emission) seems to be in agreement with recent VLT-SINFONI integral field spectroscopy observations of 17 (U)LIRGs \citep{Piq12} which show a wide morphological variety in the ionized gas component while the warm molecular gas (H$_2$ emission) is highly concentrated in the nucleus. The similarity of the sizes of the central Pa$\alpha$ emission and the stellar cusps indicate that, at least in these four LIRGs, ongoing star formation may still be building cusps. Although the small number of sources for this study does not allow us to derive any decisive conclusions, the fact that no unique morphological star-formation feature can be associated with the stellar cups might suggest that either a larger range of nuclear star-formation morphologies and dynamical processes can build up a cusp, or that star-formation evolves through different morphological stages over the time-scale of the cusp formation. In particular star formation in the central region of NGC~1614 has been studied in detail using Br$\gamma$, H$_2$, Pa$\alpha$ observations, and near-to mid-IR spectroscopy \citep[e.g.][]{Hei99, Kot01, Alo01, Ima10} which revealed a star-forming ring of $\sim$300--600~pc diameter around the nucleus with no evidence of AGN activity \citep{Ols10}. \cite{Vai12} recently detected a broad inner ring of 3.3~$\mu$m polycyclic aromatic hydrocarbon (PAH) emission, suggesting an outward propagating ring of star formation. A scenario of outward propagating star formation would explain the lack of star-formation within the nuclear stellar cusp since most of the gas is likely already consumed or expelled from the inside of the stellar cusp. However, only a larger sample of high-resolution images of star-formation tracers and gas/stellar kinematics can answer in detail the question how the current star-formation and gas dynamics are linked to the build-up of the stellar cusp.

\subsection{Spatial Resolution Dependences and Unresolved Components}
\label{subsec:systest}

It is important to test the robustness of our results against possible systematic biases in our measurements. The advantage of the GOALS LIRG sample is its completeness in terms of all-sky coverage and IR flux selection ($f_\nu(60\mu m) \geq 5.24$ and log[$L_{IR}/L_{\odot}]>11.4$), being the largest sample of local LIRGs studied so far. However, some factors can possibly have an impact on the measured cusp properties, in particular AGN light contribution and finite spatial resolution.\par

The first case, AGN activity, might affect the observed NIR luminosity and hence one would expect a possible contribution to the measured cusp luminosity and slope. In principle any point source (unresolved component) is taken into account by subtracting the HST point-spread-function (PSF) from the images during the fitting process via GALFIT (see \S \ref{subsec:galfit}). Hence a possible contribution of AGN activity to the cusp light is not very likely since the measured cusps are resolved (typical sizes of 100--500~pc) while the main dust heating by AGN (500-2000~K) is generated  on scales of (10-100)~pc \citep[e.g.,][]{Soi01}.\par

\begin{figure*}
\begin{center}
\includegraphics[scale=0.36]{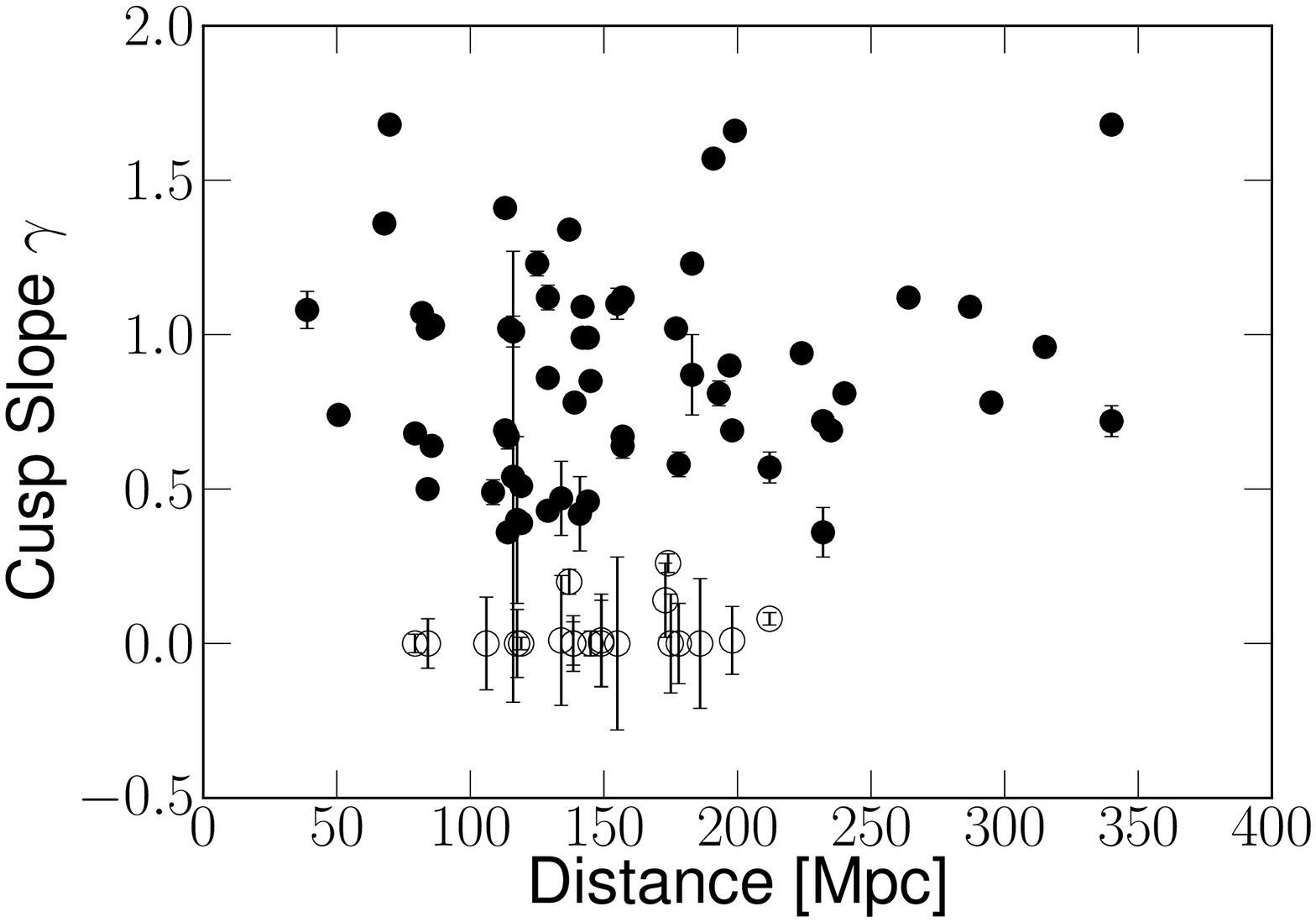}
\includegraphics[scale=0.36]{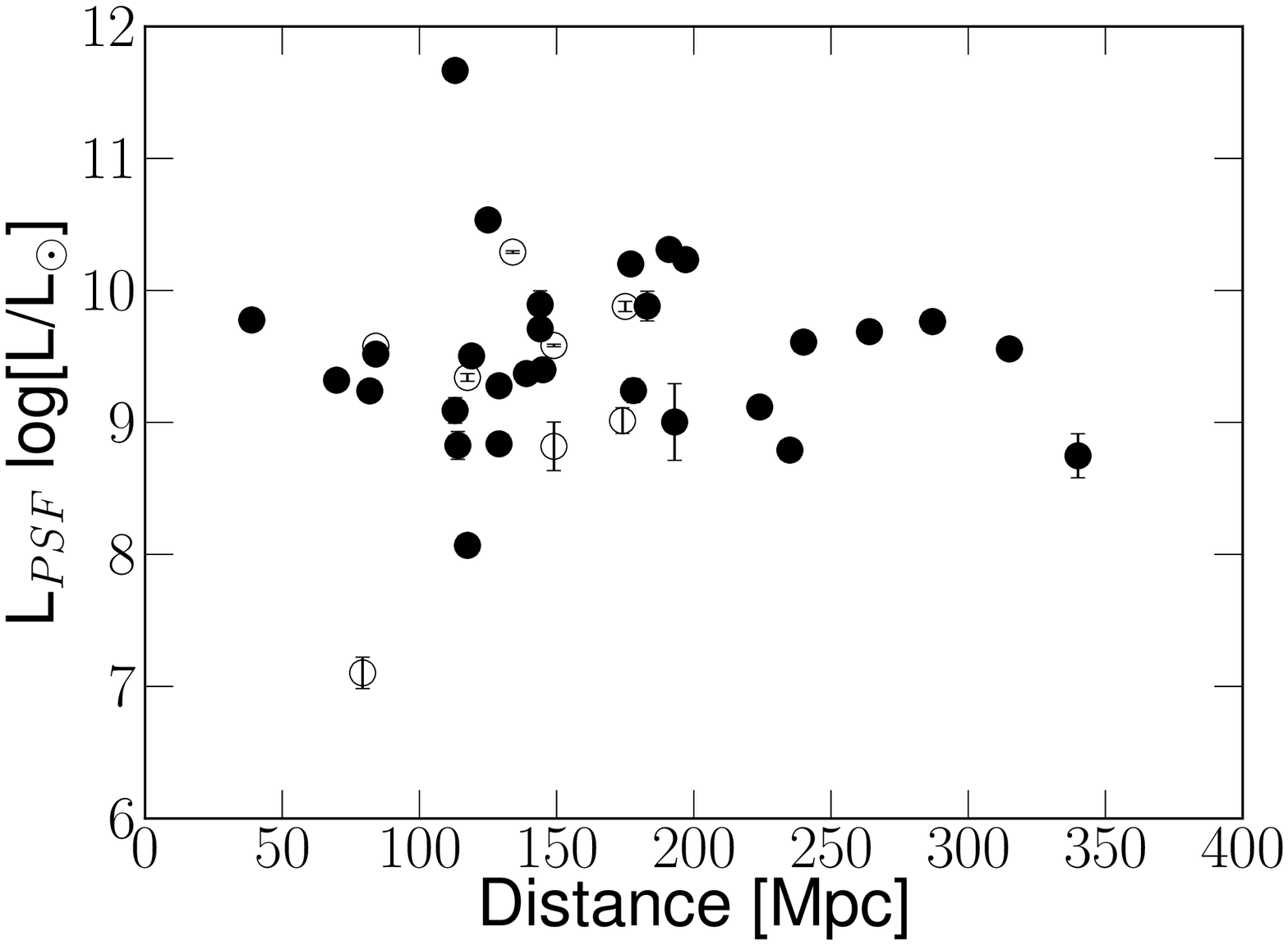}
\includegraphics[scale=0.36]{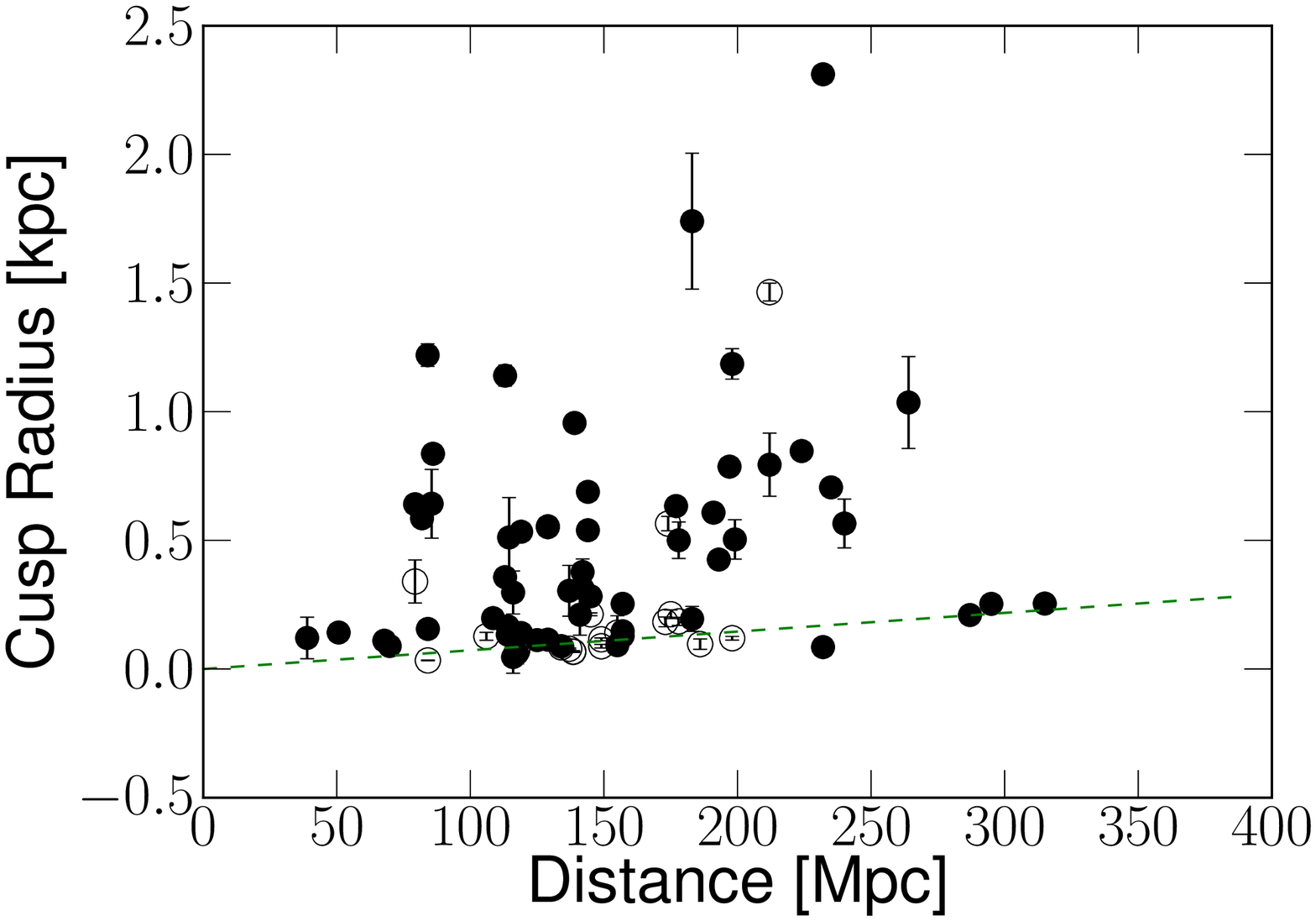}
\includegraphics[scale=0.36]{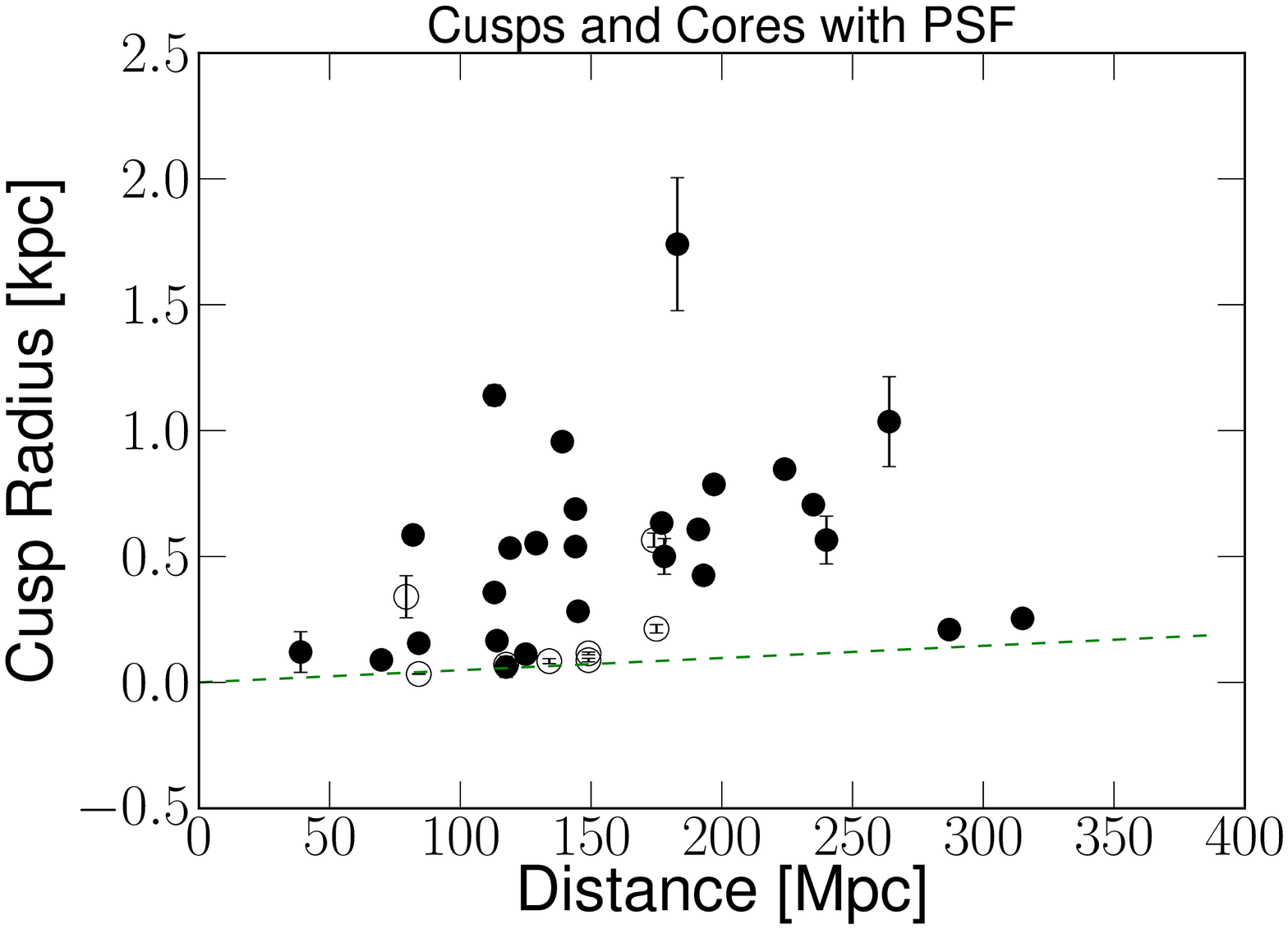}
\caption{Nuker parameters as a function of distance to the LIRGs; top left: the cups slope ($\gamma$), top right: the central unresolved component  (HST PSF), bottom left: the break radius, bottom right: the break radius for sources with a PSF. Cusps ($\gamma>0.3$) and cores ($\gamma<0.3$) are indicated with filled and open circle markers, respectively. The dashed line indicates the spatial resolution limit of the HST.}
\label{fig:test_dist}
\end{center}
\end{figure*}

\begin{figure*}
\begin{center}
\includegraphics[scale=0.45]{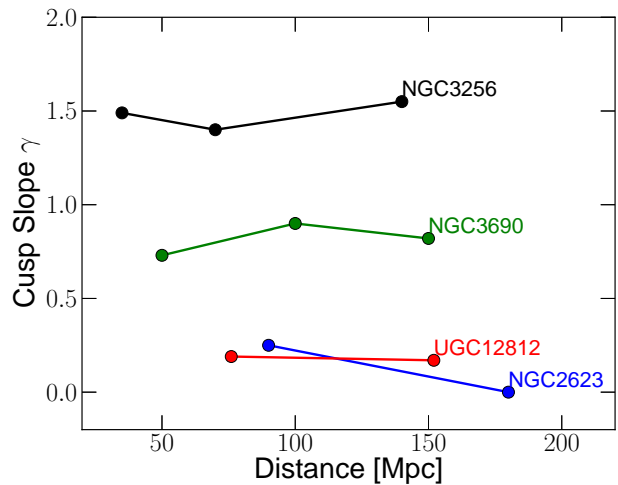}
\caption{Systematic test of possible distance (resolution) dependencies of the cusp slope parameter for four different galaxies. The GALFIT input images have been smoothed and resized in order to model the galaxies as they would appear at different distances.}
\label{fig:test_smooth}
\end{center}
\end{figure*}

The second case, a spatial size dependency, is more difficult to rule out, since the spatial resolution and hence the scale on which the cusps are resolved, depends on the distance of the LIRGs (ranging from ~30--350~Mpc for our sample). Given the small physical sizes of the nuclear cusps, it is important to verify that finite spatial resolution is not biasing our results. This is particularly important since more IR luminous LIRGs are typically found at larger distances than less luminous systems. In Fig.~\ref{fig:test_dist} we plot the cusp slope $\gamma$, the cusp break radius, and the luminosity of the central unresolved component (PSF)  as a function of distance for our sample. We cannot find any significant dependency as a function of the distance, indicating that distance related effects do not play a significant role statistically. 
Although this rules out a statistical dependence, we have also simulated directly how the nuclear slope would change as a function of distance for a representative set of LIRGs.
The four LIRGs tested (NGC~3256, NGC~3690, NGC~2623, and UGC~12812) are all nearby ($<$ 100Mpc), have no strong PSF component, and display different cusp slopes (ranging from $\gamma=0.2$ to $\gamma=1.5$). We have smoothed and resized the images using a Gaussian beam and image regridding to simulate how these galaxies would appear at larger distances, changing both, their spatial resolution and angular size of the galaxies. After that we applied the same GALFIT procedure for each simulated galaxy to obtain the cusp parameters as a function of distance. For consistency we have fitted the same physical sky region for each simulation. As shown in Fig.~\ref{fig:test_smooth}, the cusp slope $\gamma$ varies typically within a range of $\pm$(5--15\%) as a function of distance but no significant trend towards larger or smaller slopes with distance is found. The largest change appears for NGC~2623 from $\gamma=0.25$ to $\gamma=0.0$ if its distance were to double. The break-radius is slightly more sensitive to the spatial resolution with changes of 10--50\%, but with no trend towards larger or smaller radius as a function of spatial resolution. These results show that the high, but limited spatial resolution of HST cannot cause us to detect cusps in our LIRGs where none exist, which is accordance with our expectations.\par
 
On the contrary, if the cusp is unresolved, one would expect that its slope becomes flat ($\gamma=0$) since the cusp excess light is already taken into account by the central PSF or it is just poorly constrained. For instance, in the case of a finite resolution limit, \cite{Lau07} has shown that a Nuker slope of $\gamma<0.5$ can still have steep cusps if measured as the local slope at the resolution limit of the HST. Moreover, we cannot rule out that $\gamma$ might be additionally suppressed by the PSF component. This implies that the cusp fraction and luminosity in our sample is a lower limit. This is relevant for studying a possible relationship between AGN and cusp light as discussed in \S~\ref{subsec:compsb} and is also important for  the comparison to the slope in elliptical galaxies (see \S~\ref{subsec:ellipticals}).

\clearpage

\section{Discussion}
\label{sec:dis}

\subsection{The Formation of Nuclear Cusps: Stellar Masses and Implication for SF Time-scales}
\label{subsec:cuspmass}
One of the main questions related to the formation of nuclear cusps is how much stellar mass is generated in the center via gas dissipation. As shown in \S~\ref{subsec:lir} we find a strong dependency between the total IR luminosity L$_{IR}$ of a galaxy and its cusp NIR luminosity: All galaxies above log[L$_{IR}/L_{\odot}$]=11.9 have cusp luminosities of L$_{cusp}\gtrsim 10^{10}$L$_\odot$ with an average value $L_{cusp}/L_{\odot}=[3.8\pm1.9]\times10^{10}$, which is roughly five times larger than the average value of lower luminous LIRGs. Given our findings that all high-IR luminous galaxies are major mergers and that we see a similar increase of the cusp luminosity as a function of merger stage (from $L_{cusp}\sim 1\times 10^{10}$~L$_\odot$ to 3.2$\times 10^{10}$~L$_\odot$), we conclude that this increase in cusp luminosity $\Delta$L$_{cusp}$ is very likely due to merger-induced gas dissipation building up a nuclear stellar cusp.\par

The average increase in cusp luminosity during a major merger ($\Delta$~L$_{cusp}$)  is $\sim$2.5$\times10^{10}$~L$_\odot$, which corresponds to a build-up of stellar mass $\langle\Delta M_{cusp}\rangle$ of $(7\pm3.5)\times10^{9}M_\odot$ if we assume a standard mass-to-light ratio in the H-band $\Upsilon_{H}=0.3\pm0.15$. The applied $\Upsilon_{H}$ and its uncertainty is derived from STARBURST99 simulations \citep{Lei99} using Geneva High and Padova AGB tracks, Kroupa IMF, and instantaneous star formation with a mass of $10^6$~M$_\odot$ and solar metallicity. The lower limit of $\Upsilon_{H}=0.15$ is derived from the assumption that the NIR is dominated by young stars (4~Myrs) while the upper limit of $\Upsilon_{H}=0.45$ is a mix of young, intermediate (50~Myrs), and old stellar populations (few 100~Myrs) contributing equally to the NIR light. This ratio is consistent with a) derived stellar masses of our LIRGs from SED fitting \citep{U12} using the models of \cite{Bru03}, and b) recent estimates of dynamical masses obtained for the central 1~kpc in LIRGs \citep{Hin06}. For the latter case we compared the total H-band luminosity (excluding central PSF) within the central 1~kpc in the five galaxies that overlap with the sample of \cite{Hin06} and assume that the dynamical mass in the center is dominated by stars.\par

One important question is how long the current starburst will last. Merger simulations predict a strong starburst with a SFR of 50--500 M$_\odot$~yr$^{-1}$ during the final coalescence of the two gas-rich nuclei \citep[see e.g.][]{Mih94b, Com02, Spi05a, Spi05b, Cox08, Hop08b, Bou11}. However, the predicted time-scales typically show a large variation and uncertainties (from 10~Myrs to a few 100~Myrs), depending on which model, initial conditions, gas fraction and feedback processes are used. We have shown in \S~\ref{subsec:ms} that most of the cusp mass is built up between merger stage 3 (two nuclei in common envelope) and 4/5 (final coalescence of nuclei). This corresponds to an average timescale of $\sim$500~Myrs, which is based on the projected nuclear separation distance and mass ratios of the two merging galaxies \citep[see][]{Haa11}.\par 

Even so, it is possible that star-formation occurs in several bursts during this period. In principle, our observations should be able to provide important constraints given that the stellar cusp mass increase is equivalent to the integrated SFR over time during the late-stages of a major merger. If we assume an average SFR during the final starburst phase when most of the mass of the cusp is built up, the timescale of this starburst phase can be approximated by $t_{SB}=\langle \Delta M_{cusp} \rangle/ \langle SFR_{SB} \rangle$. This assumption may not necessarily be valid for low IR luminosity LIRGs and non-mergers, which have smaller cusp luminosities and we cannot rule out the possibility that there is a significant cusp contribution that has formed at an earlier phase of their galaxy evolution. However, our cusp mass estimates are based on the most luminous galaxies in our sample (ULIRGs, log[L$_{IR}$/L$_{\odot}$]$>$12.0; note that cusp contribution of low-IR luminous galaxies is subtracted in $\Delta M_{cusp}$) which are all major mergers at primarily their late stage and believed to build up most of the cusp mass during one major merger event with most of the remaining gas expelled or ionized.
The average SFR for our galaxies with log[L$_{IR}$/L$_{\odot}$]$>$11.9 (mean log[L$_{IR}$/L$_{\odot}$]=12.1) is $\sim$220~M$_\odot\;yr^{-1}$, which is based on the SFR-L$_{IR}$ luminosity relation \citep{Ken98}. We have applied a correction for extended L$_{IR}$ emission, which is typically 50\% \citep{Dia10, Dia11}, see also \S~\ref{subsec:compsb}. Given $\langle \Delta M_{cusp} \rangle$ of $(7\pm3.5)\times10^{9} M_\odot$ (see above) and assuming a constant SFR, we estimate an average timescale for the final starburst phase of $\sim60\pm30$~Myr to build up $\langle \Delta M_{cusp} \rangle$. This is essentially the cusp doubling timescale at the current SFR. Note that $\langle\Delta M_{cusp} \rangle$ includes only the stellar cusp mass that is observed so far, which does not rule out any future star-formation fueled by the remaining nuclear gas reservoir, which can be significant.  On the other hand, if we average the star formation over a timescale of 500~Myrs (from merger stage 3 to 4/5), when most of the cusp build-up occurs, we derive an average SFR$_{500}$ of $\langle \Delta M_{cusp} \rangle$/$t_{500}$=(14$\pm$7)~M$_\odot\;yr^{-1}$.

\subsection{On the Nature of Nuclear Cusps in Non-interacting LIRGs}
\label{subsec:non-merger}

One puzzling finding of this study is that all 7 non-interacting LIRGs exhibit nuclear cusp light profiles, which is a slightly larger fraction than for interacting LIRGs (see \S~\ref{subsec:dist}). 
Although there are only 7 non-interacting LIRGs in our sample, pre-merger LIRGs that show no interaction features (likely a mix between non-interacting and early stage mergers) also have a larger fraction of cusp galaxies than the mid- and late stage mergers.  However, even more interesting is that these non-interacting LIRGs have a significant cusp component at all and cusp luminosities similar to late-stage merger. 
In principle there is no reason why these nuclear dissipational processes can not occur in galaxies that are not participating in major mergers. However the exact mechanism that leads to the central starburst and build-up of a cusp  in these galaxies is still uncertain.  Moreover, is the nuclear cusp formed during the current phase of star-formation or in a previous phase of star-formation due to an earlier merger event? If the latter is the case, an intriguing possibility is that the presence of a progenitor nuclear cusp may actually provide the physical conditions for the recent nuclear starburst which adds cusp mass through dissipation. One possibility would be that a progenitor cusp could provide the critical gravitational potential well to compress the gas and trigger instabilities in the gas that leads to dissipation and a starburst.\par

Other possible mechanisms might be minor mergers or secular evolution processes. Minor mergers can channel gas into the central region of a galaxy and may trigger a nuclear starburst, but with an efficiency that declines in approximately linear fashion with the mass ratio \citep{You07, Hop08a}. Secular evolution processes are suggested to build up pseudo-bulges and dense central concentrations of gas and star formation due to bar or oval stellar distribution which exert gravitational torques on the gas and might trigger gas inflow \citep[see e.g.,][]{Lau02, Gar05, Boo07, Haa09}. However, measurements of local disk galaxies show that these structures evolve on much larger time-scales of a few billion years at nuclear SFRs at least one order of magnitude below the low IR luminosity end of LIRGs \citep[for a review see][]{Kor04}. Even so, none of the non-interacting galaxies in our sample, except IC~5298, has a strong bar based on its optical and NIR light distribution. Our multi-wavelength HST images show that most of these galaxies (e.g. MCG~-03-04-014, NGC~695, VII Zw 031, IC~5298) have non disturbed disks with strong dust and stellar spiral arms from the outer disk to the very center. \par

We can also rule out the scenario that most of the apparent non-interacting LIRGs are in a more quiescent post-merger phase where stellar tidal tails and streams have faded away: Merger simulations \citep[e.g.]{Her93, Her93b, Spi05b, Cox08, Wan12} have shown that tidal tails in major mergers remain typically present (luminosity $>$1\% of disk) until $>$10 dynamical time-scales (corresponding to a post-merger timescale of $\geq$2~Gyrs). The fact that tidal tails are not fading away quickly after the coalescence of the nuclei is supported by our HST observations in the optical light which show no significant decrease of the ratio of tidal features to disk light \citep{Kim12}. In contrast, the post-starburst timescale is expected to last only a few dynamical time-scales, much shorter than the timescale until tidal tails fade away.

\subsection{Increasing Compactness along the Merger Sequence - Comparison to high-redshift Studies}
Recent deep high-resolution ground- and space-based surveys have revealed that a significant fraction of massive galaxies at z $>$ 1 are surprisingly compact \citep[e.g.,][]{Dad05, Tru06, Tru07, Cim08, Fra08, Dam09, Tar11}  with effective radii well below the local galaxy size-mass relation. Within the last three years even deeper observations (in particular CANDELS) provided mounting evidence for the truly compact nature of many high-redshift galaxies \citep{Dok10, Rya12, Szo12, Che12, Kar12, Bru12}. In particular \cite{Bru12} have decomposed the structure of the most massive galaxies ($\sim$200 galaxies) in the CANDELS-UDS field (HST F160-band images) with photometric redshifts $1 < z < 3$ using single S\'{e}rsic and bulge/disk models. This study finds that in particular bulge-dominated objects show evidence for a growing bimodality in the size-mass relation with increasing redshift. The fraction of bulges consistent with the local size-mass relation is 10--25\% at $1<z<3$ and compact bulges have effective radii a factor of roughly 4 smaller at z $>$ 2  than local ellipticals of comparable mass. In the local universe such massive galaxies are primarily bulge-dominated, while galaxies at higher redshift (z $>$ 2) tend to be more disk-dominated.\par 

Given the argument that LIRGs are very abundant and dominate the total IR energy density emitted at red-shifts of  $1<z<3$ \citep{Elb02, LeF05, Cap07, Bri07, Mag09, Ber11}, a possible link between the morphological evolution of galaxies during their LIRG phase and the compactness of galaxies observed at high-z might be suggestive. As shown in recent merger simulations by \cite{Hop08a} a gas inflow of $\sim$10\% of the total gas shrinks the effective bulge radius to about half its previous size and subsequently enhances the bulge surface brightness. \cite{Haa11} have shown that the effective bulge radius decreases about a factor of 2.5, on average, from non-interacting/early stage to mid-stage merger and again a factor of  2.2 from mid-stage to early stage merger. The bulge surface brightness increases on average about a factor of 5 from non-interacting/early stage to mid-stage merger and a factor of 10 from mid-stage to late stage merger. In this study we confirm the increase in compactness given the increase of the nuclear surface density within one kpc (which is independent from bulge or cusp radius) and cusp surface density about a factor of 2 from early/mid-stage to late-stage merger. Note that the unresolved component has been already subtracted in all our studies.
This suggests that both components of a merger remnant, one formed through violent relaxation of progenitor stars \citep[dissipationless, spheroid, see][]{Haa11} and the other through gas dissipation (cusp; see density in Fig.~\ref{fig:ms_cusp}), become more compact towards the late merger stage due to gas infall.  Whether triggered by major mergers or by other instabilities, a large gas infall and nuclear starbursts are also expected in high-z galaxies, which might be a possible explanation for the dramatic increase in compactness in these sources.  
 
\subsection{Are LIRGs evolving into Elliptical Galaxies: Comparison of Nuclear Properties as Tracer of their Evolutionary Linkage}
\label{subsec:ellipticals}
One important question is what fraction of present day ellipticals could be formed by gas-rich major mergers that have local gas fractions? The picture that mergers transform disk galaxies into massive elliptical and S0 galaxies is often suggested by observations based on the comparisons of the large-scale appearance (at radii of several kpc) of the radial light profile of old merger remnants with those of ellipticals \citep[e.g.,][]{Wri90, Jam99, Rot04, Gen01, Tac02}. However, the end result of a major merger may depend on several factors such as merger-mass ratio, gas-fraction and feedback processes due to supernovae or AGN activity \citep[e.g.][]{Spi05a, Bou05, Rob06}. Even so, fitting the large-scale profile of ongoing and recent major mergers is often not possible since the stellar distribution is still dominated by violant relaxation. This is not the case for merger nuclei which undergo a much more rapid relaxation due to the decrease of the dynamical timescale with smaller radial distance from the center of a galaxy. 
Early type galaxies show a strong bimodality: Cusp elliptical galaxies are relatively faint in optical light, rotate rapidly, have disky isophotes, host radio-quiet AGN and do not contain large amounts of X-ray-emitting gas; core elliptical galaxies are brighter in the optical, rotate slowly, have boxy isophotes, radio-loud AGN and diffuse X-ray emission. For a summary of these observational findings see \cite{Kor09, Silk09, Nip10} and references therein. This dichotomy suggests that core and cusp ellipticals went through different evolutionary processes. \textbf{Here we compare for the first time the cusp and core properties in the few remaining gas-rich major mergers still present in the local universe to those of present day core and cusp dominated ellipticals.}\par

We have chosen the sample of \cite{Lau07} as a reference sample of early-type galaxies to which we compare the cusp properties of our LIRG sample. The sample compiled by \cite{Lau07} is the largest sample of 219 early type galaxies and has been derived by homogenizing a variety of studies \citep{Lau95, Fab97, Qui00, Rav01, Res01, Lai03, Lau05}, primarily observed in V- and H-band. Even though that this sample combines optical and NIR observations, it is unlikely that this affects the nuclear slope $\gamma$ because \cite{Sei02} have found that the NIR structural properties and $\gamma$ parameter of spiral galaxies are very similar to those obtained in the optical wavelength region for the same systems. Note that elliptical galaxies must be even less affected by any color-dependence of $\gamma$ than spiral galaxies, which usually show more local color-variations.
The \cite{Lau07} sample is the best match to our analysis because 1) all observations are taken with HST at high angular resolution, 2) the cusp parameters are derived using the Nuker law parameters, and 3) it overlaps with the host galaxies luminosity and distance range of our LIRG sample. For the hosts  magnitude we have taken the catalogued 2MASS flux H-band which is a fair proxy of the stellar mass of the galaxies and is not biased by dust extinction. The 2MASS K-band might be less affected by red supergiants, but has likely more contribution from warm dust emission and is by far not as sensitive in magnitude as the H-band. Using the catalogued 2MASS values rather than converting the visible magnitudes in \cite{Lau07} and our measured H-band fluxes, we can ensure that the comparison between both samples is not biased by instrumental and measurement method and includes only the galaxies within the same mass range. Moreover, we have applied a distance cut of  $>$20~Mpc for the Lauer sample, which sorts out most of the small galaxies which are not in the mass range of the LIRGs (M$_H>-21$). The mean H-band absolute magnitude is very similar with -24.27~mag and -24.29~mag for the \cite{Lau07} and LIRG sample, respectively. For consistency we have compared the optical magnitudes measured by \cite{Lau07} with the catalogued 2MASS H-band magnitudes within our magnitude range and found a very good linear relation.\par 

\begin{figure}
\begin{center}
\includegraphics[scale=0.45]{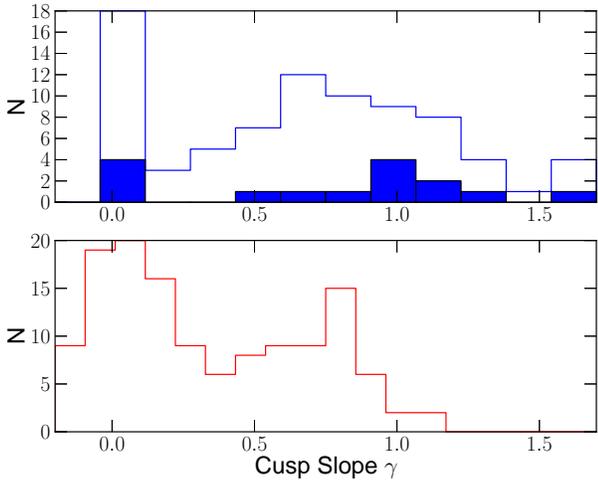}
\caption{Histogram of the Nuker gamma parameter for LIRGs (top panel) and early type galaxies (bottom panel; sample from Lauer et al. 2007) within the same total luminosity range (-25.5$<$M$_H<$-22.5). Late stage mergers are indicated as filled histogram in the top panel.}
\label{fig:hist_ell_lirgs}
\end{center}
\end{figure}

In Fig.~\ref{fig:hist_ell_lirgs} we compare the cusp slope ($\gamma$) distribution between LIRGs and ellipticals averaged over the same magnitude range. This comparison reveals the following:\textbf{ Within the -23 $<$M$_H<$-25.5 range, which contains more than 98\% of the LIRGs in our sample, we find a significantly larger ratio of cusp ($\gamma>0.3$) to core ($\gamma<0.3$)  galaxies for LIRGs (cusp/core$\geq$3.2) than for early-type galaxies (cusp/core$\approx$0.7).} This means that at least 76\% of LIRGs are cusp sources while only 41\% of early type galaxies harbour cusps within the same magnitude range. The mean and median values of the host absolute magnitude M$_H$ for both sample is given in Table~\ref{tab:comp}. The difference between LIRGs and ellipticals is likely larger since there might be a signifiant fraction of cusps unresolved with NICMOS (resolution of 0.15\arcsec in comparison to 0.05\arcsec for WFPC2 in F555W) and subsequently taken into account by the PSF component that is fitted with GALFIT. \par

\begin{figure}
\begin{center}
\includegraphics[scale=0.95]{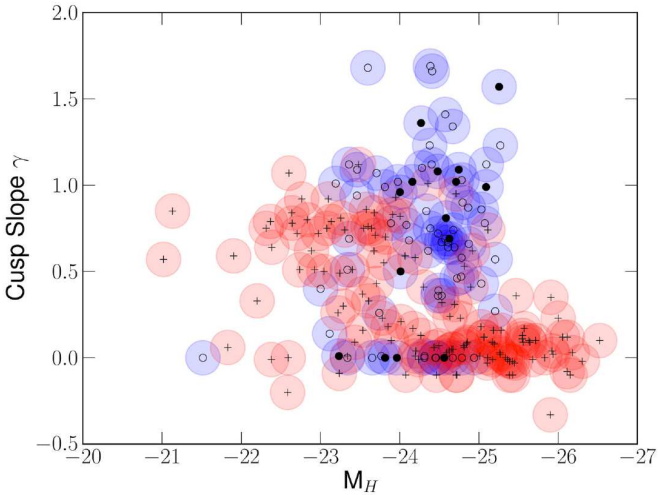}
\caption{Overview of the Nuker gamma parameter as a function of the total absolute H-band magnitude (2MASS). While early type galaxies \citep[red, plus marker][]{Lau07} show a strong dependence of the presence of cusp and cores as a function of host magnitude, LIRGs (blue, circle marker) do not follow this trend and represent a distinct population. Late stage mergers are indicated with a filled circle.}
\label{fig:comp_cusp}
\end{center}
\end{figure}

In Fig.~\ref{fig:comp_cusp} we compare $\gamma$ as a function of the host absolute magnitude M$_H$ between LIRGs and early type galaxies. The strong dependence of $\gamma$ in early type galaxies on visible magnitude \citep{Lau07} is also apparent in the NIR (H-band). \textbf{However, LIRGs do not follow this trend and form a clearly distinct population in the M$_H$-$\gamma$ diagram: Most of the LIRGs with cusps have larger H-band luminosities (on average one order in magnitude) than cusp ellipticals.} 
The distribution of LIRGs in the M$_H$-$\gamma$ diagram does not depend on whether they are non-merger, early stage merger or late stage merger, with the exception that all non-merger LIRGs are cusp galaxies (as already pointed out in \S.~\ref{subsec:ms}). Moreover the distribution seems also not to depend on the variation in spatial resolution (i.e. we find a similar distribution if only LIRGs are taken into account with distances of less than 100~Mpc). This means for most of the LIRGs for which we found a cusp, we would have expected to find a core given the galaxies' luminosity and distribution of ellipticals.\par

One intuitive explanation for the significantly larger cusp fraction in LIRGs than in elliptical galaxies would be the destruction of cusps after the LIRG phase has passed. If this would be the case a significant fraction of cusp LIRGs would evolve into massive core galaxies since the host mass of cusp LIRGs and core ellipticals is roughly the same. Regarding Fig.~\ref{fig:hist_ell_lirgs}, this would mean that half of cusp LIRGs would move down in cusp slope to end up as core ellipticals. However, a destruction or flattening of the nuclear stellar cusp cannot be easily accomplished without any major disruption. Theoretical models predict that nuclear stellar cusps remain preserved in the central light profile of a galaxy during a wet merger event \citep{Hop09b}.
In particular cusp destruction via massive BH binaries has been proposed, but it can only remove a stellar mass of the order of the combined BH mass \citep[see e.g.][for a review]{Mer05}, which is much smaller than the cusp masses in our sample. In late stage LIRGs the mean cusp mass is $\sim3 \times 10^{10}$~L$_\odot$ versus a mean bulge mass of $\sim15 \times 10^{10}$~L$_\odot$ in the H-band \citep[see][]{Haa11}; the BH mass is $\sim$400 times smaller than the bulge mass estimated by the H-band luminosity \citep{Mar03}. Therefore, either cusp destruction via BH binaries is much more efficient (mass removal of at least 10 times the combined BH mass) than previously assumed or another mechanism destroys the cusp after the LIRG phase has passed.\par

One mechanism that could possibly lead to a strong reduction of the cusp and the formation of a core is a gas-poor ('dry') merger of two cusp galaxies \citep[see][]{Lau07, Hop09b, Kor09}. Indeed observations of elliptical galaxies have shown evidence for a correlation between a central stellar light deficit $L_{def}$ in core ellipticals with the mass of the central stellar BH and the bulge velocity dispersion of their host galaxies \citep{Kor09b}, suggesting that cores are scoured by BH binaries. Interestingly this study does not find any correlation between BH mass and the excess light in cusp ellipticals, which suggests that scouring by BH binaries is only efficient in dry (gas-poor) mergers, which are believed to form core ellipticals, and not in wet (gas-rich) mergers that form nuclear cusps. Given our results, this would imply that most LIRGs in the early universe must have experienced two subsequent evolutionary processes before they could end up as core ellipticals: First, most of their gas must have been expelled, and secondly, they must have been involved in at least one more major merger event with another gas-poor galaxy. The fact that present day massive ellipticals have primarily cores suggests that these two steps, gas ejection and subsequent gas-poor re-merger, must have happened during a very short episode in cosmic time since a long-term evolution would have lead to a mix with gas-rich galaxies and couldn't have destroyed their cusps efficiently.\par

Another scenario would be that star formation was quickly shut down in gas-rich galaxies (e.g. due to feedback from supernovae, hot stars, and AGN) before gas could accumulate in the very center of massive ellipticals, which would prevent a central cusp from forming and build up a stellar core distribution instead. The reason that these feedback processes are less strong in today's LIRGs may lie in the fact that gas fractions and merger densities were much higher at cosmic times when most massive ellipticals formed than in local LIRGs. Evidence that the cusp build-up might stop in some galaxies before the final coalescence of the nuclei comes from a detailed multi-wavelength study of Markarian~266 \citep[here named NGC~5256 north and south;]{Maz12} which has two AGN with kiloparsec-scale separation and shows evidence that the dust outflow phase can begin in a LIRG well before the galaxies fully coalesce.\par

Not only massive core ellipticals, but also most cusp ellipticals are not remnants of gas-rich major mergers like the local LIRG population, which are on average one order of magnitude more luminous in the NIR light (i.e. more massive) than cusp ellipticals. One possible explanation for this mass discrepancy is that cusp ellipticals formed at an early phase of the universe when most galaxies were smaller and have not reached the masses as observed for today's gas rich major mergers. Once formed, these cusp ellipticals must have not gained significant mass via subsequent merger, unlike core ellipticals which likely evolved through several subsequent dry merger events. The reason why we find almost no cusp LIRGs today at smaller masses could be possibly explained by the fact that galaxies in the early universe had significantly larger gas fractions, eventually enabling them to reach critical gas pressure rates and instabilities to trigger central starbursts at much smaller host galaxy masses.


\section{Summary}
\label{sec:sum}
The formation of nuclear stellar cusps requires powerful compact starbursts, which makes (U)LIRGs one of the most representative galaxy population for studying the actual formation of cusps in extreme starburst galaxies. We have analysed the nuclear stellar structure and cusp properties using 2-dimensional fitting of the Nuker profile for a complete sample of 85 (U)LIRGS (with 11.4$<$log[L$_{IR}/L_{\odot}$]$<$12.5) in the GOALS sample based on HST NIR imaging, and have applied mid-IR diagnostics for AGN/starburst characterization and SFR estimations. The main results are summarized as follows:

\begin{itemize}
\item Nuclear stellar cusps are resolved in 76\% of (U)LIRGs while the remaining 24\% have flat nuclear profiles (core galaxies) with nuclear slopes $<0.3$. The cusp strength and luminosity increase with far-IR luminosity (excluding AGN), confirming models that recent starburst activity is associated with the build-up of stellar cusps. In particular, all galaxies above log[L$_{IR}/L_{\odot}$]=11.9 have cusp luminosities of ($L_{cusp}\gtrsim 10^{10}$~L$_\odot$) with an average value $L_{cusp}/L_{\odot}=[3.8\pm1.9]\times10^{10}$, which is roughly five times larger than for lower IR luminous galaxies (log[L$_{IR}/L_{\odot}\sim11.5$]). An average increase in cusp luminosity $\Delta$L$_{cusp}\sim$2.5$\times10^{10}$~L$_\odot$ is found as a function of far-IR luminosity and merger stage, which corresponds to a build-up of stellar mass $\Delta M_{cusp}$ of $(7\pm3.5)\times10^{9}M_\odot$. Most of the cusp build-up seems to occur within a few 100~Myrs during the final coalescence of the nuclei.  If we assume that most of the cusp mass is built up during one starburst, a timescale of $\sim60\pm30$~Myr would be required based on the average SFR in LIRGs.

\item The nuclear stellar surface brightness profiles of local (U)LIRGs are very different from those of present day's early-type galaxies with comparable masses. Our comparison between (U)LIRGs and a large sample of local elliptical galaxies (within the same host galaxy mass range) reveals (a) a significant larger cusp/core ratio in (U)LIRGs ($\gtrsim$3.2:1) than in ellipticals ($\lesssim$0.7:1), and (b) (U)LIRGs do not follow the cusp to core dependency of ellipticals as a function of host galaxy mass (M$_H$-$\gamma$ diagram) and clearly represent a distinct population. The majority of the cusp (U)LIRGs have on average host luminosities similar to core ellipticals, which is one order in magnitude larger than for cusp ellipticals. Moreover, gas-rich late stage mergers show no indications of a cusp destruction by BH-binary scouring during the coalescence of their nuclei. 
These results suggest that the progenitors of present day's massive ellipticals must have either expelled most of their gas during a brief episode in cosmic time and destroyed their cusps in subsequent gas-poor re-merger events, or that star formation was quickly shut down before gas could efficiently accumulate in the very center to build up a cusp (e.g. via feedback from supernovae, hot stars, and AGN due to higher gas fractions at high-z). The reason why there are almost no cusp LIRGs observed at smaller masses could be possibly explained by the fact that galaxies in the early universe had significantly larger gas fractions, eventually enabling them to reach critical gas pressure rates and instabilities to trigger central starbursts at much smaller host galaxy masses.

\item With increasing cusp strength and luminosity along the merger process, the nuclear structure becomes more compact which is observed as an increasing cusp surface density (factor of 5--10) and total nuclear surface density within a radius of 1~kpc (factor of 4--5). This result is in agreement with the significant increase of the bulge surface density shown in \cite{Haa11}, suggesting that both components of a merger remnant, one formed through violent relaxation of progenitor stars (dissipationless, spheroid) and the other through gas dissipation (cusp), become more compact towards the late merger stage due to gas infall, which might be a possible explanation for the dramatic increase in bulge compactness in high-redshift studies (e.g. CANDELS).  



\item Comparison of the nuclear structure in four cusp LIRGs observed in the HST H-band to corresponding HST Pa$\alpha$ emission line images reveals that the Pa$\alpha$ emission has various morphologies (nuclear star-forming rings, nuclear minispiral, and concentrated emission), suggesting that there is apparently no single morphological star-formation feature associated with the build-up of cusps. However, all four galaxies have a strong Pa$\alpha$ emission component with roughly the same size as the cusp, which demonstrates that the current star-formation is still linked up with the build-up of the cusp.

\item Evidence for ultra compact nuclear starbursts is found in $\sim$13\% of LIRGs, which have a strong unresolved central light component but no significant evidence of an AGN.

\end{itemize}

The authors wish to thank Robert Joseph for a careful review of the manuscript and many valuable suggestions to improve this paper. The authors are very grateful to Tod R. Lauer for helpful discussions on the characterisation of cusp and cores and the comparison between LIRGs and elliptical galaxies. This research has made use of the NASA/IPAC Extragalactic Database (NED) and the Infrared Science Archive (IRSA) which are operated by the Jet Propulsion Laboratory, California Institute of Technology, under contract with the National Aeronautics and Space Administration. Support for this work was provided through grant HST~GO 11235.01-A by NASA from the Space Telescope Science Institute, which is operated by the Association of Universities for Research in Astronomy, Inc., under NASA contract NAS 5-26555. V. Charmandaris would like to acknowledge partial support from the EU FP7 Grant PIRSES-GA-2012-316788. L. Armus and D.B. Sanders acknowledge the hospitality of the Aspen Center for Physics, which is supported by the National Science Foundation Grant No. PHY-1066293.

\clearpage
\newpage
\onecolumn 

\begin{flushleft}
\begin{footnotesize}
\begin{longtable}{@{}lrrlrrl@{}}
\caption{Sample Overview}\\
\hline \hline
\multicolumn{1}{c}{\textbf{Name}}  & \multicolumn{1}{c}{\textbf{RA}} & \multicolumn{1}{c}{\textbf{DEC}} & \multicolumn{1}{c}{\textbf{log($L_{IR}/L_{\odot}$)}} & \multicolumn{1}{c}{\textbf{D}} &  \multicolumn{1}{c}{\textbf{HST ID}} &\multicolumn{1}{c}{\textbf{Instrument}}\\
\multicolumn{1}{c}{}& \multicolumn{1}{c}{(J2000)} & \multicolumn{1}{c}{(J2000)} & \multicolumn{1}{c}{} & \multicolumn{1}{c}{[Mpc]} & \multicolumn{1}{c}{}&\multicolumn{1}{c}{} \\
\hline
\endfirsthead
\multicolumn{7}{c}%
{{\bfseries \tablename\ \thetable{} -- continued from previous page}}\\

\hline
\endhead

\hline \multicolumn{7}{r}{{Continued on next page}} \\ 
\endfoot

\endlastfoot
NGC~0034&00:11:06.61&-12:06:27.4&11.49&84.1&7268&NICMOS\\
ARP~256N&00:18:50.12&-10:21:42.6&11.48&117.5&11235&NICMOS\\
ARP~256S&00:18:50.90&-10:22:37.0&11.48&117.5&11235&NICMOS\\
MCG~+12-02-001&00:54:04.20&+73:05:06.0&11.5&69.8&10169 &NICMOS\\
IC~1623E&01:07:46.49&-17:30:22.5&11.71&85.5&7219&NICMOS\\
IC~1623W&01:07:47.42&-17:30:25.9&11.71&85.5&7219&NICMOS\\
MCG~-03-04-014&01:10:08.90&-16:51:10.0&11.65&144&11235&NICMOS\\
CGCG~436-030&01:20:02.70&+14:21:43.0&11.69&134&11235&NICMOS\\
IRAS~01364-1042&01:38:52.90&-10:27:11.0&11.85&198&11235&NICMOS\\
III~ZW~035&01:44:30.52&+17:06:07.9&11.64&119&11235&NICMOS\\
NGC~0695&01:51:14.20&+22:34:57.0&11.68&139&11235&NICMOS\\
MRK~1034W&02:23:18.91&+32:11:19.2&11.64&145&11235&NICMOS\\
MRK~1034E&02:23:22.00&+32:11:50.0&11.64&145&11235&NICMOS\\
UGC~02369S&02:54:01.77&+14:58:14.2&11.67&136&11235&NICMOS\\
UGC~02369N&02:54:01.81&+14:58:35.6&11.67&136&11235&NICMOS\\
IRASF~03359+1523&03:38:46.95&+15:32:54.5&11.55&152&11235&NICMOS\\
ESO~550-IG025N&04:21:19.87&-18:48:36.9&11.51&138.5&11235&NICMOS\\
ESO~550-IG025S&04:21:20.06&-18:48:56.3&11.51&138.5&11235&NICMOS\\
NGC~1614&04:33:59.91&-08:34:44.3&11.65&67.8&9726&NICMOS\\
ESO~203-IG001&04:46:49.42&-48:33:31.1&11.86&235&11235&NICMOS\\
VII~Zw~031&05:16:46.60&+79:40:13.0&11.99&240&9726 &NICMOS\\
IRAS~05223+1908&05:25:16.30&+19:10:45.0&11.65&128&11235&WFC3\\
IRAS~06076-2139&06:09:45.81&-21:40:23.7&11.65&165&11235&WFC3\\
ESO~255-IG007N&06:27:22.04&-47:10:42.0&11.9&173&11235&NICMOS\\
ESO~255-IG007S&06:27:23.12&-47:11:03.1&11.9&173&11235&NICMOS\\
AM~0702-601N&07:03:24.10&-60:15:23.0&11.64&141&11235&NICMOS\\
AM~0702-601S&07:03:28.63&-60:16:42.9&11.64&141&11235&NICMOS\\
2MASX~-J07273754-0254540&07:27:37.55&-02:54:54.1&12.39&400&11235&WFC3\\
2MASX~J08370182-4954302&08:37:01.80&-49:54:30.0&11.62&210&11235&NICMOS\\
NGC~2623&08:38:24.00&+25:45:17.0&11.6&84.1&7219&NICMOS\\
ESO~060-IG016&08:52:30.77&-69:01:59.8&11.82&210&11235&NICMOS\\
IRAS~-F08572+3915&09:00:25.40&+39:03:54.0&12.16&264&9726 &NICMOS\\
IRAS~09111-1007&09:13:37.61&-10:19:24.8&12.06&246&11235&WFC3\\
UGC~04881&09:15:55.1&+44:19:55.00&11.74&178&11235&WFC3\\
UGC~05101&09:35:51.40&+61:21:11.0&12.01&177&9726&NICMOS\\
IRAS~-F10173+0828&10:19:59.9&+08:13:34.00&11.86&224&11235&WFC3\\
NGC~3256&10:27:51.03&-43:54:18.2&11.64&38.9&9735&NICMOS\\
IRAS~-F10565+2448&10:59:18.20&+24:32:37.0&12.08&197&7219 &NICMOS\\
Arp~148&11:03:53.20&+40:50:57.0&11.62&158&11235&WFC3\\
IRAS~-F11231+1456&11:25:45.00&+14:40:36.0&11.64&157&9726 &NICMOS\\
IC~2810&11:25:49.52&+14:40:06.9&11.64&157&11235&WFC3\\
NGC~3690W&11:28:30.91&+58:33:45.2&11.93&50.7&9726&NICMOS\\
NGC~3690E&11:28:33.50&+58:33:45.2&11.93&50.7&9726&NICMOS\\
IRAS~-F12112+0305&12:13:45.90&+02:48:39.0&12.36&340&7219&NICMOS\\
WKK~0787&12:14:22.10&-56:32:33.0&11.65&114.5&11235&NICMOS\\
VV~283&13:01:50.8&+04:20:0.00&11.68&175&11235&WFC3\\
ESO~507-G070&13:02:52.30&-23:55:18.0&11.56&106&11235&WFC3\\
NGC~5010&13:12:26.30&-15:47:52.00&11.50&45&11235&WFC3\\
WKK~2031&13:15:06.30&-55:09:23.0&12.32&144&11235&NICMOS\\
UGC~08335W&13:15:31.15&+62:07:44.4&11.81&142&11235&NICMOS\\
UGC~08335E&13:15:35.29&+62:07:27.5&11.81&142&11235&NICMOS\\
UGC~08387&13:20:35.30&+34:08:22.0&11.73&110&7219 &NICMOS\\
NGC~5256&13:38:17.79&+48:16:35.1&11.56&129&7328&NICMOS\\
NGC~5257&13:39:52.97&+00:50:22.5&11.56&129&11235&NICMOS\\
NGC~5258&13:39:57.68&+00:49:49.80&11.56&129&11235&WFC3\\
UGC~08696&13:44:41.90&+55:53:12.1&12.21&173&9726&NICMOS\\
NGC~5331S&13:52:16.17&+02:06:01.2&11.66&155&11235&NICMOS\\
NGC~5331N&13:52:16.54&+02:06:28.8&11.66&155&11235&NICMOS\\
IRAS~-F14348-1447&14:37:38.20&-15:00:24.0&12.39&387&7219&NICMOS\\
IRAS~-F14378-3651&14:40:58.91&-37:04:31.8&12.23&315&7896 &NICMOS\\
UGC~09618S&14:57:00.34&+24:36:24.7&11.74&157&11235&NICMOS\\
UGC~09618N&14:57:00.70&+24:37:01.5&11.74&157&11235&NICMOS\\
VV~705&15:18:06.30&+42:44:37.0&11.92&183&11235&WFC3\\
ESO~099-G004&15:24:58.20&-63:07:34.0&11.74&137&11235&NICMOS\\
IRAS~-F15250+3608&15:26:59.30&+35:58:37.0&12.08&254&7219&NICMOS\\
UGC~09913&15:34:57.20&+23:30:10.9&12.28&87.9&9726&NICMOS\\
NGC~6090&16:11:40.60&+52:27:25.0&11.58&137&7219 &NICMOS\\
2MASXJ~16191179-0754026&16:19:11.80&-07:54:03.0&11.62&128&11235&NICMOS\\
ESO~069-IG006N&16:38:11.84&-68:26:10.4&11.98&212&11235&NICMOS\\
ESO~069-IG006S&16:38:13.48&-68:27:19.3&11.98&212&11235&NICMOS\\
IRAS~16399-0937&16:42:40.20&-09:43:14.0&11.63&114&11235&NICMOS\\
NGC~6240&16:52:58.70&+02:24:04.0&11.93&116&7219&NICMOS\\
IRASF~17132+5313&17:14:20.24&+53:10:30.8&11.96&232&11235&NICMOS\\
IRAS~-F17138-1017&17:16:35.60&-10:20:38.0&11.49&84&10169&NICMOS\\
IRAS~-F17207-0014&17:23:22.20&-00:17:02.0&12.46&198&7219&NICMOS\\
IRAS~18090+0130&18:11:38.42&+01:31:38.8&11.65&134&11235&NICMOS\\
IC~4689&18:13:40.28&-57:44:53.5&11.62&81.9&11235&NICMOS\\
IRAS~18293-3413&18:32:41.22&-34:11:26.7&11.88&86&11235&NICMOS\\
NGC~6670W&18:33:33.77&+59:53:15.6&11.65&129.5&11235&NICMOS\\
NGC~6670E&18:33:37.70&+59:53:23.0&11.65&129.5&11235&NICMOS\\
NGC~6786S&19:10:53.90&+73:24:37.0&11.49&113&11235&NICMOS\\
UGC~11415&19:10:53.90&+73:24:37.7&11.49&113&11235&NICMOS\\
NGC~6786N&19:11:04.44&+73:25:37.4&11.49&113&11235&NICMOS\\
ESO~593-IG008&19:14:30.98&-21:19:06.8&11.93&222&11235&NICMOS\\
IRAS~-F19297-0406&19:32:22.28&-04:00:01.1&12.45&395&7896&NICMOS\\
IRAS~19542+1110&19:56:35.39&+11:19:03.0&12.12&295&11235&NICMOS\\
IRAS~20351+2521&20:37:17.80&+25:31:38.0&11.61&151&11235&NICMOS\\
II~ZW~096&20:57:23.94&+17:07:39.5&11.94&161&11235&NICMOS\\
ESO~286-IG019&20:58:26.80&-42:39:00.0&12.06&193&11235&NICMOS\\
IRAS~21101+5810&21:11:29.82&+58:23:07.2&11.81&174&11235&NICMOS\\
ESO~239-IG002&22:49:39.90&-48:50:58.0&11.84&191&11235&NICMOS\\
IRAS~-F22491-1808&22:51:49.30&-17:52:24.0&12.2&351&7219&NICMOS\\
NGC~7469  / IC5283&23:03:17.88&+08:53:39.3&11.65&70.8&11235&NICMOS\\
ESO~148-IG002&23:15:46.79&-59:03:13.0&12.06&199&7896&NICMOS\\
IC~5298&23:16:00.71&+25:33:24.0&11.6&119&11235&NICMOS\\
ESO~077-IG014&23:21:04.44&-69:12:54.8&11.76&186&11235&NICMOS\\
NGC~7674&23:27:56.70&+08:46:45.0&11.56&125&11235&NICMOS\\
IRASF~23365+3604&23:39:01.30&+36:21:09.8&12.2&287&11235&NICMOS\\
IRAS~23436+5257&23:46:05.59&+53:14:01.0&11.57&149&11235&NICMOS\\
UGC~12812W / MRK0331&23:51:18.73&+20:34:42.9&11.5&79.3&11235&NICMOS\\
UGC~12812E  / MRK0331&23:51:26.80&+20:35:10.0&11.5&79.3&11235&NICMOS\\
\hline
\label{tab:obs}
\end{longtable}
\footnotesize{Overview of the properties of our 85 LIRG systems (101 pointings). Column (1): Source Name from NED, Column (2): right ascension (J2000), Column (3): source declination (J2000), Column (4): The luminosity distance in Mpc  \citep[adopted from][]{Arm09}, Column (5): The total infrared luminosity in log$_{10}$ Solar units, Column (6): The data origin given by the ID of the observational program, Column (7): The HST Instrument used to obtain the 1.6$\mu$m data fit in this paper.}
\end{footnotesize}
\end{flushleft}
\clearpage
\newpage



\begin{flushleft}
\begin{scriptsize}
\begin{longtable}{@{}lrrcrrrrrrc@{}}
\caption{Nuclear Properties and Merger Classification}\\
\hline \hline
\multicolumn{1}{c}{\textbf{Name}}&\multicolumn{1}{c}{\textbf{RA$_{Cusp}$}}&\multicolumn{1}{c}{\textbf{DEC$_{Cusp}$}}& \multicolumn{1}{c}{\textbf{MS}}& \multicolumn{1}{c}{\textbf{L$_{IR}$}}& \multicolumn{1}{c}{\textbf{$\gamma$}}& \multicolumn{1}{c}{\textbf{$R_{cusp}$}}& \multicolumn{1}{c}{\textbf{$L_{cusp}$}}& \multicolumn{1}{c}{\textbf{L$_{nuc}$}}&\multicolumn{1}{c}{\textbf{$L_{PSF}$}}& \multicolumn{1}{c}{\textbf{SB}}\\
\multicolumn{1}{c}{}&\multicolumn{1}{c}{\textbf{(J2000)}}& \multicolumn{1}{c}{\textbf{(J2000)}}& \multicolumn{1}{c}{}&\multicolumn{1}{c}{\textbf{log10[L$_\odot$]}}&\multicolumn{1}{c}{}& \multicolumn{1}{c}{\textbf{[arcsec]}}&\multicolumn{1}{c}{\textbf{log10[$L_\odot$]}}&\multicolumn{1}{c}{\textbf{log10[$L_\odot$]}}&\multicolumn{1}{c}{\textbf{log10[$L_\odot$]}}&\multicolumn{1}{c}{}\\
\hline
\endfirsthead
\multicolumn{10}{c}%
{{\bfseries \tablename\ \thetable{} -- continued from previous page}}\\
\hline
\endhead

\hline \multicolumn{10}{r}{{Continued on next page}} \\ 
\endfoot
\hline \hline
\endlastfoot
NGC~0034 &00:11:06.545 &-12:06:27.24&5& 11.43& 1.02 $\pm$ 0.01 & 0.38 $\pm$ 0.05 & 10.33 $\pm$ 0.01 & 11.0 $\pm$ 0.01 & 9.52 $\pm$ 0.02 & cen\\ 
ARP~256N &00:18:50.130 &-10:21:41.69& 3&10.45 & 0.4 $\pm$ 0.27 & 0.11 $\pm$ 0.12 & 7.52 $\pm$ 1.19 & 10.4 $\pm$ 0.04 & 8.07 $\pm$ 0.06 & off\\ 
ARP~256S &00:18:50.872 &-10:22:36.56&3& 11.44 & 0.0 $\pm$ 0.11 & 0.12 $\pm$ 0.06 & $\leq$7.3  & 10.66 $\pm$ 0.04 & 9.34 $\pm$ 0.03 & cen\\ 
MCG~+12-02-001 &00:54:04.005&+73:05:05.40&3& 11.5 & 1.68 $\pm$ 0.02 & 0.26 $\pm$ 0.06 & 9.86 $\pm$ 0.06 & 11.0 $\pm$ 0.01 & 9.32 $\pm$ 0.04 & cen\\ 
IC~1623E &01:07:46.553&-17:30:22.61&3& 11.71 & 0.64 $\pm$ 0.02 & 1.55 $\pm$ 0.37 & 9.57 $\pm$ 0.21 & 10.06 $\pm$ 0.2 & $\leq$6.6  & off\\ 
MCG~-03-04-014 &01:10:08.916 &-16:51:09.58&0& 11.65 & 0.46 $\pm$ 0.01 & 0.77 $\pm$ 0.06 & 9.94 $\pm$ 0.01 & 10.97 $\pm$ 0.01 & 9.71 $\pm$ 0.02 & cen\\ 
CGCG~436-030 &01:20:02.622&+14:21:42.35&2& 11.69 & 0.01 $\pm$ 0.21 & 0.13 $\pm$ 0.06 & $\leq$7.36  & 10.56 $\pm$ 0.05 & 10.29 $\pm$ 0.01 & cen\\ 
IRAS~01364-1042  &01:38:52.855&-10:27:11.41&5& 11.85 & 0.01 $\pm$ 0.11 & 0.12 $\pm$ 0.06 & $\leq$7.41  & 10.48 $\pm$ 0.03 & $\leq$7.33 & cen\\ 
III~ZW~035 n1 &01:44:30.554&+17:06:09.18&3& 11.64 & 0.0 $\pm$ 0.02 & 0.21 $\pm$ 0.05 & $\leq$7.37 & 10.61 $\pm$ 0.01 & $\leq$6.89 & cen\\ 
III~ZW~035 n2 &01:44:30.537&+17:06:08.84&3& 11.64 & 0.51 $\pm$ 0.01 & 0.24 $\pm$ 0.05 & 9.17 $\pm$ 0.01 & 10.41 $\pm$ 0.01 & $\leq$6.89 & off\\ 
NGC~0695 &01:51:14.334&+22:34:56.01&0& 11.68 & 0.78 $\pm$ 0.01 & 1.42 $\pm$ 0.09 & 10.28 $\pm$ 0.03 & 10.75 $\pm$ 0.01 & 9.37 $\pm$ 0.03 & cen\\ 
MRK~1034W &02:23:18.956&+32:11:18.73&2& 11.16 & 0.85 $\pm$ 0.01 & 0.4 $\pm$ 0.1 & 9.76 $\pm$ 0.13 & 10.86 $\pm$ 0.04 & 9.4 $\pm$ 0.04 & cen\\ 
MRK~1034E &02:23:21.950&+32:11:48.83&2& 11.47 & 0.0 $\pm$ 0.04 & 0.3 $\pm$ 0.06 & $\leq$8.13  & 11.13 $\pm$ 0.02 & $\leq$7.06 & cen\\ 
ESO~550-IG02N &04:21:19.976&-18:48:39.31&2& 11.27 & 0.0 $\pm$ 0.09 & 0.1 $\pm$ 0.06 & $\leq$7.24  & 10.66 $\pm$ 0.03 & $\leq$7.02  & cen\\ 
ESO~550-IG02S &04:21:20.018&-18:48:57.14&2& 11.13& 0.0 $\pm$ 0.07 & 0.1 $\pm$ 0.05 & $\leq$7.29  & 10.85 $\pm$ 0.02 & $\leq$7.02 & cen\\ 
NGC~1614 &04:34:00.015&-08:34:45.11&5& 11.66 & 1.36 $\pm$ 0.01 & 0.33 $\pm$ 0.05 & 10.41 $\pm$ 0.01 & 10.94 $\pm$ 0.01 & $\leq$6.4  & cen\\ 
ESO~203-IG001 &04:46:49.534&-48:33:29.90&3& 11.86 & 0.69 $\pm$ 0.01 & 0.62 $\pm$ 0.06 & 9.83 $\pm$ 0.02 & 10.51 $\pm$ 0.01 & 8.79 $\pm$ 0.08 & cen\\ 
VII~Zw~031 &05:16:46.482&+79:40:12.84&0& 11.99 & 0.81 $\pm$ 0.02 & 0.49 $\pm$ 0.08 & 10.28 $\pm$ 0.17 & 11.00 $\pm$ 0.07 & 9.61 $\pm$ 0.06 & cen\\ 
ESO~255-IG007S &06:27:21.754&-47:10:35.74&3& 11.78 & 0.14 $\pm$ 0.12 & 0.22 $\pm$ 0.07 & 8.91 $\pm$ 0.13 & 10.97 $\pm$ 0.05 & $\leq$7.21  & cen\\ 
AM~0702-601N &07:03:24.258&-60:15:22.49&1& 11.46 & - & - & - & - & 11.28 $\pm$ 0.01 & cen\\ 
AM~0702-601S &07:03:28.626&-60:16:44.66&1& 11.19 & 0.42 $\pm$ 0.12 & 0.31 $\pm$ 0.12 & 9.21 $\pm$ 0.5 & 10.53 $\pm$ 0.25 & $\leq$7.03  & cen\\ 
IRAS~08355-4944 &08:37:01.881&-49:54:30.58&3& 11.62 & - & - & - & - & 10.47 $\pm$ 0.03 & cen\\ 
NGC~2623 &08:38:24.099&+25:45:16.70&5& 11.6 & $\geq$0.0 & $\leq$0.08  & $\geq$6.96  & 10.63 $\pm$ 0.02 & 9.58 $\pm$ 0.01 & cen\\ 
IRAS-F~08572+3915 &09:00:25.361&+39:03:55.31&3& 12.16 & 1.12 $\pm$ 0.02 & 0.81 $\pm$ 0.19 & 10.01 $\pm$ 0.18 & 10.28 $\pm$ 0.12 & 9.69 $\pm$ 0.02 & cen\\ 
IRAS~09111-1007E&09:13:38.837&-10:19:19.72&1& 11.39 & 0.01 $\pm$ 0.1 & 1.63 $\pm$ 0.11 & $\leq$8.5 & 9.98 $\pm$ 0.05 & $\leq$7.52  & cen\\ 
IRAS~09111-1007W&09:13:36.449&-10:19:29.95&1& 11.95 & - & - & - & 10.97 $\pm$ 0.02 & 9.78 $\pm$ 0.05 & cen\\ 
UGC~4881N &09:15:55.518&+44:19:58.04&2& 11.58& 0.0 $\pm$ 0.13 & 0.21 $\pm$ 0.07 & $\leq$7.32 & 10.65 $\pm$ 0.03 & $\leq$7.24  & cen\\ 
UGC~4881S &09:15:54.685&+44:19:51.46&2& 11.24 & 0.58 $\pm$ 0.04 & 0.58 $\pm$ 0.13 & 9.65 $\pm$ 0.15 & 10.61 $\pm$ 0.06 & 9.24 $\pm$ 0.09 & cen\\ 
UGC~05101 &09:35:51.631&+61:21:11.85&5& 12.02 & 1.02 $\pm$ 0.01 & 0.74 $\pm$ 0.06 & 10.81 $\pm$ 0.01 & 11.29 $\pm$ 0.01 & 10.2 $\pm$ 0.01 & cen\\ 
IRAS-F~10173+0828 &10:20:00.207&+08:13:33.78&0& 11.86 & 0.94 $\pm$ 0.01 & 0.78 $\pm$ 0.06 & 10.22 $\pm$ 0.01 & 10.65 $\pm$ 0.01 & 9.12 $\pm$ 0.06 & cen\\ 
NGC~3256 &10:27:51.254&-43:54:14.00&5& 11.64 & 1.08 $\pm$ 0.06 & 0.64 $\pm$ 0.48 & 9.92 $\pm$ 0.89 & 11.16 $\pm$ 0.28 & 9.78 $\pm$ 0.08 & cen\\ 
IRAS-F~10565+2448 &10:59:18.146&+24:32:34.57&2& 12.08 & 0.9 $\pm$ 0.01 & 0.82 $\pm$ 0.06 & 10.67 $\pm$ 0.01 & 11.13 $\pm$ 0.01 & 10.23 $\pm$ 0.01 & cen\\ 
IRAS-F11231+1456 &11:25:45.073&+14:40:35.96&1& 11.46 & 0.64 $\pm$ 0.04 & 0.17 $\pm$ 0.07 & 9.42 $\pm$ 0.11 & 10.9 $\pm$ 0.05 & $\leq$7.13  & cen\\ 
NGC~3690E &11:28:33.688&+58:33:46.36&3& 11.58 & 0.74 $\pm$ 0.01 & 0.58 $\pm$ 0.07 & 9.28 $\pm$ 0.03 & 10.41 $\pm$ 0.02 & $\leq$6.15 & cen\\ 
IRAS-F~12112+0305S &12:13:45.903&+02:48:39.28&4& 12.1 & 1.68 $\pm$ 0.01 & 0.61 $\pm$ 0.20 & 10.56 $\pm$ 0.35 & 10.67 $\pm$ 0.28 & $\leq$7.8 & cen\\ 
IRAS-F12112+0305N &12:13:46.022&+02:48:41.67&4& 12.05 & 0.72 $\pm$ 0.05 & 0.31 $\pm$ 0.25 & 9.58 $\pm$ 1.06 & 10.44 $\pm$ 0.27 & 8.75 $\pm$ 0.17 & cen\\ 
WKK~0787 &12:14:22.097&-56:32:33.28&0& 11.65 & 1.02 $\pm$ 0.01 & 0.92 $\pm$ 0.33 & 10.18 $\pm$ 0.31 & 10.8 $\pm$ 0.14 & $\leq$6.85  & cen\\ 
VV~283 &13:01:50.294&+04:20:00.51&5& 11.68 & 0.0 $\pm$ 0.16 & 0.25 $\pm$ 0.07 & $\leq$7.68 & 10.83 $\pm$ 0.04 & 9.88 $\pm$ 0.04 & cen\\ 
ESO~507 &13:02:52.378&-23:55:17.54&6& 11.56 & 0.0 $\pm$ 0.15 & 0.25 $\pm$ 0.08 & $\leq$7.07& 10.69 $\pm$ 0.04 & $\leq$6.79 & cen\\ 
WKK~2031 &13:15:06.338&-55:09:22.71&5& 12.32 & 0.99 $\pm$ 0.01 & 0.99 $\pm$ 0.06 & 10.87 $\pm$ 0.02 & 11.3 $\pm$ 0.01 & 9.89 $\pm$ 0.1 & cen\\ 
UGC~08335W &13:15:30.800&+62:07:45.26&2& 11.5 & 1.09 $\pm$ 0.02 & 0.55 $\pm$ 0.12 & 9.93 $\pm$ 0.14 & 10.67 $\pm$ 0.06 & $\leq$7.04  & cen\\ 
UGC~08335E &13:15:35.028&+62:07:28.84&2& 11.5 & 0.99 $\pm$ 0.01 & 0.45 $\pm$ 0.07 & 10.24 $\pm$ 0.04 & 10.96 $\pm$ 0.02 & $\leq$7.04  & cen\\ 
NGC~5256S &13:38:17.272&+48:16:32.10&3& 11.36 & 0.43 $\pm$ 0.01 & 0.89 $\pm$ 0.07 & 10.0 $\pm$ 0.01 & 10.98 $\pm$ 0.01 & 8.84 $\pm$ 0.06 & cen\\ 
NGC~5256N &13:38:17.762&+48:16:41.17&3& 11.13 & 0.86 $\pm$ 0.01 & 0.88 $\pm$ 0.1 & 10.05 $\pm$ 0.04 & 10.7 $\pm$ 0.03 & 9.28 $\pm$ 0.02 & cen\\ 
NGC~5257 &13:39:52.933&+00:50:24.51&2& 11.31 & 1.12 $\pm$ 0.04 & 0.18 $\pm$ 0.06 & 9.47 $\pm$ 0.06 & 10.59 $\pm$ 0.02 & $\leq$6.96  & off\\ 
NGC~5258 &13:39:57.681&+00:49:50.89&2& 11.32& 0.49 $\pm$ 0.04 & 0.38 $\pm$ 0.06 & 8.85 $\pm$ 0.17 & 10.5 $\pm$ 0.05 & $\leq$6.81  & off\\ 
NGC~5331S &13:52:16.202&+02:06:05.20&3& 11.54 & 0.0 $\pm$ 0.28 & 0.19 $\pm$ 0.14 & $\leq$7.0  & 10.84 $\pm$ 0.06 & $\leq$7.12 & cen\\ 
NGC~5331N &13:52:16.425&+02:06:31.15&3& 11.02 & 1.1 $\pm$ 0.05 & 0.12 $\pm$ 0.06 & 9.63 $\pm$ 0.08 & 10.88 $\pm$ 0.02 & $\leq$7.12  & cen\\ 
IRAS-F~14348-1447N &14:37:38.401&-15:00:21.13&4& 12.1& 0.0 $\pm$ 0.09 & 0.1 $\pm$ 0.06 & $\leq$7.62  & 10.75 $\pm$ 0.03 & 9.92 $\pm$ 0.02 & cen\\ 
IRAS-F~14348-1447S &14:37:38.288&-15:00:24.14&4& 12.1 & 0.97 $\pm$ 0.01 & 0.38 $\pm$ 0.08 & 10.03 $\pm$ 0.07 & 10.52 $\pm$ 0.04 & $\leq$7.91  & cen\\ 
IRAS-F~14378-3651 &14:40:59.010&-37:04:31.92&6& 12.23 & 0.96 $\pm$ 0.01 & 0.17 $\pm$ 0.05 & 10.28 $\pm$ 0.03 & 10.99 $\pm$ 0.02 & 9.56 $\pm$ 0.05 & cen\\ 
UGC~09618S &14:57:00.325&+24:36:24.00&1& 11.0 & 1.12 $\pm$ 0.03 & 0.19 $\pm$ 0.06 & 9.09 $\pm$ 0.08 & 10.46 $\pm$ 0.01 & $\leq$7.13  & cen\\ 
VV~705N &15:18:06.138&+42:44:45.29&4& 11.85 & 1.23 $\pm$ 0.02 & 1.96 $\pm$ 0.35 & 10.63 $\pm$ 0.16 & 10.69 $\pm$ 0.1 & 9.88 $\pm$ 0.11 & cen\\ 
VV~705S &15:18:06.349&+42:44:38.34&4& 11.12 & 0.87 $\pm$ 0.13 & 0.22 $\pm$ 0.1 & - & - & -7.26 $\pm$ 0.01 & cen\\ 
ESO~099-G004N &15:24:57.941&-63:07:29.68&3& 11.74 & 1.34 $\pm$ 0.03 & 0.46 $\pm$ 0.2 & 10.16 $\pm$ 0.37 & 10.64 $\pm$ 0.25 & $\leq$7.01 & cen\\ 
IRAS~F15250+3608 &15:26:59.425&+35:58:37.22&5& 12.08 & -& - & - & 10.73 $\pm$ 0.16 & 9.71 $\pm$ 0.02 & cen\\ 
NGC~6090 &16:11:40.914&+52:27:27.21&4& 11.58 & 0.2 $\pm$ 0.04 & 0.12 $\pm$ 0.01 & 8.08 $\pm$ 0.03 & 10.67 $\pm$ 0.01 & $\leq$7.01 & cen\\ 
ESO~069-IG006N &16:38:11.870&-68:26:08.19&2& 11.97 & 0.08 $\pm$ 0.02 & 1.45 $\pm$ 0.03 & 9.89 $\pm$ 0.02 & 11.02 $\pm$ 0.01 & $\leq$7.39  & cen\\ 
ESO~069-IG006S &16:38:13.469&-68:27:16.83&2& 10.48& 0.57 $\pm$ 0.05 & 0.77 $\pm$ 0.17 & 10.0 $\pm$ 0.17 & 10.66 $\pm$ 0.1 & $\leq$7.39 & cen\\ 
IRAS~16399-0937S &16:42:40.177&-09:43:18.77&3& 11.0 & 0.67 $\pm$ 0.04 & 0.3 $\pm$ 0.08 & 9.21 $\pm$ 0.12 & 10.54 $\pm$ 0.05 & 8.83 $\pm$ 0.1 & off\\ 
IRAS~16399-0937N &16:42:40.141&-09:43:13.18&3& 11.62 & 0.36 $\pm$ 0.03 & 0.24 $\pm$ 0.06 & 8.84 $\pm$ 0.03 & 10.41 $\pm$ 0.02 & $\leq$6.85  & cen\\ 
NGC~6240S &16:52:58.884&+02:24:03.28&4& 11.63 & 1.01 $\pm$ 0.05 & 0.53 $\pm$ 0.2 & 10.85 $\pm$ 0.33 & 11.44 $\pm$ 0.24 & $\leq$6.87  & cen\\ 
NGC~6240N &16:52:58.933&+02:24:04.92&4& 11.63 & 0.54 $\pm$ 0.73 & 0.08 $\pm$ 0.16 & 8.92 $\pm$ 6.05 & 11.02 $\pm$ 0.49 & $\leq$6.87  & cen\\ 
IRASF~17132+5313N &17:14:20.449&+53:10:31.96&2& 11.65 & 0.72 $\pm$ 0.01 & 2.05 $\pm$ 0.07 & 10.9 $\pm$ 0.01 & 10.96 $\pm$ 0.01 & $\leq$7.47 & cen\\ 
IRASF~17132+5313S &17:14:19.800&+53:10:28.86&2& 11.65 & 0.36 $\pm$ 0.08 & 0.08 $\pm$ 0.05 & 8.97 $\pm$ 0.08 & 10.61 $\pm$ 0.03 & $\leq$7.47  & cen\\ 
IRAS-F~17138-1017 &17:16:35.791&-10:20:38.81&6& 11.49 & 0.5 $\pm$ 0.01 & 3.0 $\pm$ 0.16 & 10.21 $\pm$ 0.03 & 10.74 $\pm$ 0.01 & $\leq$6.58  & off\\ 
IRAS-F~17207-0014 &17:23:21.951&-00:17:00.74&5& 12.46 & 0.69 $\pm$ 0.01 & 1.24 $\pm$ 0.11 & 10.44 $\pm$ 0.05 & 10.86 $\pm$ 0.02 & $\leq$7.33  & cen\\ 
IRAS~18090+0130 &18:11:38.415&+01:31:40.01&2& 11.52 & 0.47 $\pm$ 0.12 & 0.14 $\pm$ 0.08 & 9.0 $\pm$ 0.26 & 10.9 $\pm$ 0.08 & $\leq$6.99  & cen\\ 
IC~4689 &18:13:40.387&-57:44:54.12&2& 10.88 & 1.07 $\pm$ 0.01 & 1.48 $\pm$ 0.09 & 10.16 $\pm$ 0.03 & 10.72 $\pm$ 0.01 & 9.24 $\pm$ 0.03 & cen\\ 
IRAS~18293-3413 &18:32:41.135&-34:11:27.54&1& 11.88 & 1.03 $\pm$ 0.01 & 2.01 $\pm$ 0.13 & 10.73 $\pm$ 0.04 & 11.13 $\pm$ 0.02 & $\leq$6.61 & cen\\ 
NGC~6786S &19:10:53.836&+73:24:36.70&2& 11.23 & 0.69 $\pm$ 0.01 & 2.08 $\pm$ 0.12 & 10.21 $\pm$ 0.03 & 10.64 $\pm$ 0.02 & 9.09 $\pm$ 0.1 & cen\\ 
NGC~6786N &19:11:04.295&+73:25:33.40&2& 11.15 & 1.41 $\pm$ 0.01 & 0.65 $\pm$ 0.06 & 10.3 $\pm$ 0.02 & 10.92 $\pm$ 0.01 & $\leq$7.10 & cen\\ 
IRAS~19542+1110 &19:56:35.785&+11:19:05.09&0& 12.12 & 0.78 $\pm$ 0.02 & 0.18 $\pm$ 0.05 & 10.44 $\pm$ 0.02 & 11.24 $\pm$ 0.01 & $\leq$7.68  & cen\\ 
ESO~286-IG019 &20:58:26.798&-42:39:00.21&5& 12.06 & 0.81 $\pm$ 0.04 & 0.45 $\pm$ 0.06 & 9.99 $\pm$ 0.05 & 10.76 $\pm$ 0.02 & 9.0 $\pm$ 0.29 & cen\\ 
IRAS~21101+5810 &21:11:29.296&+58:23:08.02&2& 11.81 & 0.26 $\pm$ 0.03 & 0.67 $\pm$ 0.08 & 9.3 $\pm$ 0.04 & 10.39 $\pm$ 0.04 & 9.01 $\pm$ 0.1 & cen\\ 
ESO~239-IG002 &22:49:39.889&-48:50:58.14&5& 11.84 & 1.57 $\pm$ 0.01 & 0.66 $\pm$ 0.09 & 10.7 $\pm$ 0.05 & 10.94 $\pm$ 0.04 & 10.31 $\pm$ 0.02 & cen\\ 
ESO~148-IG002 &23:15:46.744&-59:03:15.74&4& 12.06 & 1.66 $\pm$ 0.01 & 0.52 $\pm$ 0.13 & 10.55 $\pm$ 0.15 & 10.79 $\pm$ 0.11 & $\leq$7.33& cen\\ 
IC~5298 &23:16:00.679&+25:33:24.03&0& 11.6 & 0.39 $\pm$ 0.02 & 0.92 $\pm$ 0.08 & 9.9 $\pm$ 0.03 & 10.86 $\pm$ 0.02 & 9.5 $\pm$ 0.07 & cen\\ 
ESO~077-IG014 &23:21:05.354&-69:12:47.18&2& 11.56 & 0.0 $\pm$ 0.21 & 0.11 $\pm$ 0.07 & $\leq$7.25 0.43 & 11.01 $\pm$ 0.06 & $\leq$7.28 & cen\\ 
NGC~7674 &23:27:56.715&+08:46:44.41&2& 11.54 & 1.23 $\pm$ 0.04 & 0.19 $\pm$ 0.06 & 9.9 $\pm$ 0.06 & 10.78 $\pm$ 0.02 & 10.53 $\pm$ 0.01 & cen\\ 
IRASF~23365+3604 &23:39:01.270&+36:21:08.56&5& 12.2 & 1.09 $\pm$ 0.02 & 0.15 $\pm$ 0.06 & 9.98 $\pm$ 0.1 & 10.8 $\pm$ 0.05 & 9.76 $\pm$ 0.02 & cen\\ 
IRAS~23436+5257N &23:46:05.496&+53:14:01.42&4& 11.4 & 0.0 $\pm$ 0.14 & 0.12 $\pm$ 0.06 & $\leq$7.12 & 10.67 $\pm$ 0.03 & 9.58 $\pm$ 0.01 & cen\\ 
IRAS~23436+5257S &23:46:05.704&+53:13:56.71&4& 11.1 & 0.01 $\pm$ 0.15 & 0.16 $\pm$ 0.06 & $\leq$7.5  & 10.4 $\pm$ 0.03 & 8.82 $\pm$ 0.18 & cen\\ 
UGC~12812W &23:51:18.673&+20:34:41.58&4& 9.32 & 0.0 $\pm$ 0.03 & 0.88 $\pm$ 0.27 & $\leq$6.75 & 10.05 $\pm$ 0.01 & 7.1 $\pm$ 0.12 & cen\\ 
UGC~12812E & 23:51:26.738&+20:35:10.39&1& 11.5& 0.68 $\pm$ 0.01 & 1.67 $\pm$ 0.06 & 10.33 $\pm$ 0.01 & 10.94 $\pm$ 0.01 & $\leq$6.53  & cen\\ 
\hline
\label{tab:galfit}
\end{longtable}
\footnotesize{Column (1): Source Name from NED, Column (2): right ascension (J2000) in deg, Column (3): declination (J2000) in deg, Column (4): The merger stage classification, Column (5): The L$_{IR}$ luminosity computed for each merger component (see text in \S~\ref{subsec:lir}), Column (6): The central slope of the Nuker profile $\gamma$, Column (7): The effective break radius (cusp/core radius) between central and outer power law of Nuker profile in arcsec, Column (8): The cusp luminosity or the upper limit based on the error in $\gamma$, Column (9): The nuclear NIR luminosity integrated over the central area with a radius of 1~kpc, Column (10): The NIR luminosity of the central unresolved component (HST PSF), Column (11): The identification whether the starburst peak (based on MIPS24$\mu$m emission) is in the center of the galaxy or offset.}
\end{scriptsize}
\end{flushleft}

\clearpage

\newpage
\twocolumn
\begin{table}
\caption{Cusp Fraction versus Merger Stage}
\begin{scriptsize}
\begin{tabular}{@{}lrrrrrr@{}}
\hline
Merger Stage & MS~0 & MS~1 & MS~2 & MS~3 & MS~4 & MS~5 \\
\hline 
Number of Galaxies&7&7&23&17&11&15\\
Number of Cusps&7&6&15&12&8&11\\
Cusp Fraction [\%]&100 &86&65&71&73&73\\
\hline
\end{tabular}
\label{tab:ms}
\end{scriptsize}
\end{table}

\begin{table}
\caption{Comparison of absolute host H-band magnitude between LIRGs and Ellipticals.}
\begin{tabular}{@{}lcc@{}}
\hline
 & LIRGs & Ellipticals \\
\hline 
median (mean) M$_H$ cusps:& -24.5 (-24.4) & -23.5 (-23.5)\\
median (mean) M$_H$ cores:& -24.3 (-24.0) & -25.1 (-24.9) \\
\hline
\end{tabular}
\label{tab:comp}
\end{table}


\appendix

\section[]{Characterisation of the Elliptical Galaxy Reference Sample and the $\gamma$-M$_{host}$ diagram}
\label{app:a}

\begin{figure}
\begin{center}
\includegraphics[scale=0.46]{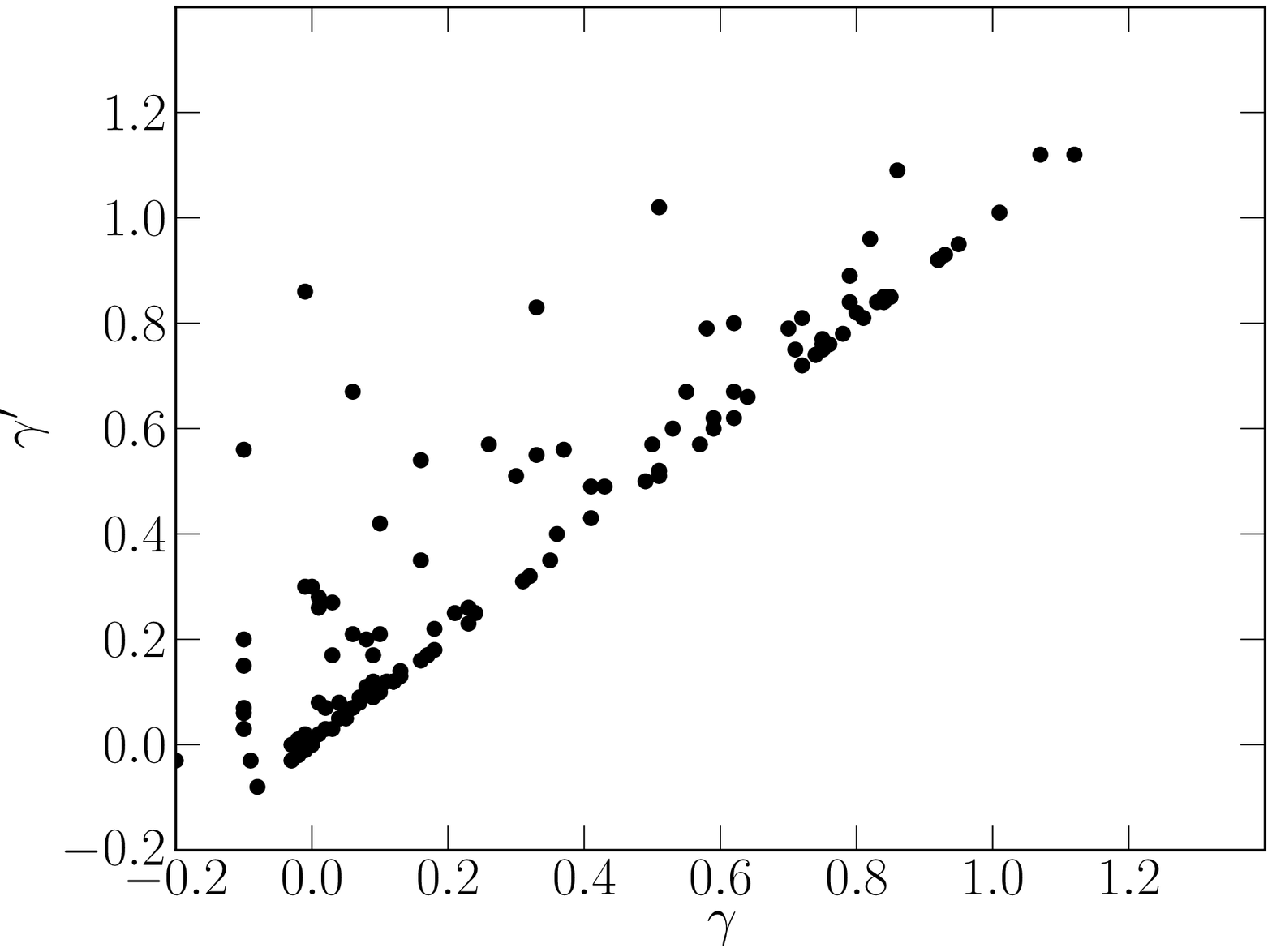}
\caption{Comparison of the local logarithmic slope at the HST resolution limit $\gamma^\prime$ and the Nuker gamma parameter $\gamma$ for our reference sample of elliptical galaxies (measurements by Lauer et al. 2007).}
\label{fig:gamma_prime}
\end{center}
\end{figure}

In \S~\ref{subsec:ellipticals} we have chosen the sample of \cite{Lau07} as reference sample of early-type galaxies to which we compare the cusp properties of our LIRG sample. The original  $\gamma^\prime$-M$_V$ diagram published in \cite{Lau07} shows a strong bimodality between cusp and core galaxies. \cite{Lau07} have derived the cusp slope in two different ways, one by fitting the Nuker law parameter $\gamma$ as in our study, and the other one by calculating the local logarithmic slope at the HST resolution limit, defined in their study as $\gamma^\prime$. For HST WFPC2 in F555W, $\gamma^\prime$ was measured at 0.05\arcsec from the center.  Since the nuclear slope is monotonically decreasing as the radius is going towards zero, this may be regarded as an upper limit on the true slope. The difference between $\gamma$ and $\gamma^\prime$ for our reference sample of elliptical galaxies is shown in Fig.~\ref{fig:gamma_prime}. Within our magnitude range we find that 15\% of the ellipticals have a difference in $\gamma^\prime-\gamma$ of more than 0.1  with an average $\gamma^\prime-\gamma$ of 0.22. This is only relevant for our study to the extent that it will shift some of the core elliptical galaxies into the transition range between cusps and cores ($0<\gamma<0.5$), which reduces the bimodal behaviour in the distribution of cusp and core galaxies as seen in \cite{Lau07}. However, less than 5\% of the elliptical galaxies would end up as cusp rather than core galaxies, and hence we find a similar strong correlation between the slope $\gamma$ and the host magnitude as demonstrated in Fig.~\ref{fig:comp_cusp}.



\section[]{Examples of GALFIT Maps and Derived Radial Profiles}
\label{app:b}
\begin{figure}
\begin{center}
\begin{minipage}[b]{8cm}
\includegraphics[scale=0.56]{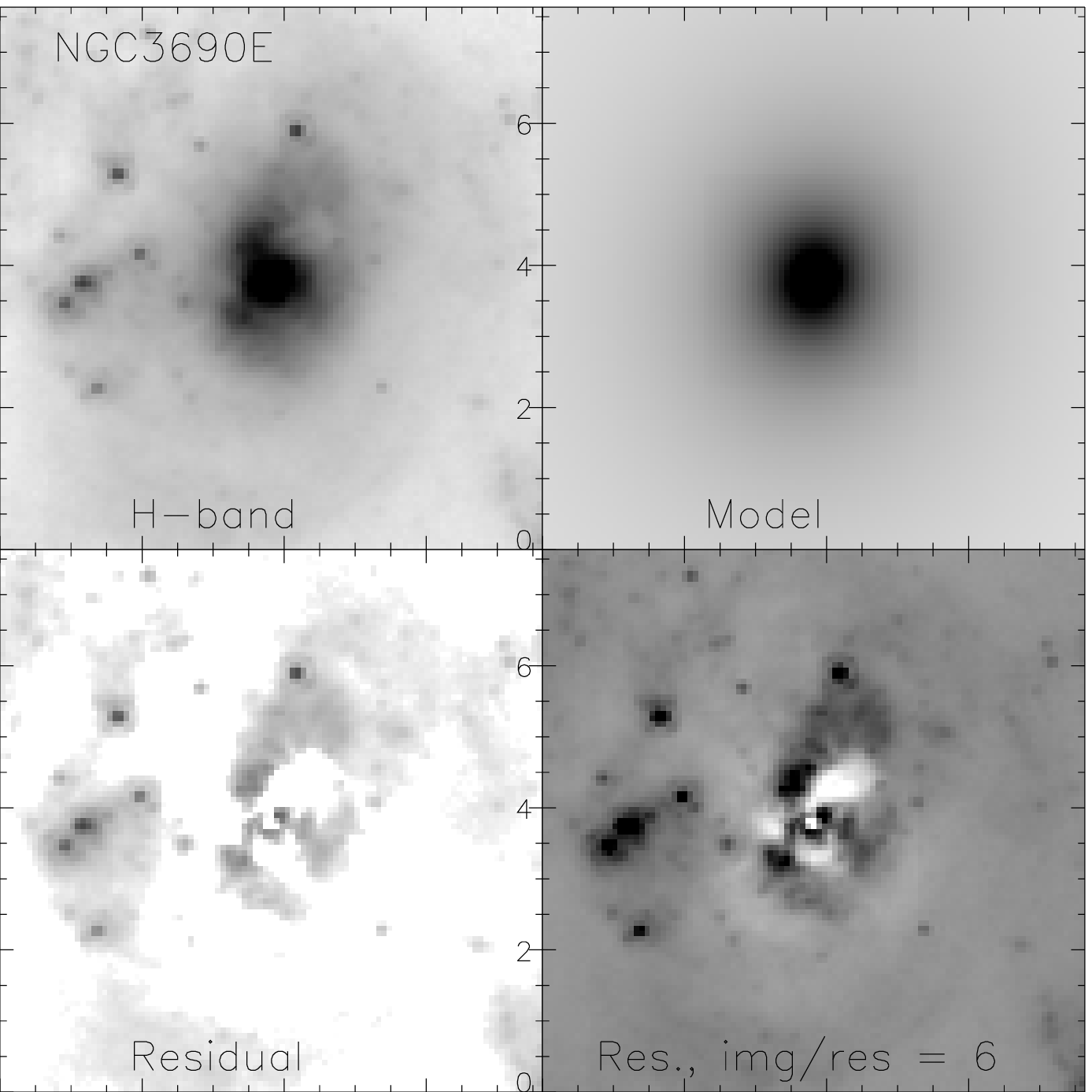}\\
\hspace{1.2cm}

\includegraphics[scale=0.56]{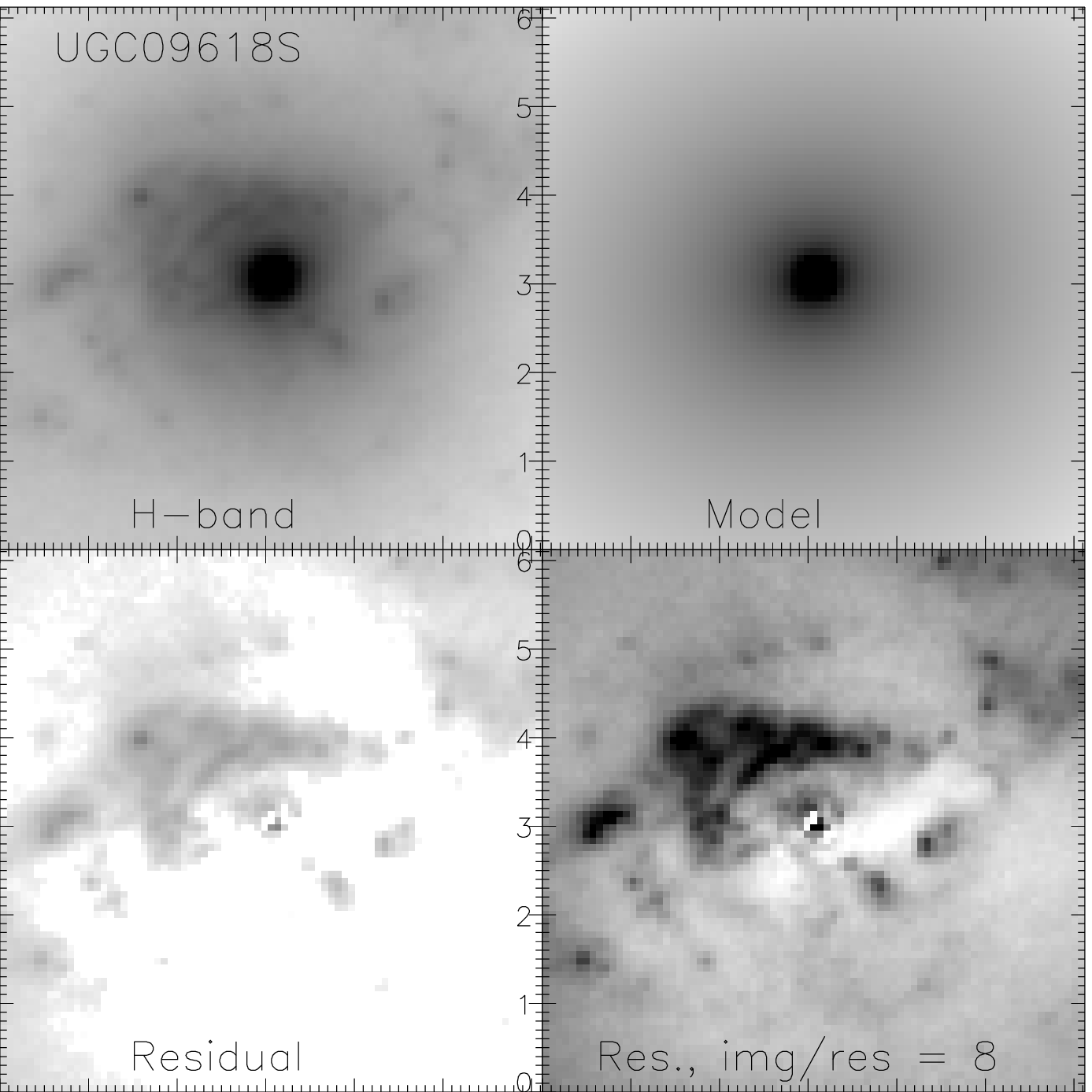}
\end{minipage}
\begin{minipage}[b]{1.0cm}
\end{minipage}
\begin{minipage}[b]{8cm}
\includegraphics[scale=0.56]{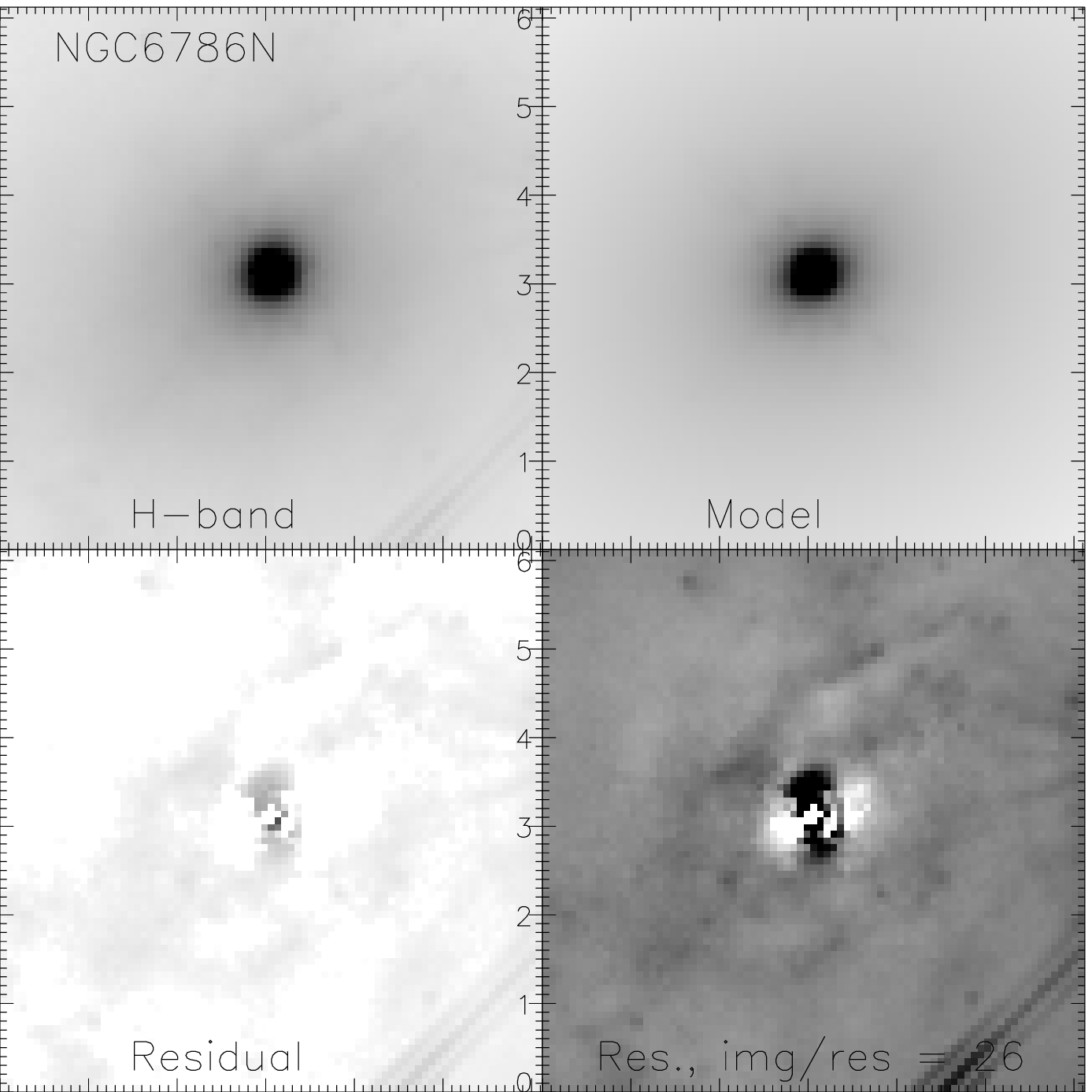}\\
\hspace{1.2cm}

\includegraphics[scale=0.56]{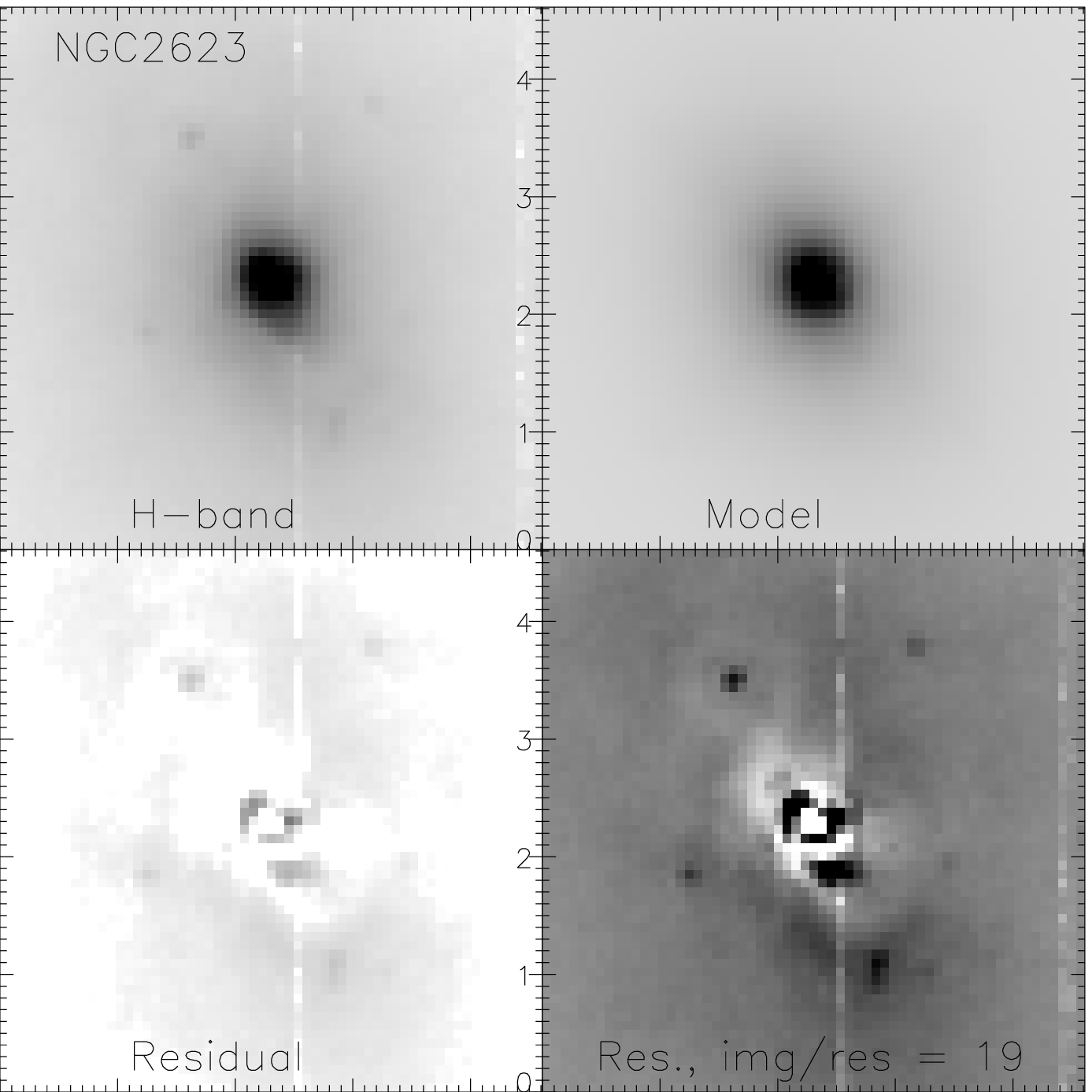}
\end{minipage}
\caption{Examples of GALFIT output images for NGC~3690 East, NGC~6786 North, UGC~9618 South, and NGC~2623 (scaled with the square root of intensity). In each panel: Top left: Region of NICMOS image that is fitted. Top right: Galfit model. Bottom left: Residual image (shows the difference between model and NICMOS image) with the same brightness level as the NICMOS image. Bottom right: Residual image scaled to its maximum brightness (the brightness ratio of NICMOS image to scaled residual map is shown as well at the bottom) to highlight the faint diffuse emission, spiral structure, and clusters remaining in the model-subtracted data. The axes are in units of arcsec and the figures are orientated in the observational frame.}
\label{fig:example_cusp}
\end{center}
\end{figure}

\begin{figure}
\begin{center}
\includegraphics[scale=0.49]{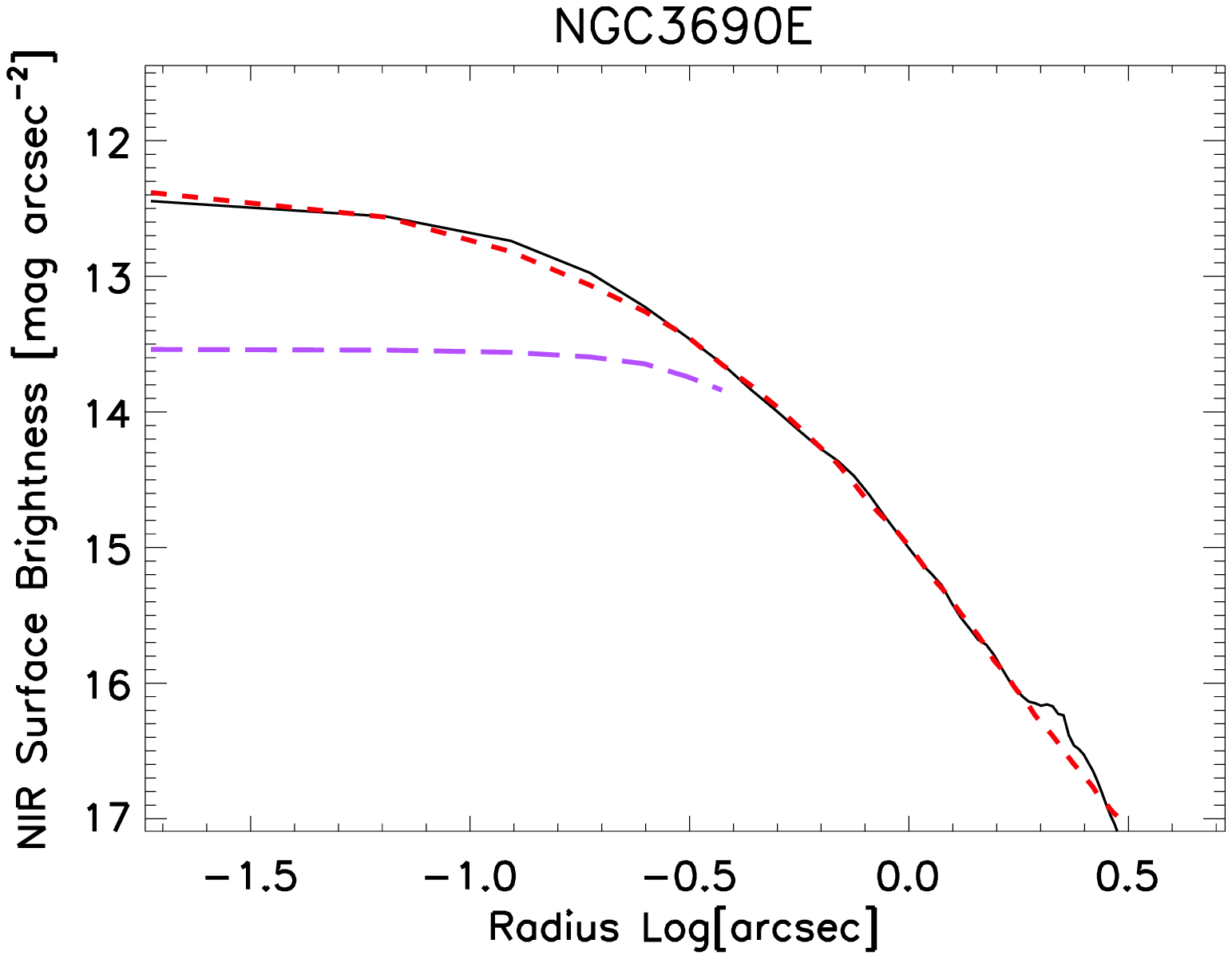}
\includegraphics[scale=0.49]{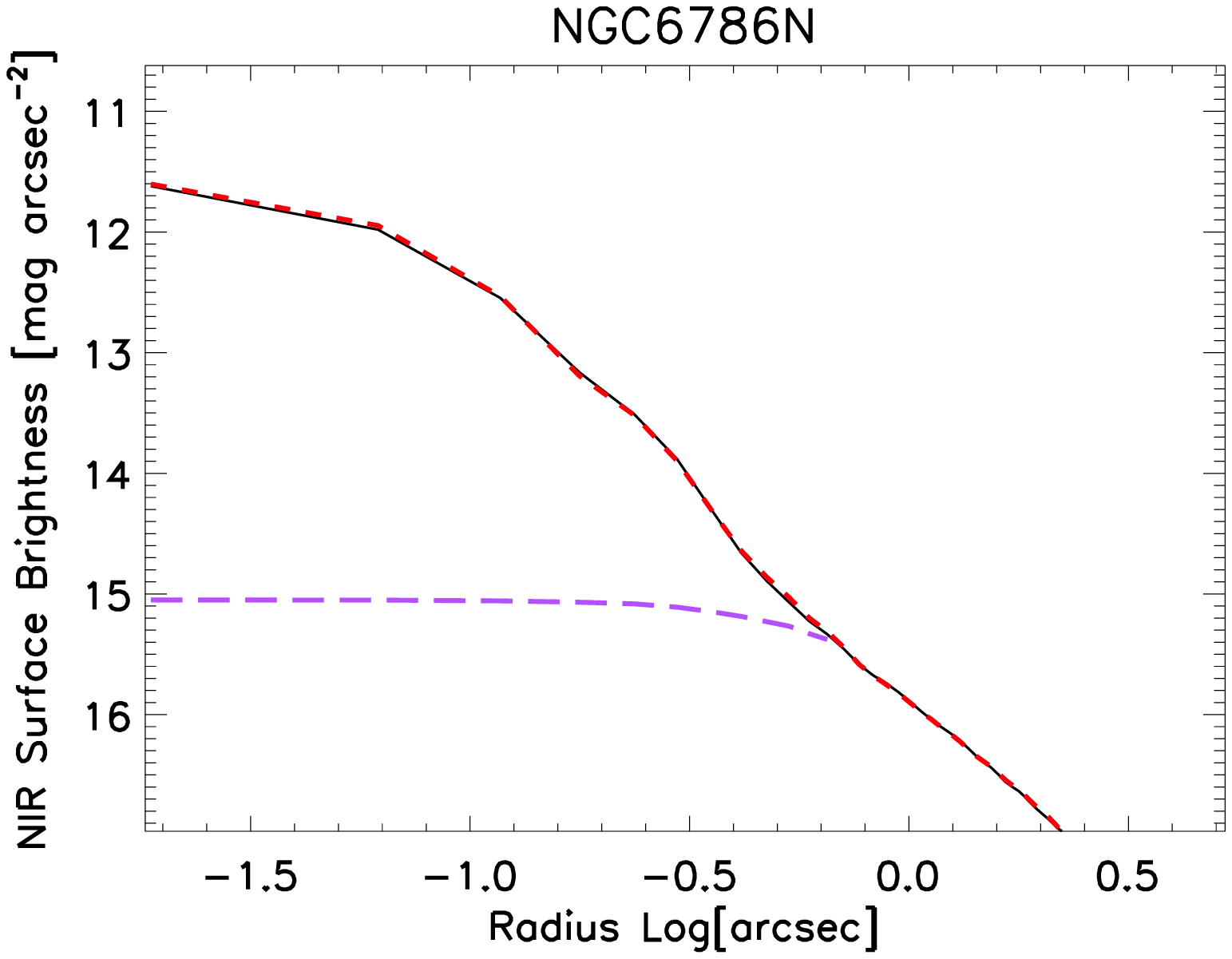}
\includegraphics[scale=0.49]{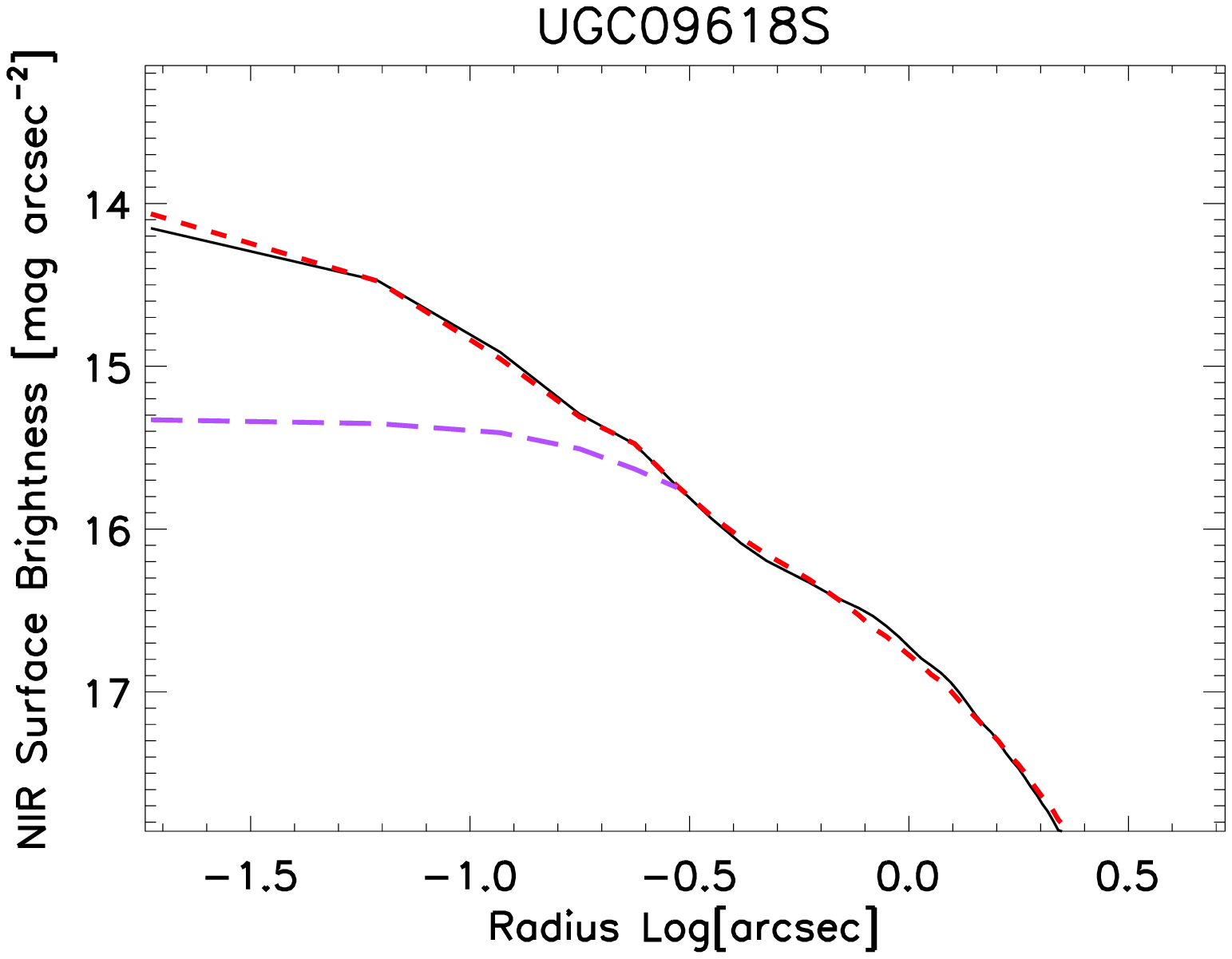}
\includegraphics[scale=0.49]{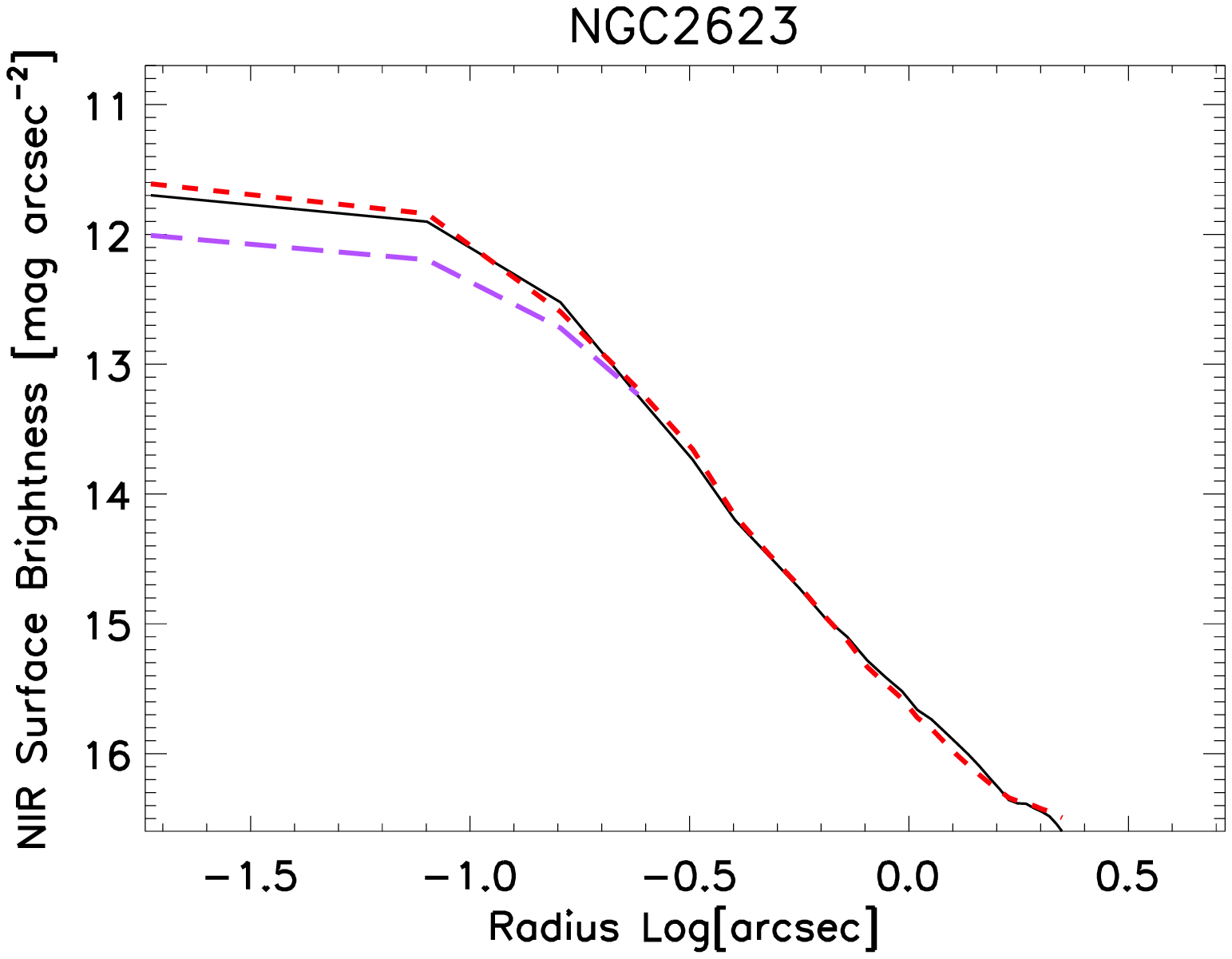}
\caption{The logarithmic radial profiles of NGC~3690 East (intermediate cusp source), NGC~6786 North (strong cusp source), UGC~9618 South (strong cusp source), and NGC~2623 (core source with PSF). The solid black line is the actual data as observed with HST NICMOS at 1.6$\mu$m and the short dashed red line the radial profile of the GALFIT 2-D model. The long dashed blue line shows a fit of the outer envelope with no cusp component, except for the core galaxy NGC~2623 where only the unresolved (PSF) component is excluded.}
\label{fig:example_profile}
\end{center}
\end{figure}

\begin{figure}
\begin{center}
\includegraphics[scale=1.0]{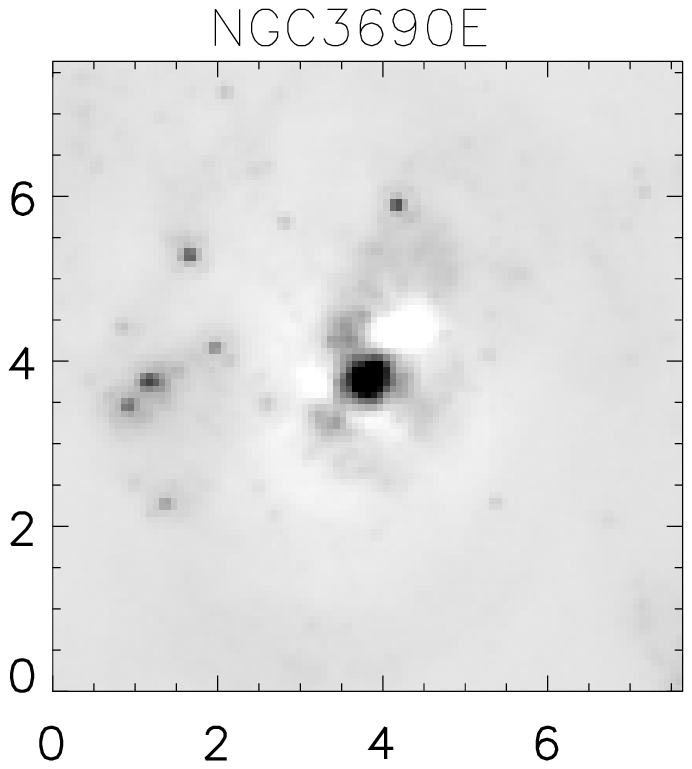}
\includegraphics[scale=1.0]{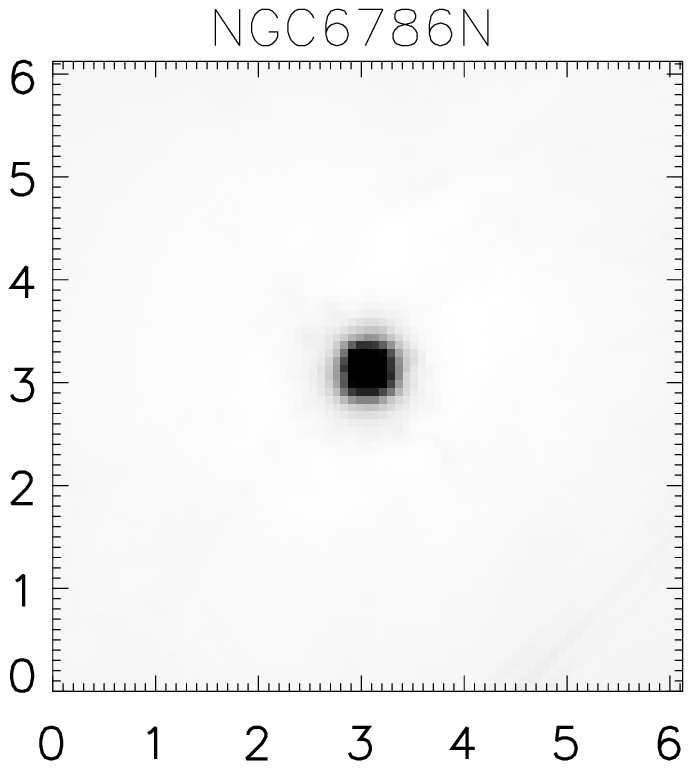}
\includegraphics[scale=1.0]{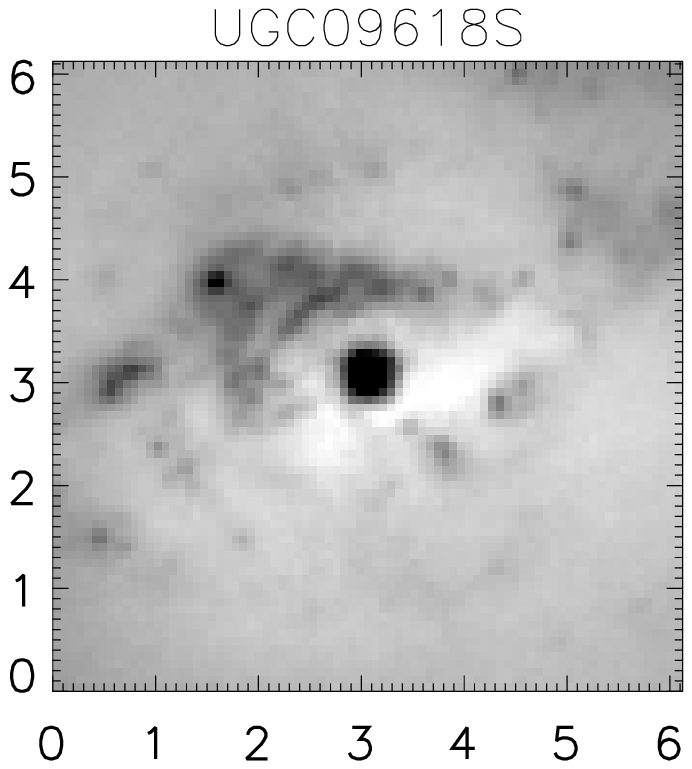}
\includegraphics[scale=1.0]{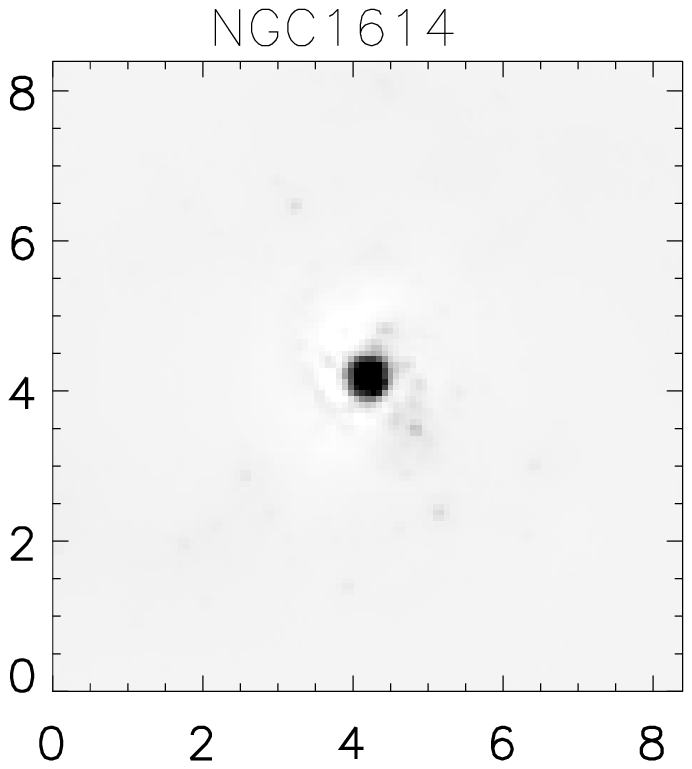}
\caption{The cusp NIR emission derived from the residual map of the observed data after subtracting the GALFIT model of the outer envelope with no cusp component. The example galaxies are NGC~3690 East, NGC~6786 North, UGC~9618 South, and NGC~1614. The axes are in units of arcsec and the figures are orientated in the observational frame. Residual emission that is spatially not associated with the cusp has typically a surface brightness of less than 10\%  cusp of the cusp surface brightness.}
\label{fig:example_cuspimage}
\end{center}
\end{figure}

Here we present four typical examples of our sample, ranging from strong cusp sources ($\gamma>1.$; NGC~6786 North, UGC~9618), to intermediate cusp ($\gamma\sim0.7$; NGC~3690 East) and core source with a strong unresolved component (NGC~2623) as shown in Fig.~\ref{fig:example_cusp}, Fig.~\ref{fig:example_profile}, and Fig.~\ref{fig:example_cuspimage}.

\newpage
\onecolumn 

\clearpage
\bsp
\label{lastpage}
\end{document}